\documentclass[11pt]{article}
\pdfoutput=1
\usepackage{jheppub}
\usepackage{float, extarrows, tikz-cd}
\usepackage{graphicx}
\usepackage{slashed}
\usepackage{tabularx,ragged2e}
\usepackage{amssymb}
\usepackage{subcaption}
\usepackage{amsmath,amssymb}
\usepackage{caption}
\usepackage{xcolor}
\usepackage{dsfont}
\usepackage{verbatim}
\usepackage{mathtools, xcolor,ytableau, amsfonts,tikz}
\usepackage{graphicx}
\let\olddiv\div

\usepackage{physics}
\newcolumntype{C}{>{\Centering\arraybackslash}X}

\numberwithin{equation}{section}

\newcommand{\mO}{\mathcal{O}}
\newcommand{\pd}{\partial}

\def\le{\left}
\def\ri{\right}
\newcommand\ov{\over}
\newcommand\p{\ensuremath{\partial}}
\newcommand{\es}[2] {\begin{equation} \label{#1} \begin{split} #2 \end{split} \end{equation}}

\def\<{\langle}
\def\>{\rangle}

\newcommand\ep{\varepsilon}

\newcommand\lam{\lambda}

\newcommand\Om{\Omega}
\newcommand\ga{{\ensuremath{{\gamma}}}}
\newcommand\Ga{{\ensuremath{{\Gamma}}}}
\newcommand\de{{\ensuremath{{\delta}}}}
\newcommand\De{{\ensuremath{{\Delta}}}}

\author[a,b]{Gabriel Cuomo,}   
\author[a,b]{Zohar Komargodski,}        
\author[a]{M\'ark Mezei }                           
              \affiliation[a]{Simons Center for Geometry and Physics, SUNY, Stony Brook, NY 11794, USA}      
              \affiliation[b]{C. N. Yang Institute for Theoretical Physics, Stony Brook University, Stony Brook, NY 11794, USA}                                                              
                                            
\emailAdd{gcuomo@scgp.stonybrook.edu}					 \emailAdd{zkomargodski@scgp.stonybrook.edu}  
     \emailAdd{mmezei@scgp.stonybrook.edu}

\title{Localized magnetic field in the \texorpdfstring{$O(N)$}{O(N)} model}

\abstract{
We consider the critical $O(N)$ model in the presence of an external magnetic field localized in space.  This setup can potentially be realized in quantum simulators and in some liquid mixtures. The external field can be understood as a relevant perturbation of the trivial line defect,  and thus triggers a defect Renormalization Group (RG) flow.  In agreement with the $g$-theorem,  the external localized field leads at long distances to a stable nontrivial defect CFT (DCFT) with $g<1$. We obtain several predictions for the corresponding DCFT data in the epsilon expansion and in the large $N$ limit.  The analysis of the large $N$ limit involves a new saddle point and, remarkably, the study of fluctuations around it is enabled by recent progress in AdS loop diagrams.  Our results are compatible with results from Monte Carlo simulations and we make several predictions that can be tested in the future.}

\begin{document}

\maketitle

\section{Introduction and summary}

Line operators play several roles in quantum field theory. 
First, in the context of zero temperature quantum physics, they describe point-like impurities in space. Second, the expectation values of line operators are used to diagnose phases of theories \cite{Wilson:1974sk}. Finally, topological line operators can be interpreted as (potentially non-invertible) symmetry generators, see \cite{Chang:2018iay} and references therein.

Historically, line operators played a very important role in the development of Quantum Field Theory (QFT). The Kondo problem that arose in the study of magnetic impurities in metals, has paved the way to the renormalization group and also to important developments in integrability; see \cite{Affleck:1995ge} for a nice review. 

Our focus here will be on line operators in theories which are critical (conformal) in the bulk. Therefore, our setup is a bulk CFT in $d$ spacetime dimensions and a one-dimensional line operator. The line operator undergoes a renormalization group flow so that at long distances one  expects to typically find a critical line operator. When the bulk is given by a $d$ dimensional CFT and the line operator is conformal as well, the full system with an insertion of a single straight line operator preserves $$SL(2,\mathbb{R})\times SO(d-1)$$ symmetry.\footnote{A conformal line operator may have transverse spin but we will not discuss such examples here.} This is the definition of a Defect Conformal Field Theory (DCFT).

In general, one can consider line operators which in the ultraviolet are described by a certain DCFT$_\text{UV}$ and in the infrared by some different DCFT$_\text{IR}$. The renormalization group flow is triggered by a relevant  defect operator with $\Delta<1$  in the ultraviolet.\footnote{Or a marginally relevant defect operator with $\Delta=1$.} Remember that throughout this process, the bulk remains at the same fixed point. 

Following classic results in $d=2$ \cite{Affleck:1991tk,Friedan:2003yc,Casini:2016fgb}, it was recently understood that there is a renormalization group monotone that governs such flows in any $d$ \cite{Cuomo:2021rkm} (see \cite{Beccaria:2017rbe,Kobayashi:2018lil} for previous discussions). Denote the expectation value of the circular loop of radius $R$ by $g(MR)$. 
Here $M$ is some mass scale associated to the renormalization group flow on the defect. 
 A subtlety is that both in the ultraviolet and in the infrared a cosmological constant term on the defect, which can be thought of as the defect mass, can contribute a term linear in $R$ to $g(MR)$.  This introduces some slight scheme dependence into $g(MR)$.
 This scheme dependence can be taken care of by acting with the operator $\left(1-R{\partial\over \partial R}\right)$, and defining a defect entropy function by 
 \es{sDef}{
s(MR)=\left(1-R{\partial\over \partial R}\right)\log g (MR)\,.
} 
$s(MR)$ is indeed a scheme independent physical observable. This turns out to be monotonically decreasing with $R$ and satisfies a gradient formula. In $d=2$ the defect entropy $s(MR)$ controls multiple physical quantities besides the expectation value of circular defects: It encodes  a universal contribution from the impurity to the thermodynamic entropy, and its fixed point values were conjectured by Affleck and Ludwig to obey $g_\text{UV}\geq g_\text{IR}$ \cite{Affleck:1991tk}.\footnote{For the inequality $g_\text{UV}\geq g_\text{IR}$ to make sense, one potentially needs to subtract a linear in $R$ cosmological constant term, i.e. one should really write $s_\text{UV}\geq s_\text{IR}$. We are sometimes careless about the distinction between $g$ and $s$ at fixed points.} In $d=2$ (and only in $d=2$), $g$ can also be viewed as the contribution to the vacuum entanglement entropy due to the impurity.

In this note we explore a particular class of renormalization group flows on line defects which correspond to activating external fields. These are perhaps the simplest possible line defects in QFT. Suppose that DCFT$_\text{UV}$ is completely trivial. That means that the line operator is just the unit operator. On such a defect, the defect operators are just the usual bulk operators restricted to the defect. Hence, nontrivial renormalization group flows can be triggered by integrating bulk operators with $\Delta<1$ on a line:
\begin{equation}\label{genl}
e^{-h\int d\tau \, {\mathcal{O}(\tau)}  }~,
\end{equation}
where $h$ is some coefficient with positive mass dimension. The infrared limit of this line defect is necessarily nontrivial since the function $s(h^\gamma R)$ is monotonically decreasing with $R$ (the exponent is given by $\gamma={1\over 1-\Delta}$). For $R\to 0$ we have $s=0$ as appropriate for trivial line defect and for $R\to \infty$ we must have $s<0$ and hence the line defect must be  nontrivial.

The line defect~\eqref{genl} can be experimentally realized in a lattice system by turning on the background field coupling to $\mathcal{O}$ on a few neighboring lattice sites, so that it is localized in space. This then leads to some behavior at long distances (and long times) which should be consistent with the properties of the infrared DCFT. At intermediate distances and times one can probe the renormalization group flow from the trivial DCFT$_\text{UV}$ to the nontrivial DCFT$_\text{IR}$.  We expect the DCFT$_\text{IR}$ to be stable, namely not to have any relevant deformation. Hence the defect renormalization flow will reach the IR fixed point without any tuning.

Defects of the kind \eqref{genl} that additionally break the internal symmetry group of the theory are also of particular interest in Monte Carlo simulations of systems exhibiting second order phase transitions. This is because introducing a symmetry breaking defect improves the precision in the analysis of the ordered phase, and in the detection of the critical point \cite{Assaad:2013xua}. In this context, it is particularly relevant to identify the dimension of the leading irrelevant perturbation in the DCFT$_\text{IR}$ that controls the corrections to scaling at long distances  \cite{2017PhRvB..95a4401P}.

In Wilson-Fisher theories there is a natural candidate for a line operator~\eqref{genl}, where $h$ is the background magnetic field coupling to the order parameter $\phi$. Consider the $O(N)$ symmetric Wilson Fisher theories in $2<d\leq 4$, where the fundamental field is $\phi_a$, with $a=1,...,N$. Then we can consider the line operator 
\begin{equation}\label{WFl}
e^{-h\int d\tau {\phi_1(\tau)}  }~,
\end{equation}
where we have chosen the preferred direction in internal space ``1'', breaking the symmetry to $O(N-1)$ in the presence of this defect.

For instance, in the realization of the $O(3)$ Wilson-Fisher critical theory as a quantum antiferromagnet of spin $1/2$ particles on a square lattice, the line operator~\eqref{WFl} naturally arises by introducing a staggered background magnetic field over a few lattice sites, see \cite{sachdev2008quantum} for a review.  It was pointed out in \cite{2017PhRvB..95a4401P} that for $N=1$ the defect \eqref{WFl} is also experimentally realizable  in a mixture of two liquids at the demixing critical point by introducing a suitably shaped colloidal impurity  \cite{LAW2001159,2003svcm.book..237F}. The $N=1$ (Ising) quantum critical point has also been recently realized in a large-scale programmable quantum simulator made of neutral atoms \cite{Ebadi:2020ldi}.\footnote{We thank Subir Sachdev for discussions.} Therefore we expect that some of our predictions should be experimentally testable in $d=3$.

To put our work in context,  the defect \eqref{WFl} was studied via Monte Carlo simulations in \cite{2014arXiv1412.3449A,2017PhRvB..95a4401P}. As we will see, the results of these works are nicely compatible with ours.   Previous field theoretical studies appeared in \cite{PhysRevLett.84.2180,Allais:2014fqa},\footnote{We thank Simone Giombi for bringing ref.~\cite{Allais:2014fqa} to our attention.} but their results are inconclusive.  The findings of \cite{PhysRevLett.84.2180} rely on an unjustified mean field approximation and differ from ours. In \cite{Allais:2014fqa} the authors considered the one-point function of the bulk order parameter in the presence of the defect \eqref{WFl} in the epsilon expansion and at large $N$, as we will also do here. They reached the surprising conclusion that these two simplifying limits do not commute. As we explain in the summary we reach a different conclusion: while we agree with their $\varepsilon$-expansion results (that we also extend to several other observables), we find different results at large $N$ and conclude that the two limits commute. 

We also mention that several other papers previously considered line defects in the $O(N)$ model. These mostly focused on spin impurities, which involve additional degrees of freedom localized on the line \cite{sachdev1999quantum,vojta2000quantum,PhysRevB.61.4041,Sachdev:2001ky,Sachdev:2003yk,PhysRevLett.96.036601,PhysRevB.75.224420,Liu:2021nck}. Other works focused on symmetry (twist) defects \cite{Billo:2013jda,Gaiotto:2013nva,Bianchi:2021snj,Giombi:2021uae,Gimenez-Grau:2021wiv} (which are not genuine line defects, since they are attached to a nontrivial topological surface).  Finally, we note that correlation functions with two insertions of the line defect \eqref{WFl} were recently considered in \cite{Soderberg:2021kne} for a free bulk theory and, after an earlier version of this work appeared, in \cite{Rodriguez-Gomez:2022gbz} for the bulk interacting theory.

\subsection{Summary}

The main aim of this paper is to shed light on the infrared properties of the defect~\eqref{WFl}. First, some general observations about the infrared DCFT are in order:

\begin{itemize}
\item Since $g_\text{IR}<1$ and the infrared defect is nontrivial, it must have two protected defect operators: the displacement operator of dimension $2$ and since the $O(N)$ symmetry is explicitly broken to $O(N-1)$ symmetry by the defect it must also have a ``tilt'' operator in the vector representation of $O(N-1)$ of dimension 1. 
\item We expect the DCFT$_\text{IR}$ to have no relevant operators. 
\item The problem makes sense in $d = 2$, where, for instance for $N=1$,~\eqref{WFl} is the integrated spin operator of the Ising$_{2d}$ model, where we can identify the infrared DCFT with one of the known conformal interfaces of the Ising$_{2d}$ model. For $N\geq 2$, we mostly leave the details of the $d=2$ limit to the future, but we will quote some results which we will use.
\item  In the $d=4$ case  the bulk is free and one can easily solve for the renormalization group flow of~\eqref{WFl}.
\end{itemize}

 Beyond the aforementioned special cases one does not hope for an exact solution, although we will see that the $N\to\infty$ limit admits an almost analytical solution.

We attack the problem on two fronts: the epsilon  and the large $N$ expansion. For the epsilon expansion we use the fact that the perturbation of the trivial DCFT$_\text{UV}$ is only weakly relevant in $d=4-\varepsilon$ dimensions with $\varepsilon \ll1$. This allows us to compute many properties of DCFT$_\text{IR}$ perturbatively. 
 For the large $N$ limit in generic $d$  we find that the line operator~\eqref{WFl} leads to a {\it new classical RG trajectory}\footnote{The saddle point is different than the one proposed in \cite{Allais:2014fqa}. This difference is the origin of the previously mentioned discrepancy between our results and the ones of \cite{Allais:2014fqa}.} which we analyze in great detail, including the fluctuations around this new classical trajectory.  This leads to an explicit determination of several scaling dimensions and one-point functions of the DCFT$_\text{IR}$.

Let us now summarize the results in these two limits, which are nicely consistent with each other and lead to a coherent picture for the properties of the infrared DCFT for finite $N$ and $2<d<4$. 

\begin{figure}[t!]
\centering
\includegraphics[scale=0.8]{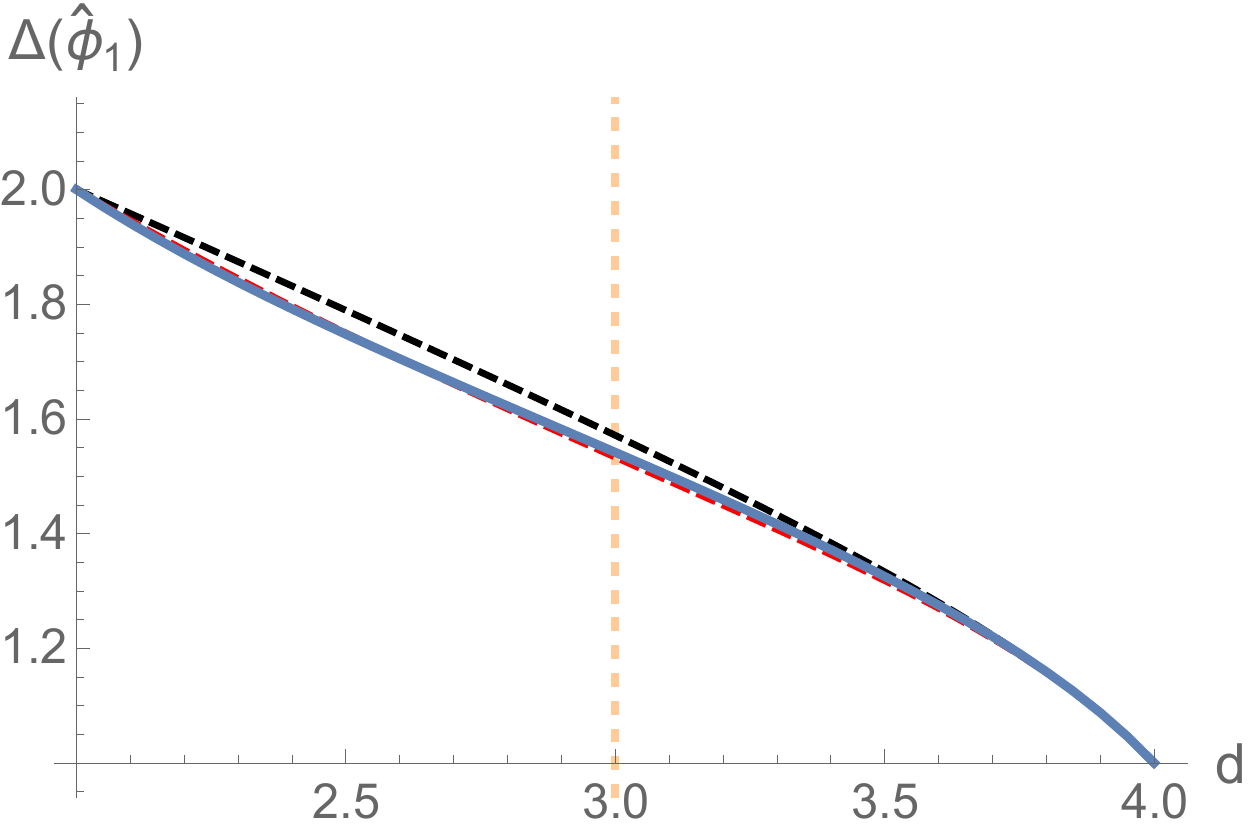}
\caption{We plot the leading order in large $N$ result for $\Delta(\hat{\phi_1})$ (blue) as a function of the dimension $d$. We compare it to two Pad\'e approximants (red and black dashed) obtained from the combination of the large $N$ limit of the two loop $4-\ep$ expansion result and of the $2+\tilde\ep$ expansion result that $\De(\hat\phi_1)\vert_{2d}=2$. }\label{fig:intro}
\end{figure}

Our most precise prediction concerns with the scaling dimension of the lowest dimension $O(N-1)$ singlet nontrivial operator in the DCFT$_\text{IR}$, which is   associated with the operator $\phi_1$ defining the defect \eqref{WFl}. In the epsilon expansion we obtain our most precise estimate in sec.~\ref{SubSecD2} by interpolating with a Pad\'e approximant in between the result of a two-loop calculation in the epsilon expansion and the exact value in $d=2$.  
We also compute the leading large $N$ result in sec.~\ref{SubSecNspectrum}.  For this analysis, the simplest description is provided by mapping the flat space DCFT to AdS$_2\times S^{d-2}$ through a Weyl transformation. The most sophisticated computation that we perform involves a scalar bubble diagram, whose evaluation is made possible by recent breakthroughs in AdS loop diagrams~\cite{Fitzpatrick:2010zm,Penedones:2010ue,Fitzpatrick:2011hu,Giombi:2017hpr,Carmi:2018qzm,Carmi:2019ocp}.

Denoting with $\hat{\phi}_1$ the corresponding defect operators, our findings in $d=3$ are:
\begin{equation}\label{Eq_Result}
\Delta(\hat{\phi}_1)=\begin{cases}
1.55 \pm 0.14&\varepsilon-\text{exp.}\,,\\
1.542...  &N\rightarrow \infty\,.
\end{cases}
\end{equation}
The first line in eq.  \eqref{Eq_Result} is the result of the epsilon expansion and the $d=2$ data alluded to above, while the second line follows from the large $N$ expansion.

We may compare the $N=1$ result with the formerly mentioned Monte Carlo studies \cite{2014arXiv1412.3449A,2017PhRvB..95a4401P}:
\begin{equation}\label{Eq_ResultMC}
\text{Monte Carlo:}\qquad
\Delta(\hat{\phi}_1)\stackrel{N=1}{=}
\begin{cases}
1.60(5) & $\cite{2017PhRvB..95a4401P} $ \\
1.52(6) & $\cite{2017PhRvB..95a4401P} $ \\
1.40(3) & $\cite{2014arXiv1412.3449A} $\,.
\end{cases}
\end{equation}
The first two values refer to the two measures reported in \cite{2017PhRvB..95a4401P} and nicely agree with our results (and with each other), while the last line was obtained in \cite{2014arXiv1412.3449A}\footnote{We extracted the result by identifying the corrections to scaling in eq.~(16) of \cite{2014arXiv1412.3449A} with the effect of the leading irrelevant operator as explained in \cite{2017PhRvB..95a4401P}.} and is compatible with our estimated uncertainty in \eqref{Eq_Result}, but in slight tension with the results of \cite{2017PhRvB..95a4401P}.

The result in eq. \eqref{Eq_Result} is independent of $N$ to the reported precision. This observation is broadly consistent with preliminary Monte Carlo results for $N=3$ reported in \cite{2017PhRvB..95a4401P}, which suggest $\Delta(\hat{\phi}_1)\approx 1.3 \olddiv 1.6$.
In figure~\ref{fig:intro} we also compare the large $N$ result with that of a two-loop calculation in the epsilon expansion, demonstrating  their non-trivial perfect agreement in the overlapping regime of validity.

In secs.~\ref{SubSecOther},~\ref{SubSecPhi1N} and~\ref{SubSecNspectrum} we present results concerning one-point functions of bulk primary operators as well as the scaling dimensions of other defect operators, to one-loop order in the epsilon expansion and to leading order in the large $N$ limit.  Whenever a comparison was possible, we found perfect agreement. As an illustration,  for the normalized one-point function of the operator $\phi_1$ we obtained:
\begin{equation}\label{eq_ResultA}
\frac{\langle\phi_1(0,\mathbf{x})\rangle}{\sqrt{\langle\phi_1(\infty)\phi_1(0)\rangle_{h=0}}}=\frac{a_{\phi}}{|\mathbf{x}|^{\Delta(\phi)}}\,,
\end{equation}
where $|\mathbf{x}|$ is the distance from the defect, $\Delta(\phi)$ the (bulk) scaling dimension of $\phi_a$ and
\begin{equation}
a_{\phi}^2=\begin{cases}\displaystyle
\frac{N+8}{4}+\varepsilon
\frac{  (N+8)^2 \log 4+N^2-3 N-22}{8 (N+8)}+\mO\left(\varepsilon^2\right)
&d=4-\varepsilon\,,\\
\displaystyle 0.55813\,N+ \mO\left(N^0\right) & d=3\text{ and }N\rightarrow \infty\,.
\end{cases}
\end{equation}
Incidentally, we notice that the large $N$ limit of the $\varepsilon$-expansion result is numerically close to the correct one all the way to $\varepsilon=1$ (see fig.~\ref{fig:aphi1}). Unfortunately there are no Monte Carlo results available for comparison of the normalized one-point function \eqref{eq_ResultA} (or other observables).\footnote{The unnormalized one-point function was studied in \citep{2014arXiv1412.3449A} to validate the DCFT scaling law \eqref{eq_ResultA}; unfortunately, the normalization of the two-point function in the absence of the defect is not reported.}

Finally we also studied the $g$-function of the defect. In sec.s~\ref{SubSecEpResults} and~\ref{SubSecNg} we obtain that its value at the fixed point is
\begin{equation}
\log g=\begin{cases}
\displaystyle-
\frac{N+8}{16}\varepsilon+\mO\left(\varepsilon^2\right) &d=4-\varepsilon\\[0.6em]
-  0.153673 \,N+\mO\left(N^0\right) & d=3 \text{ and }N\rightarrow\infty
\end{cases}
\end{equation}
$\log g$ is negative in agreement with the $g$-theorem. Notice that for $N\rightarrow\infty$ we get that $g\to0$ exponentially.

We finally mention that in Appendix \ref{AppAway} we analyze the defect in the ordered phase, which is gapless for $N>1$. There we show that, in $d=3$, the defect coupling is marginally irrelevant and thus leads to a logarithmic correction to the one-point function of the bulk order parameter, as well as to other defect correlators. These findings might be relevant for future Monte Carlo studies utilizing symmetry breaking defects as in \cite{Assaad:2013xua,2017PhRvB..95a4401P}. It would be nice to observe this logarithmic correction in the future.

The rest of the paper is organized as follows. In sec.~\ref{SecWarmUp} we study the defect \eqref{WFl} when the bulk theory is given by a free massless scalar, in which case the theory can be solved exactly. In sec.~\ref{SecEpsilon} we study the DCFT for the Wilson-Fisher fixed point in the $\varepsilon$-expansion. In sec.~\ref{SecLargeN} we finally discuss the large $N$ limit of the DCFT. We conclude with a brief outlook in sec.~\ref{sec:outlook}. In appendix~\ref{AppAway} we discuss the case of a non-conformal ordered bulk, while appendix~\ref{App_G_eps} contains technical details on the calculation of the $g$-function in $4-\varepsilon$ dimensions.

\section{Warm-up: localized magnetic field in free massless theory}\label{SecWarmUp}

Perhaps the simplest possible nontrivial line defect in QFT is (in Euclidean signature)
\begin{equation}\label{scalard}
e^{-h\int d\tau \phi}~,
\end{equation}  
where $\phi$ is a free field in the bulk with action $S=\int d^dx\, \frac12(\partial\phi)^2$
and $h$ has dimensions of mass$^{2-d/2}$, which means that $h$ is a relevant perturbation of the trivial line defect in $d<4$ and an irrelevant perturbation of the trivial line defect in $d>4$. In $d=4$ it is exactly marginal \cite{Kapustin:2005py}. For a general approach to conformal defects in free scalar theories see \cite{Lauria:2020emq,Nishioka:2021uef}.

It is straightforward to solve for the RG flow triggered by $h$ in $d<4$ since the bulk is free. We consider the action 
\begin{equation}
S=\int d^dx \,\frac12(\partial\phi)^2+h\int_\gamma d\sigma \phi~,
\end{equation}
where $\gamma^\mu(\sigma)$ is the worldline of the impurity and $\sigma$ is a normalized coordinate on the worldline. In $d>2$, the equation of motion is solved by 
\begin{equation}
\phi_{cl}(x)=-\frac{h}{(d-2)\Omega_{d-1}}\int_\gamma d\sigma \frac{1}{ |x-\gamma(\sigma)|^{d-2}}~,\qquad \Omega_{d-1} = {2\pi^{d/2} \over \Gamma(d/2)}~,
\end{equation}
where $\Omega_{d-1}$ is the volume of the $d-1$-dimensional sphere.

Since the theory is free, the fluctuations around $\phi_{cl}(x)$ are insensitive to the existence of the line defect. Therefore, if we consider a circular line defect of radius $R$ we can compute the defect partition function, normalized by the partition function of the theory without the defect, as a function of $R$ by simply plugging the classical solution back into the action. 
We therefore find 
\begin{equation}\label{freeg}
\begin{split}
g&\equiv\log \left(Z_{{\rm defect}}/ Z_{\rm{bulk}}\right)
 ={\pi h^2R^{4-d}\over (d-2)\Omega_{d-1}}\int_0^{2\pi} d\phi {1\over \left[4\sin^2(\phi/2) \right]^{d/2-1}}
\\[0.55em]
&=\pi h^2R^{4-d}\frac{  \Gamma \left(\frac{3}{2}-\frac{d}{2}\right) \Gamma \left(\frac{d}{2}-1\right)}{2^{d-1}\pi^{\frac{d-1}{2}}\Gamma \left(2-\frac{d}{2}\right)}\,.
\end{split}
\end{equation}
The pole in $d=3$ can be renormalized by the cosmological constant counterterm since this divergence is linear in $R$. This divergence in $d=3$ will cancel out from the defect entropy, which we will soon compute.\footnote{More can be said about the special case of $d=3$. Since the renormalization group flow on the defect is triggered by an operator of dimension $1/2$, this example admits a couplings-space conformal anomaly of the type studied in~\cite{Gomis:2015yaa,Schwimmer:2018hdl,Schwimmer:2019efk}. Indeed, the radius dependence of $g$ in $d=3$ contains, in addition to the pure cosmological constant counterterm, a physical piece $\sim h^2 R\log R$, which upon rescaling $R$, due to the logarithm, generates the cosmological constant counterterm. This is the signature of a conformal anomaly in the space of couplings.  We thank A. Schwimmer and S. Theisen for suggesting this point of view. }  The pole in $d=5$ on the other hand must be renormalized by an extrinsic curvature counterterm, which is allowed in $d=5$, since there we have perturbed the trivial line defect by an irrelevant operator.

\begin{figure}
\centering
\includegraphics[scale=0.8]{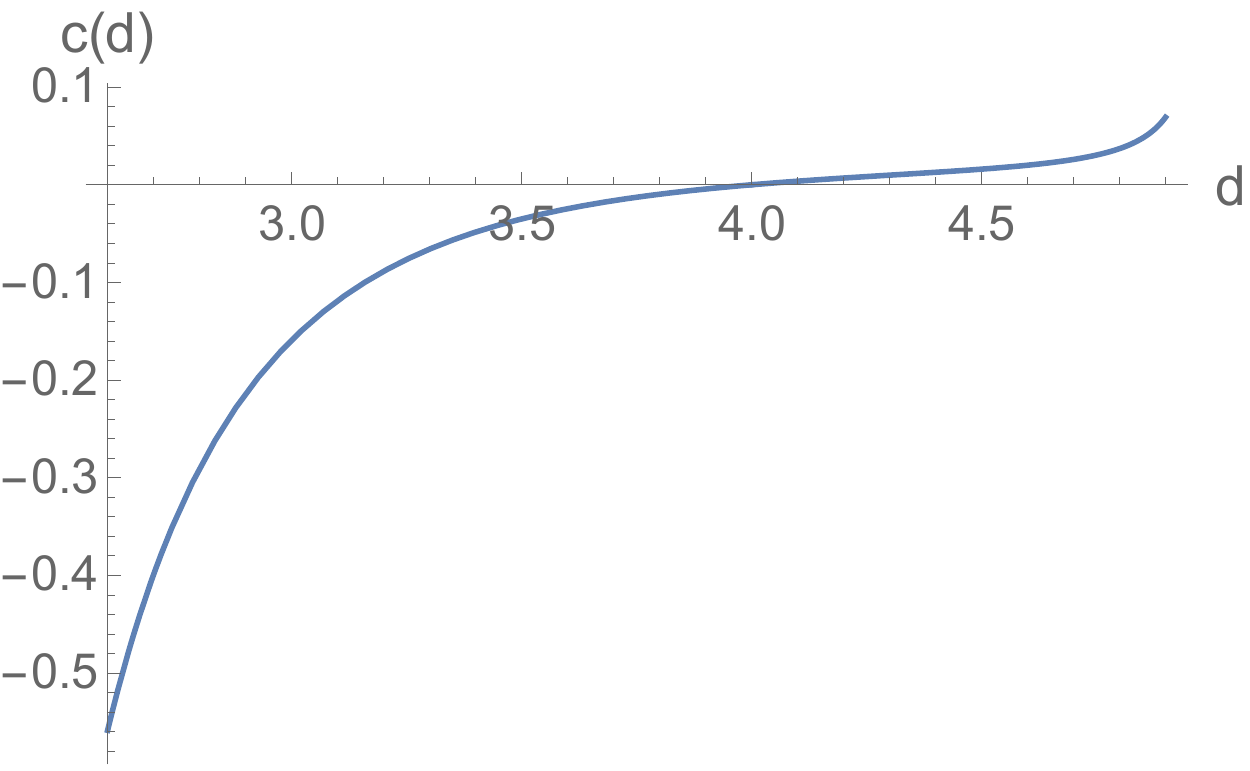}
\caption{Plot of the function $c(d)$ (in red) in eq.~\eqref{dent}.}
\label{plotCd}
\end{figure}

We can extract the defect entropy $s$ from $\log g$ by applying the operator $\le(1-R{\pd\over \pd R}\ri)$, hence, we find 
\begin{equation}\label{dent}
s=\pi h^2R^{4-d}c(d)\,,
 \end{equation}
where we defined
\begin{equation}
c(d)\equiv-\frac{  \Gamma \left(\frac{5}{2}-\frac{d}{2}\right) \Gamma \left(\frac{d}{2}-1\right)}{2^{d-2}\pi^{\frac{d-1}{2}}\Gamma \left(2-\frac{d}{2}\right)}\,.
\end{equation}
The function $c(d)$ is shown in fig.~\ref{plotCd}. It is negative for $d<4$, it vanishes in $d=4$ and it is positive for $4<d<5$; at $d=5$ it has a pole associated with the aforementioned extrinsic curvature counterterm, that, unlike the cosmological constant,  is not subtracted by the differential operator $\le(1-R\frac{\pd}{\pd R}\ri)$.
For concreteness, in $d=3$ we find $s=-{1\over 2\pi^{3/2}}
\pi h^2R$. We see that the defect entropy in the ultraviolet ($R\to0$) is vanishing as expected (since the line defect is trivial) and in the infrared ($R\to\infty$) $s\to-\infty$ which is consistent with the $g$-theorem \cite{Cuomo:2021rkm}, as $s$ monotonically decreases. Somewhat pathologically, the flow never terminates in an infrared DCFT. The same is true for all $2<d<4$. This  behavior of the renormalization group flow on the line defect is perhaps only possible in theories with a moduli space of bulk vacua.  In CFTs with finitely many degrees of freedom and a single vacuum, we expect that the renormalization group flow on line defects will  terminate in a healthy DCFT with $g>0$ (or $s=$finite).\footnote{Perhaps relatedly, lower (and upper) bounds on $g$ in $2d$ BCFTs were obtained in \cite{Friedan:2012jk,Collier:2021ngi}, under the assumption of a sufficiently large gap in the spectrum of bulk operators.} We note that in large $N$ theories a natural behavior is $s= - N\,s_0+\dots$ (an example of which we will see soon), and hence formally $s\to -\infty$ as $N\to \infty$. This divergence is however unrelated to that of the free theory and we would not characterize it as pathological.

The situation in $d=4$ is that the defect entropy $s$ is independent of $h$ and we get $s=0$ for all $h$. In $d=4$ $h$ should be thought of as an exactly marginal parameter on the defect and the fact that $s$ is independent of it is consistent with the $g$-theorem.
Various other quantities, such as the energy-momentum tensor away from the defect and the one-point function of $\phi$ do depend on $h$ \cite{Kapustin:2005py} and hence this exactly marginal parameter is not entirely trivial. 

We end this section by providing an explicit check of the gradient formula recently proven in \cite{Cuomo:2021rkm}. This states that the dependence of the defect entropy on the size of the loop $R$, or equivalently on the RG scale, is determined by the following equation:
\begin{equation}\label{eq_gradient_formula_WarmUP}
R\frac{\pd s}{\pd R}=-\int  d\sigma_1 \int d\sigma_2\,\langle T_D(\sigma_1) T_D(\sigma_2)\rangle_c\left[1-\cos\left(
\frac{\sigma_1-\sigma_2}{R}\right)\right]\,,
\end{equation}
where $T_D$ is the defect stress tensor and $\langle T_D(\sigma_1) T_D(\sigma_2)\rangle_c$ is its connected two-point function.  In the simple model \eqref{scalard} the defect stress tensor and its two-point function can be computed exactly:
\begin{equation}\label{eq_gradient_check_WarmUp}
T_D(\sigma)=\beta_{h}\phi\left(\gamma(\sigma)\right)\quad
\implies\quad
\langle T_D(\sigma_1) T_D(\sigma_2)\rangle_c=\frac{\beta_{h}^2}{(d-2)\Omega_{d-1}R^{d-2}}\, \frac{1}{\left[4\sin^2\left(\frac{\phi_1-\phi_2}{2}\right)\right]^{\frac{d-2}{2}}}\,,
\end{equation}
where $\beta_{h}=\frac{d-4}{2}h$ is the beta function of the coupling.  Using eq.~\eqref{eq_gradient_check_WarmUp} to evaluate the integral on the right hand side of eq.~\eqref{eq_gradient_formula_WarmUP} we find:
\begin{equation}
-\int  d\sigma_1 \int d\sigma_2\,\langle T_D(\sigma_1) T_D(\sigma_2)\rangle_c\left[1-\cos\left(
\frac{\sigma_1-\sigma_2}{R}\right)\right]=(4-d)\pi h^2 R^{4-d}c(d)\,.
\end{equation}
Using this result one easily sees that the defect entropy \eqref{dent} indeed satisfies eq.~\eqref{eq_gradient_formula_WarmUP}.

\section{Epsilon expansion results}\label{SecEpsilon}

\subsection{The bulk fixed point of the \texorpdfstring{$O(N)$}{O(N)} model}

We consider the $O(N)$ Wilson-Fisher model in $d=4-\varepsilon$ dimensions. The action reads
\begin{equation}\label{eq_BulkAction_epsilon}
S=\int d^dx\left[\frac{1}{2}(\pd\phi_a)^2+\frac{\lambda_0}{4!}\left(\phi_a^2\right)^2\right]\,.
\end{equation}
We tuned the mass term to zero to focus on the critical point.\footnote{In the dimensional regularization scheme that we will use here the critical point appears when there is no bare mass term in~\eqref{eq_BulkAction_epsilon}. In other regularization schemes a bare mass term could be necessary to tune to the critical point.}  The model can be studied by  expanding perturbatively in the coupling $\lambda_0$ and using the propagator
\begin{equation}\label{eq_propagator_free}
\langle \phi_a(x)\phi_b(0)\rangle_{\lambda=0}=
\frac{\delta_{ab}}{(d-2)\Omega_{d-1}|x|^{d-2}}
\equiv \delta_{ab}G(x)
\,,
\end{equation}
where $\Omega_{d-1}=\frac{2\pi^{d/2}}{\Gamma(d/2)}$ is the volume of the $d-1$-dimensional sphere. In the following we review some basic results that we will need in our analysis; further details may be found in \cite{Kleinert:2001ax}.

We work in dimensional regularization within the minimal subtraction scheme, for which the relation between the bare coupling and the renormalized (physical) one is expressed through an ascending series of poles at $\varepsilon= 0$ \cite{Collins:1984xc}:
\begin{equation}\label{eq_bare_renormalized_BULK}
\lambda_0=M^{\varepsilon}\left(\lambda+\frac{\delta\lambda}{\varepsilon}+\frac{\delta_2\lambda}{\varepsilon^2}+\ldots\right)\,,
\end{equation}
where $M$ is the sliding scale.  To two-loop order only $\delta\lambda$ and $\delta_2\lambda $ are nonzero and read \cite{Kleinert:2001ax}:
\begin{equation}
\begin{split}
\delta\lambda&=\frac{N+8}{3}\frac{\lambda^2}{(4\pi)^2}-
\frac{3N+14}{6}\frac{\lambda^3}{(4\pi)^4}
+\mO\left(\frac{\lambda^4}{(4\pi)^6}\right)\,,\\[0.7em]
\delta_2\lambda&=\frac{(N+8)^2}{9}\frac{\lambda^3}{(4\pi)^4}
+\mO\left(\frac{\lambda^4}{(4\pi)^6}\right)\,.
\end{split}
\end{equation}
From eq.~\eqref{eq_bare_renormalized_BULK} we easily extract the beta function of the coupling
\begin{equation}\label{eq_beta_BULK}
\begin{split}
\frac{\pd\lambda}{\pd\log M}\equiv\beta_{\lambda}&=
-\varepsilon\lambda+
\lambda\frac{\pd\delta\lambda}{\pd\lambda}-\delta\lambda \\
&=
-\varepsilon\lambda+\frac{N+8}{3}\frac{\lambda^2}{(4\pi)^2}
-\frac{14+3N}{3}\frac{\lambda^3}{(4\pi)^4}
+\mO\left(\frac{\lambda^4}{(4\pi)^6}\right)\,.
\end{split}
\end{equation}
The beta function \eqref{eq_beta_BULK} admits a zero $\beta(\lambda_*)=0$ at the Wilson-Fisher fixed point, for which:
\begin{equation}\label{eq_lambda_fix}
\frac{\lambda_*}{(4\pi)^2}=\frac{3 \varepsilon }{N+8}+
\frac{9 (3 N+14) \varepsilon ^2}{(N+8)^3}
+\mO\left(\varepsilon^3\right)\,.
\end{equation}
The fixed point that describes the long distance behavior of correlation functions is $O(N)$ invariant and weakly coupled for $\varepsilon \ll 1$.  Notice also that $\lambda_*\sim 1/N$ in the large $N$ limit. We focus on this fixed point in what follows.

The bare fundamental field is related to the renormalized one as
\begin{equation}\label{eq_Zphi_epsilon}
\phi_a(x)=Z_{\phi} \left[\phi_a\right](x)\,,
\end{equation}
where $\left[\phi_a\right]$ is the renormalized field and $Z_{\phi}$ is the wave-function renormalization. Similarly to eq.~\eqref{eq_bare_renormalized_BULK}, $Z_{\phi}$ is given by an ascending series of poles; to two-loop order it reads
\begin{equation}
Z_{\phi}=1-\frac{(N+2)}{72}\frac{\lambda^2 }{(4\pi)^4 \varepsilon }+
\mO\left(\frac{\lambda^3}{(4\pi)^6}\right)\,.
\end{equation}
From $Z_\phi$ we extract the anomalous dimension of the field:
\begin{equation}\label{eq_gamma_bulk}
\gamma_{\phi}=\beta_\lambda\frac{\pd \log Z_{\phi}}{\pd\lambda}=
\frac{(N+2)}{36}\frac{\lambda^2 }{(4\pi)^4 }
+
\mO\left(\frac{\lambda^3}{(4\pi)^6}\right)~.
\end{equation}
At the fixed point the anomalous dimension~\eqref{eq_gamma_bulk} is a physical quantity, which is the difference between the scaling dimension $\Delta(\phi)$ and the engineering (mean field) dimension $\frac{d-2}{2}$ for the operator $[\phi]$. Thus,
\begin{equation}\label{phid}
\Delta(\phi)=\frac{d-2}{2}+\gamma_{\phi}(\lambda_*)\,,\qquad
\gamma_{\phi}(\lambda_*)=\frac{(N+2) \varepsilon ^2}{4 (N+8)^2}+
\mO\left(\varepsilon^3\right)\,.
\end{equation}
The two-point function of the operator $[\phi]$ thus decays as $|x|^{-2\Delta(\phi)}$ at the fixed point. For future reference, we will also need the normalization of the two-point function in the particular scheme we are working with:
\begin{equation}\label{eq_NormPhi_bulk}
\langle \left[\phi_a\right](x)\left[\phi_b\right](0)\rangle=
\frac{\mathcal{N}_{\phi}^{\,2}}{|x|^{2\Delta(\phi)}}\,,\qquad
\mathcal{N}_\phi^{\,2}=\frac{1}{(d-2)\Omega_{d-1}}\left[1+\mO\left(\varepsilon^2\right)\right]\,.
\end{equation}

\subsection{The defect fixed point} 

We now consider the model \eqref{eq_BulkAction_epsilon} in the presence of a symmetry breaking perturbation localized on a one-dimensional worldline. As explained in the introduction, this amounts to modifying the action \eqref{eq_BulkAction_epsilon} via a linear term in $\phi_1$ integrated on the worldline $x=x(\tau)$. The parameter $\tau$ on the worldline is assumed to be the proper time, i.e. $|dx/d\tau|=1$. The action  is thus modified to
\begin{equation}\label{eq_DefectAction_epsilon}
S\rightarrow S+h_0\int_Dd\tau\, \phi_1\left(x(\tau)\right)\,,
\end{equation}
where the magnetic field $h_0$ explicitly breaks the $O(N)$ symmetry to $O(N-1)$ for $N>1$ and fully breaks the $\mathbb{Z}_2$ symmetry for $N=1$. $h_0$ should be interpreted as a defect coupling, and like all bare couplings, it is subject to renormalization. Physically this means that the effective external magnetic field observed at the impurity is scale dependent. We renormalize the defect coupling similarly to eq.~\eqref{eq_bare_renormalized_BULK}:
\begin{equation}\label{eq_h0_renormalization}
h_0=M^{\varepsilon/2}\left(h+\frac{\delta h}{\varepsilon}
+\frac{\delta_2 h}{\varepsilon^2}+\ldots\right)\,.
\end{equation}

In the following we will mostly be interested in the physical setup of a straight line defect, located at $\mathbf{x}=0$ in the coordinates $x^\mu=(\tau,\mathbf{x})$. Notice that, up to rotations of the order parameter, for small enough $\varepsilon$, the operator $\phi_1$ in eq.~\eqref{eq_DefectAction_epsilon} is obviously the only possible relevant defect perturbation, if in the ultraviolet we have the trivial line defect.\footnote{If we allow for additional degrees of freedom localized on the line we can model spin impurities in materials,  see e.g. \cite{PhysRevB.61.4041,sachdev1999quantum,vojta2000quantum,Sachdev:2001ky,Sachdev:2003yk,Liu:2021nck} in a similar context.} In fact, this is true also in $d=3$, in which case it is known from  Monte Carlo and bootstrap data \cite{Pelissetto:2000ek,Poland:2018epd} that $\phi_a$ is the only operator with dimension smaller than one. Therefore, for $3\leq d<4$ the model \eqref{eq_DefectAction_epsilon} describes the only relevant perturbation of the trivial line defect with $g_\text{UV}=1$.

We solved exactly the DQFT \eqref{eq_DefectAction_epsilon} at zero coupling $\lambda=0$ in sec.~\ref{SecWarmUp},  where we found that in $d<4$, the defect coupling $h$ grows indefinitely under the defect RG flow, without ever reaching an IR DCFT. We will now show that the situation is  different for the bulk interacting theory, for which the IR limit  defines a non-trivial stable DCFT with defect coupling $h_*\sim\mO(1)$. (If factors of $N$ are restored then $h_*\sim\mO(\sqrt N)$.)

\begin{figure}[t]
   \centering
     \begin{subfigure}[t]{0.19\textwidth}
         \centering
         \includegraphics[width=0.8\textwidth]{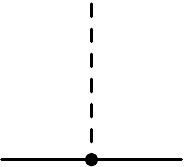}
        \caption{}\label{fig:DiagramsBetaTree}
     \end{subfigure}
    \hfill
     \begin{subfigure}[t]{0.19\textwidth}
         \centering
         \includegraphics[width=0.8\textwidth]{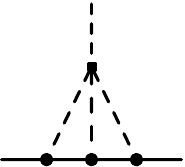}
        \caption{}\label{fig:DiagramsBeta1Loop}
     \end{subfigure}
    \hfill
      \begin{subfigure}[t]{0.19\textwidth}
         \centering
         \includegraphics[width=0.8\textwidth]{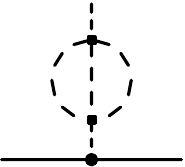}
        \caption{}\label{fig:DiagramsBeta2LoopA}
     \end{subfigure}
     \hfill
       \begin{subfigure}[t]{0.19\textwidth}
         \centering
         \includegraphics[width=0.8\textwidth]{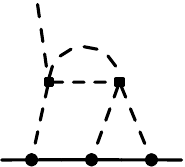}
        \caption{}\label{fig:DiagramsBeta2LoopB}
     \end{subfigure}
     \hfill
       \begin{subfigure}[t]{0.19\textwidth}
         \centering
         \includegraphics[width=0.8\textwidth]{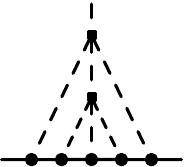}
        \caption{}\label{fig:DiagramsBeta2LoopC}
     \end{subfigure}
\caption{Diagrams contributing to the one-point function of $\phi_a$. Scalar propagators are represented by dashed lines, the defect is denoted by a solid line.  Bulk quartic vertices are denoted by squares, $h$-insertions by solid circles on the defect. Fig.~\ref{fig:DiagramsBetaTree} is the tree-level, fig.~\ref{fig:DiagramsBeta1Loop} is the one-loop contribution and the diagrams~\ref{fig:DiagramsBeta2LoopA},~\ref{fig:DiagramsBeta2LoopB} and~\ref{fig:DiagramsBeta2LoopC} are two-loop.}
\label{fig:DiagramsBeta}
\end{figure}

To demonstrate the existence of a stable infrared DCFT in the epsilon expansion we extract the coefficients of the singular terms in the relation \eqref{eq_h0_renormalization}. We compute the singular terms  by requiring that the one-point function of the renormalized fundamental field at distance $|\mathbf{x}|$ from the defect is finite in the limit $\varepsilon\rightarrow 0$ for arbitrary values of $\lambda $ and $h$:
\begin{equation}
\langle\left[\phi_a\right](0,\mathbf{x})\rangle=\text{finite}\,.
\end{equation}
The diagrams contributing to the one-point function to two-loop order are shown in fig.~\ref{fig:DiagramsBeta}.
Notice that while we work perturbatively in the bulk coupling $\lambda$ (which we know is parametrically small in the epsilon expansion), we do not assume that the defect coupling $h$ is small. This means that, at every order in perturbation theory we need to include diagrams with an arbitrary number of insertions of $h$ at the defect. This is easy in practice, as the number of possible $h$ insertions is finite at every fixed order in $\lambda$.\footnote{Alternatively, we could treat the defect as a source, solve perturbatively in $\lambda$ for the classical profile and then expand the action around the corresponding non-trivial solution. This would require regularizing the delta function in the corresponding equation of motion, allowing one to extract the terms of order $\lambda^n h^{2n+1}$ from the regularization of the classical saddle point profile \cite{Cuomo:2022xgw}. At loop level one would however
need to work with the propagator in a non-trivial symmetry breaking background. We thus found it easier to expand around $\phi_a=0$ as explained in the main text. } Using the propagator \eqref{eq_propagator_free}, a straightforward calculation leads to
\begin{align}\label{eq_h0_renormalization2}
\delta h&=\frac{\lambda}{(4\pi)^2}\frac{h^3}{12}+
\frac{\lambda^2}{(4\pi)^4}\left(\frac{N+2}{72} h-
\frac{N+8}{108} h^3-\frac{h^5}{48}\right)+
\mO\left(\frac{\lambda^3}{(4\pi)^6}\right)\,,\\[0.7em]
\delta_2h&=\frac{\lambda^2}{(4\pi)^4}
\left(\frac{N+8}{108} h^3 +\frac{h^5}{96}\right)
+\mO\left(\frac{\lambda^3}{(4\pi)^6}\right)\,.
\end{align}
Using the independence of the bare coupling $h_0$ from the sliding scale, from eq.~\eqref{eq_h0_renormalization} we find the beta-function of the defect coupling to order $\mO\left(\lambda^2\right)$:
\begin{equation}\label{eq_beta_h}
\begin{split}
\beta_h &=-\frac{\varepsilon}{2}h+\frac12h\frac{\pd\delta h}{\pd h}+\lambda\frac{\pd\delta h}{\pd \lambda}-\frac12\delta h 
\\
&=-\frac{\varepsilon}{2}h+\frac{\lambda }{(4\pi)^2}\frac{h^3}{6}+
\frac{\lambda^2}{(4\pi)^4}\left(
\frac{N+2}{36} h-\frac{N+8}{36}h^3-\frac{h^5}{12}
\right)
+\mO\left(\frac{\lambda^3}{(4\pi)^6}\right)\,.
\end{split}
\end{equation}
The beta function for $h$ was previously obtained in \cite{Allais:2014fqa}.\footnote{Notice that the couplings in \cite{Allais:2014fqa} are normalized differently.}
Notice that eqs. \eqref{eq_h0_renormalization2} and \eqref{eq_beta_h} hold for arbitrary, small, bulk coupling $\lambda$ (which does not have to be at the fixed point $\lambda_*$). Using that at the bulk fixed point $\lambda_*\sim\varepsilon$ (see eq.~\eqref{eq_lambda_fix}) we can find an attractive IR DCFT at the zero of $\beta_h$ by working perturbatively in $\ep$:
\begin{equation}\label{eq_h_fix}
h_*^2=(N+8)+
\varepsilon\frac{4 N^2+45 N+170}{2 N+16}+
\mO\left(\varepsilon^2\right)\,.
\end{equation}
We emphasize that, unlike the bulk coupling $\lambda_*$, the defect coupling at the fixed point is not small. Nevertheless its value is trustworthy, since throughout this section we were working nonperturbatively in $h$.
We remark that $h_*^2\sim N$ at large $N$; this fact will play an important role in the next section, where we discuss the solution of the model \eqref{eq_DefectAction_epsilon} in the $N\rightarrow\infty $ limit.

\subsection{The lowest dimension operators and the \texorpdfstring{$g$}{g}-function}\label{SubSecEpResults}

We now analyze the DCFT defined by eq.~\eqref{eq_h_fix}.  In this section we consider the two lowest dimension operators and the $g$-function of the theory. Additional observables are discussed in sec.~\ref{SubSecOther}.  

Let us consider first the operator spectrum. As a reminder, defect operators are naturally classified by their transformations under the conformal group $SL(2,\mathbb{R})$, their transverse spin under the $SO(d-1)$ rotations which leave the defect invariant, and their representation under the unbroken $O(N-1)$ internal group.  In general, the spectrum of dimensions of defect operators and bulk operators are different. Consequently, defect operators require their own wave-function renormalization. Below we will denote defect operators with a hat. Furthermore, $\hat a=2,...,N$ denotes a vector index of $O(N-1)$.

\begin{table}[t]
\centering
\begin{tabular}{c|ccccc}
 & dimension & $SO(d-2)$ rep. & $O(N-1)$  rep. & protected  & comments   \\ \hline
$\hat \phi_1$ & \eqref{eq_gamma1_epsilon} & scalar & singlet & no &  \\
$\hat \phi_{a}$, $a=2,...,N$ & 1 & scalar & vector & yes & only $N\geq 2$
\end{tabular}
\caption{Defect operators with scaling dimension close to $1$.}
\label{TableEps1}
\end{table}

The two lowest dimension operators on the defect are $\hat \phi_1$ and, for $N\geq 2$, the $O(N-1)$ vector $\hat \phi_{\hat{a}}$. In the ultraviolet DCFT with $h=0$ their dimensions coincide with the bulk dimension of $\phi_a$, which was quoted in~\eqref{phid}. In the infrared DCFT their scaling dimensions are summarized in table~\ref{TableEps1}.  

Most importantly, while at the UV DCFT we have $\Delta(\hat \phi_1)<1$, which is why the perturbation by $\hat \phi_1$ is relevant in the first place, in the infrared DCFT (for $d<4$), $\Delta(\hat \phi_1)>1$.
Indeed $\Delta(\hat \phi_1)$ can be computed from the derivative of the beta function \eqref{eq_beta_h}~\cite{Gubser:2008yx}:
\begin{equation}\label{eq_gamma1_epsilon}
\Delta(\hat \phi_1)=1+\left.\frac{\pd \beta_h}{\pd h}\right\vert_{h=h_*}=1+\varepsilon-\varepsilon^2
\frac{3 N^2+49 N+194}{2 (N+8)^2}
+\mO\left(\varepsilon^3\right)\,.
\end{equation}
A numerological curiosity is that the coefficient of the 2-loop correction depends very weakly on $N$. (And the coefficient of the 1-loop correction is entirely $N$ independent.)

The operator $\hat\phi_{\hat{a}}$ is quite interesting. In the UV DCFT its dimension is again smaller than 1 since the full $O(N)$ symmetry is unbroken.  However the infrared DCFT preserves only the $O(N-1)$ subgroup and $\hat\phi_{\hat{a}}$  thus becomes a displacement operator in the internal $O(N)$ space. Such operators are protected and are sometimes referred to as ``tilt'' operators:\footnote{These operators can be shown to be protected using Ward identities, see~\cite{Cuomo:2021cnb} for a comprehensive review. An intuitive argument is as follows: by the equations of motion $\hat\phi_{\hat{a}}$ parametrizes the breaking of the internal symmetry by the defect,
\begin{equation}\label{eq_Ward_breaking}
\pd_\mu J^\mu_{1 \hat{a}}=h_0\hat \phi_{\hat{a}}\delta^{d-1}_D\,,
\end{equation}
where $J^\mu_{1\hat{a}}=\phi_{1}\pd^\mu\phi_{\hat{a}}-\phi_{\hat{a}}\pd^\mu\phi_1$ are the Noether currents for the generators broken by the defect and $\delta^{d-1}_D$ is a delta function localized at the defect.  Since the bulk current has protected dimension $\Delta(J)=d-1$, the Ward identity \eqref{eq_Ward_breaking} requires the protected dimension \eqref{eq_gammaA_epsilon} for consistency. Notice that, even though eq.~\eqref{eq_Ward_breaking} was derived from classical considerations, quantum effects can only contribute (at most) with small corrections to the proportionality coefficient on the RHS of eq. \eqref{eq_Ward_breaking} and thus do not affect the argument. }
\begin{equation}\label{eq_gammaA_epsilon}
\Delta(\hat \phi_{\hat{a}})=1 \qquad
(\text{exact})\,.
\end{equation}

Notice that $\Delta(\hat \phi_{\hat{a}})=1$  does not coincide with the engineering dimension for $d<4$, nor does it coincide with the UV DCFT dimension. These facts imply nontrivial conspiracies between bulk and boundary loop corrections. We checked that diagrammatically to one-loop order the tilt operator indeed has dimension 1.

Note also that this defect is truly infrared stable; there are no symmetry preserving or symmetry breaking relevant defect operators.

We now consider the defect g-function.  As we reviewed in the introduction, the $g$-function is a useful characterization of the defect RG flow and it is is defined as the partition function in the presence of the defect  on a circle of radius $R$ normalized by the partition function without it:
\begin{equation}\label{eq_defect_gfunction}
\log g \equiv \log\left( Z^{\text{bulk}+{\text{defect}}}/ Z^{\text{bulk}}\right), 
\end{equation}
where $ Z^{\text{bulk}+{\text{defect}}}$ refers to the  partition function of the full theory including the defect, and $Z^{\text{bulk}}$ refers to the partition function of the bulk theory alone.  In computing the $g$-function we assume that the bulk is tuned to the critical point, but we work at arbitrary defect coupling.   Notice that from the definition \eqref{eq_defect_gfunction} $g=1$ for the trivial defect.

To one loop order we find the following result
\begin{equation}\label{eq_g_epsilon1}
\log g=\frac{h_0^2}{2}\int_D d\tau_1\int_D d\tau_2 \,
G\left( x(\tau_1),x(\tau_2)\right)-\frac{h_0^4 \lambda_0}{4!}
\int d^dy\left[\int_D d\tau \,G\left(y-x(\tau)\right)\right]^4+\mO\left(\lambda^2\right)\,,
\end{equation}
where the defect is placed on the curve $x^\mu(\tau)=R\left(\cos\tau/R,\sin\tau/R,0,\ldots\right)$ and the propagator is given in eq.~\eqref{eq_propagator_free}.
\begin{figure}[t]
   \centering
     \begin{subfigure}[t]{0.4\textwidth}
         \centering
         \includegraphics[width=0.5\textwidth]{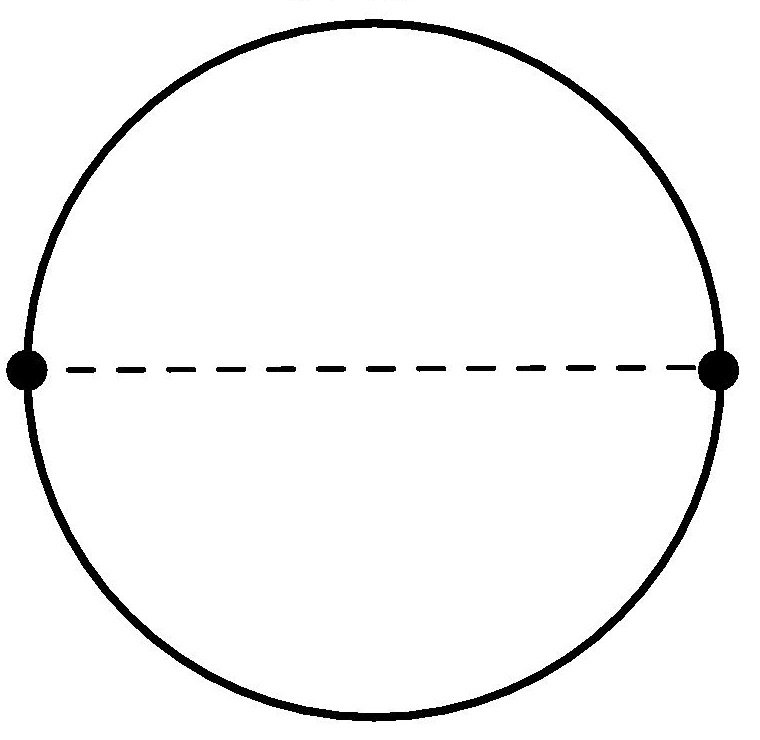}
        \caption{}\label{fig:DiagramsGTree}
     \end{subfigure}
    \hspace*{0.2cm}
     \begin{subfigure}[t]{0.4\textwidth}
         \centering
         \includegraphics[width=0.5\textwidth]{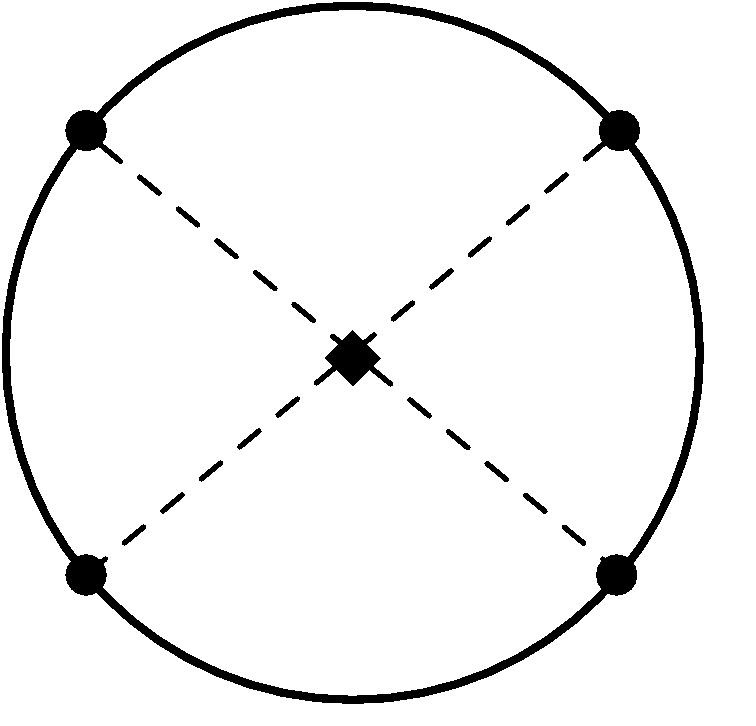}
        \caption{}\label{fig:DiagramsG1Loop}
     \end{subfigure}
\caption{Diagrams contributing to the $g$-function.  The notation is as in fig.~\ref{fig:DiagramsBeta}. Fig.~\ref{fig:DiagramsGTree} is $\mO\left(\lambda^0\right)$, while the diagram in fig.~\ref{fig:DiagramsG1Loop} is $\mO\left(\lambda\right)$.}
\label{fig:DiagramsG}
\end{figure}
The diagrams contributing to eq.~\eqref{eq_g_epsilon1} are shown in fig.~\ref{fig:DiagramsG}.
We use the following results:
\begin{equation}\label{eq_epsilonInt1}
\int_D d\tau_1\int_D d\tau_2 \,
G\left( x(\tau_1),x(\tau_2)\right)=-\frac{\varepsilon}{4}+\mO\left(\varepsilon^2\right)\,,
\end{equation}
\begin{equation}\label{eq_epsilonInt2}
\int d^dy\left[\int_D d\tau\, G\left(y-x(\tau)\right)\right]^4=-\frac{1}{16\pi^2}+\mO\left(\varepsilon\right)\,.
\end{equation}
Eq.~\eqref{eq_epsilonInt1} was previously obtained in \eqref{freeg} and it is consistent with the fact that in $d=4$ the defect coupling $h$ is exactly marginal.  The integral in eq.~\eqref{eq_epsilonInt2} is rather subtle and we detail its evaluation in appendix~\ref{App_G_eps}.
Using then the relation in eqs. \eqref{eq_h0_renormalization} and \eqref{eq_h0_renormalization2} between the bare and renormalized coupling we find
\begin{equation}\label{eq_logG_1loop}
\log g=-\frac{\varepsilon}{8}h^2+\lambda_*\frac{h^4}{768\pi^2}
+\mO\left(\lambda_*^2\right)\,,
\end{equation}
whose value at the fixed point is
\begin{equation}\label{eq_logG_1loopFix}
\log g\vert_{h=h_*}=-\frac{8+N}{16}\varepsilon+\mO\left(\varepsilon^2\right)\,.
\end{equation}
Notice that $\log g\sim -N$ at the fixed point in the large $N$ limit. However it is always finite for finite $N$, unlike the case of a free bulk theory discussed in sec.~\ref{SecWarmUp}.

We may use the relations \eqref{eq_logG_1loop} and \eqref{eq_logG_1loopFix} to verify the $g$-theorem recently proven in \cite{Cuomo:2021rkm} (see also \cite{Affleck:1991tk,Friedan:2003yc,Casini:2016fgb,Beccaria:2017rbe,Kobayashi:2018lil}).  To this aim, we recall that from the defect \eqref{eq_defect_gfunction} one obtains the defect entropy $s$ through:
\begin{equation}
s=\left(1-R\frac{\pd}{\pd R}\right)\log g\,.
\end{equation}
Using the Callan-Symanzik equation $(R\pd/\pd R+\beta_{h}\pd/\pd h)\log g=0$, we see that $\log g$ and $s$ in general coincide to the leading nontrivial order in perturbation theory and they are equal at the fixed points.\footnote{Strictly speaking this is true only in mass independent schemes,  such as the one we are using, where no cosmological constant counterterm is generated at the quantum level.}
This implies:
\begin{equation}
g\vert_{h=h_*}< g\vert_{h=0}=1\,,
\end{equation}
in agreement with eq.~\eqref{eq_logG_1loopFix}. Additionally, the defect entropy obeys the following gradient equation:
\begin{equation}\label{eq_gradient_formula}
M\frac{\pd s}{\pd M}=-\int_0^{2\pi R} d\tau_1 \int_0^{2\pi R} d\tau_2\,\langle T_D(\tau_1) T_D(\tau_2)\rangle\left[1-\cos\left(
\frac{\tau_1-\tau_2}{R}\right)\right]\,,
\end{equation}
where $T_D$ is the defect stress tensor.  We may verify this equation in perturbation theory using that $T_D=\beta_{h} \phi_a$ in eq.~\eqref{eq_gradient_formula}. Evaluating the derivative with respect to $M$ with the Callan-Symanzik equation to the first nontrivial order this gives:
\begin{equation}
\frac{\pd\log g}{\pd h}=\frac{\beta_h}{2}\,,
\end{equation}
which is satisfied by eqs. \eqref{eq_logG_1loop} and \eqref{eq_beta_h}.

\subsection{Exact results in \texorpdfstring{$d=2$}{d=2} and Pad\'e extrapolation to \texorpdfstring{$d=3$}{d=3}}\label{SubSecD2}

Before discussing more predictions from the epsilon expansion, let us pause and discuss what we can learn from the results in the previous section about the physical DCFT in $d=3$.

As often happens in the epsilon expansion, for some quantities, the two-loop correction dominates over the leading orders upon setting  $\varepsilon=1$  (i.e. three spacetime dimensions). 
For instance, in our case, the $\mO(\varepsilon^2)$ term in eq.~\eqref{eq_gamma1_epsilon} becomes larger than the leading order one upon setting $\varepsilon=1$.
The extrapolation of the scaling dimensions and the one-point function coefficients to $d=3$ should be therefore done with care.

We will improve the situation in two ways. First, in this section we will solve the $d=2$ case for $N=1$ exactly.  We will also briefly comment on the expected result for $N\geq 2$ for $d\rightarrow 2$, leaving a detailed explanation for the future. This will allow us to generate more stable numerical predictions via a Pad\'e approximation. Second,  we will solve the model at large $N$ exactly for arbitrary $d$ in the next section. Of course, in the future, higher orders in the epsilon and $1/N$ expansions can be computed and yet more precise predictions can be obtained.

For $N=1$ in $d=2$ the defect \eqref{eq_DefectAction_epsilon} corresponds to perturbing the trivial conformal interface of the Ising model with the lowest $\mathds{Z}_2$ odd primary operator, $\sigma$.  Since $\Delta(\sigma)=1/8$ this is a relevant perturbation and the endpoint of the defect RG must thus be given by a conformal interface with $g<1$. 
In fact $\sigma$ is the only relevant defect perturbation since $\epsilon$, the lowest dimension $\mathds{Z}_2$ even primary operator, can be shown to be exactly marginal \cite{PhysRevB.25.331,Oshikawa:1996ww,Oshikawa:1996dj}.  The conformal interfaces for the Ising model have been exhaustively classified in \cite{Oshikawa:1996ww,Oshikawa:1996dj}. The most stable fixed point has $g=1/2$ and it is given by the product of two ``Dirichlet" boundary conditions for the two copies (left and right) of the Ising CFT.\footnote{Since there are two Dirichlet boundary conditions in the Ising model, corresponding to spin up or down, we have to specify which of the four different ways to glue two such Dirichlet boundary conditions we choose to be the end point of our RG flow. Depending on the sign of $h_0$ in \eqref{eq_DefectAction_epsilon}, we have to choose $|+\rangle\langle+|$ or $|-\rangle\langle-|$. } This $g=1/2$ interface has no relevant perturbations and thus should describe the IR limit of the defect RG flow at hand.\footnote{Anecdotally, notice that the value $g=1/2$ in $d=2$ is not tremendously off from taking $\varepsilon=2$ and $N=1$ in \eqref{eq_logG_1loopFix}, which gives $g\simeq \exp\left[-\frac{9}{16}\times 2\right]\simeq 0.32$.} 

The Dirichlet boundary condition in the Ising CFT was solved long ago by Cardy \cite{Cardy:1989ir} building on Ishibashi's work \cite{Ishibashi:1988kg}. The spectrum of boundary operators is just given by the Virasoro descendants of the identity. The lowest dimension descendant is the displacement operator $\hat{D}$ whose dimension is $2$.

Now we fuse two Dirichlet boundary conditions and get our desired conformal interface. This has thus two defect operators of dimension $2$. One linear combination is parity odd and becomes the true displacement operator of the conformal interface, while the other one is parity even and expectedly becomes the infrared version of the defect perturbation; this is thus identified with the $d=2$ limit of the operator $\hat\phi_1$.\footnote{It is also possible to obtain the one-point functions coefficient for the (normalized) bulk primary operators $\sigma$ and $\epsilon$:
\begin{equation}
\langle \sigma(0,x_\bot)\rangle=\frac{a_{\sigma}}{|x_{\bot}|^{1/8}}\,,\qquad
\langle \epsilon(0,x_\bot)\rangle=\frac{a_{\epsilon}}{|x_{\bot}|}\,.
\end{equation}
where $x_\bot$ denotes the coordinate transverse to the interface at $x_\bot=0$. We find
\begin{equation}\label{eq_a_2d}
a_{\sigma}=2^{1/8}\,,\qquad
a_{\epsilon}= 2^{-1}\,.
\end{equation}
Eqs.~\eqref{eq_a_2d} were obtained considering the explicit expansion of the Cardy state associated with the Dirichlet boundary condition in terms of Ishibashi states as explained, e.g., in \cite{Cardy:2004hm}.}

Based on known results on symmetry breaking boundary conditions in the $d=2+\tilde{\varepsilon}$ expansion \cite{Giombi:2020rmc}, we expect $\Delta(\hat{\phi}_1)=2$ for $d\rightarrow 2$ also for $N\geq 2$. 
In the following we assume that this is indeed the case,  and validate this expectation in the large $N$ limit in the next section. We plan to give more details about the $N>2$ analysis and the special $N = 2$ case in a future publication.

\begin{figure}[t]
\centering
\includegraphics[scale=0.7]{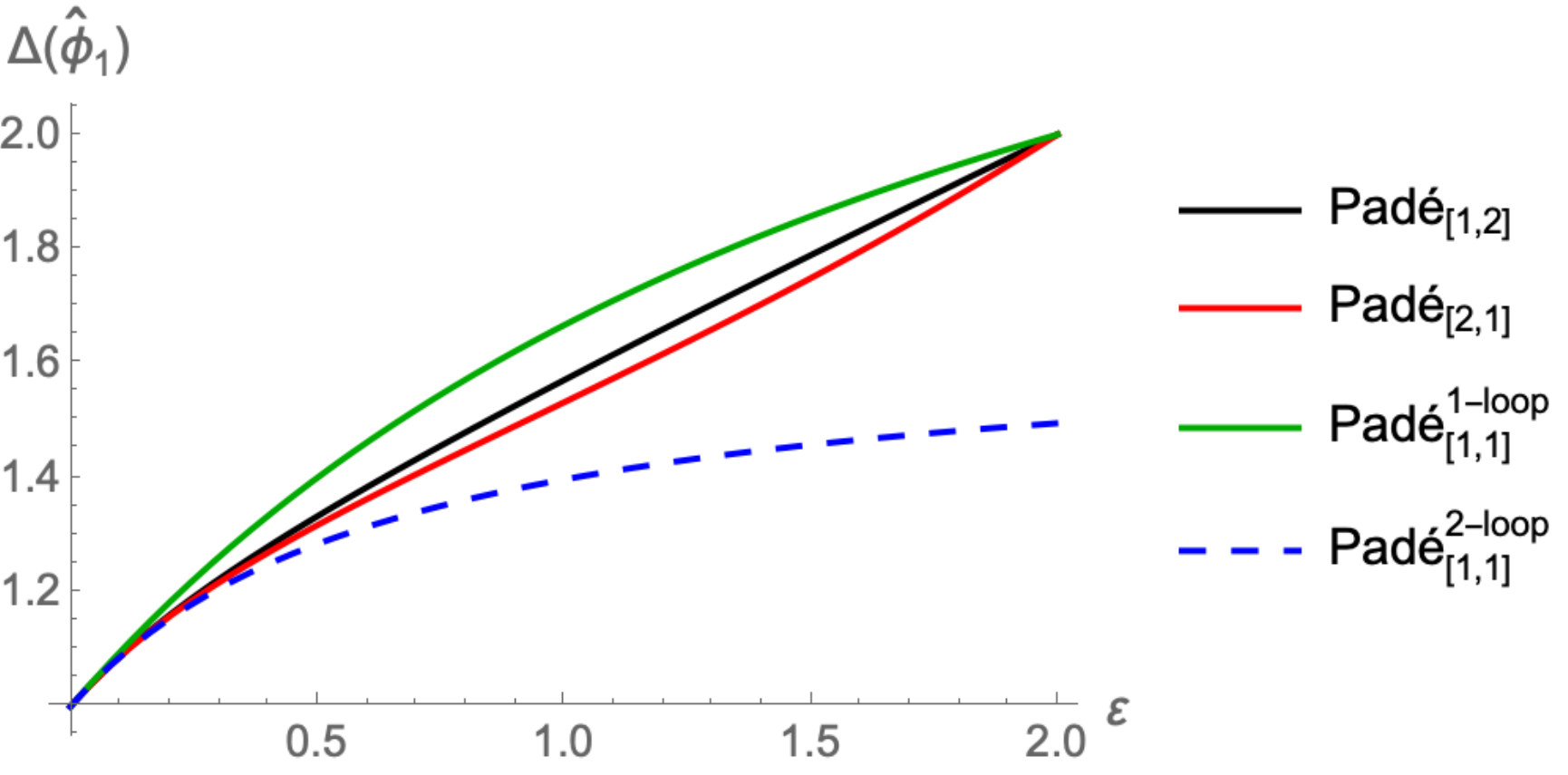}
\caption{Plot of the Pad\'e approximants for $\Delta(\hat \phi_1)$ as a function of $\varepsilon=4-d$ for $N=1$.}
\label{FigPade}
\end{figure}

We can use the prediction $\Delta(\hat \phi_1)=2$ in $d=2$ to generate  Pad\'e approximants of $\Delta(\hat{\phi}_1)$ in $2\leq d<4$ dimensions for the Ising model subject to the conditions that our Pad\'e approximants reproduce the expansion \eqref{eq_gamma1_epsilon} in $d=4-\varepsilon$ and give $\Delta(\hat\phi_1)=2 $ in $d=2$.

We thus obtain two asymmetric Pad\'e approximants $\text{Pad\'e}_{[m,n]}$ of order $m+n=3$, where $m$ and $n$ represent, respectively, the polynomial order of the numerator and the denominator. 
The results are plotted in fig.~\ref{FigPade} for $N=1$.  It turns out however that the result depends very weakly on $N$.  To estimate the error we also consider two second-order symmetric Pad\'e approximants. One that we call $\text{Pad\'e}_{[1,1]}^{\text{1-loop}}$ is obtained from the 1-loop truncation of eq.~\eqref{eq_gamma1_epsilon}, $\Delta(\hat\phi_1)\simeq 1+\varepsilon$, and the requirement $ \Delta(\hat\phi_1)\vert_{d=2}=2$. The other, that we denote $\text{Pad\'e}_{[1,1]}^{\text{2-loop}}$, is obtained from the 2-loop result in the epsilon expansion in eq.~\eqref{eq_gamma1_epsilon} without any input in $d=2$.  Also these are shown in fig.~\ref{FigPade} for $N=1$.  In practice, also these additional approximants depend very weakly on $N$. Notice however that $\text{Pad\'e}_{[1,1]}^{\text{2-loop}}\vert_{\varepsilon=2}\approx 1.5$ is quite far from the correct two-dimensional result.

The results in $d=3$ are the same for every $N$ to two digits precision and are reported in table~\ref{TablePade}.
Overall, these approximations suggest that $\Delta(\hat\phi_1)\approx 1.55$ for every $N$ within a conservative estimate of $\mO(10\%)$ uncertainty. We expect that the actual uncertainty is only $\mO(2\%)$ coming from the difference between $\text{Pad\'e}_{[2,1]}$ and $\text{Pad\'e}_{[1,2]}$. We will further validate these results in the next section, where we will study the model in the large $N$ limit obtaining strikingly similar results.

\begin{table}[t]
\centering
\begin{tabular}{c|cccc}
 & $\text{Pad\'e}_{[2,1]}$ & $\text{Pad\'e}_{[1,2]}$ & $\text{Pad\'e}_{[1,1]}^{\text{1-loop}}$ & $\text{Pad\'e}_{[1,1]}^{\text{2-loop}}$     \\ \hline
$\Delta(\hat\phi_1)$ & $1.57$ & $1.53$ & $1.67$ & $1.40$
\end{tabular}
\caption{Pad\'e approximants of $\Delta(\hat\phi_1)$ in $d=3$.  }
\label{TablePade}
\end{table}

\subsection{Other observables}\label{SubSecOther}

In this section we consider additional observables of interest in the DCFT \eqref{eq_DefectAction_epsilon} to one-loop order in the $\varepsilon$-expansion.  In particular, we consider the scaling dimension of the next six defect primary operators, whose dimension in the epsilon expansion is close to $2$,   as well as the one-point functions of the three lowest dimension bulk operators.   Our results provide some insights on the organization of the DCFT spectrum that might apply also in the physical theory in three dimensions. Additionally, we will use these results as a useful benchmark of the large $N$ analysis that we detail in the next section.  However, we refrain from discussing the limit $\varepsilon\rightarrow 1$ of our quantitative results, leaving this task for the future together with the calculation of higher order terms in the expansion.  The reader who is not interested in the detailed predictions discussed here may skip directly to sec.~\ref{SecLargeN}.

Before focusing on observables which are specific to the DCFT, we remind the reader about a few basic facts about bulk composite operators that will be needed in what follows. In particular, we consider the two primary operators with engineering (mean field) scaling dimension $d-2$. One operator is the $O(N)$ singlet $\phi_a^2$ and, for $N\geq 2$, we also have the traceless-symmetric combination $T_{ab}=\phi_a\phi_b-\delta_{ab}\,\phi_c^2/N$.
These are multiplicatively renormalized as in eq.~\eqref{eq_Zphi_epsilon}:
\begin{equation}
\phi_a^2(x)=Z_{\phi^2}\left[\phi_a^2\right](x)\,,\qquad
T_{ab}(x)=Z_T\left[T_{ab}\right](x)\,.
\end{equation}
The wave-function renormalizations and the anomalous dimensions at the fixed point to one-loop order read\footnote{Higher order results for $\gamma_T$ can be found, e.g.,  in \cite{Wallace_1975,Calabrese:2002bm}.}
\begin{eqnarray}\label{eq_Z2_bulk1}
& \displaystyle  Z_{\phi^2}=1-\frac{\lambda  (N+2)}{48 \pi ^2 \varepsilon }
+\mO\left(\frac{\lambda^2}{(4\pi)^4}\right)
\quad &
\implies\quad\gamma_{\phi^2}\stackrel{\lambda=\lambda_*}{=}
\frac{N+2}{N+8}\varepsilon+\mO\left(\varepsilon^2\right)\,,\\
&
\displaystyle 
Z_{T}=1-\frac{\lambda }{24 \pi ^2 \varepsilon }
+\mO\left(\frac{\lambda^2}{(4\pi)^4}\right)
\quad &
\implies\quad\gamma_{T}\stackrel{\lambda=\lambda_*}{=}
\frac{2}{N+8}\varepsilon
+\mO\left(\varepsilon^2\right)\,. \label{eq_Z2_bulk2}
\end{eqnarray}
The anomalous dimensions are obtained as in eq.~\eqref{eq_gamma_bulk}. Finally we also report the normalization of the corresponding two-point functions:
\begin{align}
&\langle \left[\phi^2_a\right](x)\left[\phi^2_b\right](0)\rangle=
\frac{\mathcal{N}_{\phi^2}^{\,2}}{|x|^{2\left(d-2+\gamma_{\phi^2}\right)}}\,,\\[0.7em]
&\langle \left[T_{ab}\right](x)\left[T_{cd}\right](0)\rangle=\frac12
\left(\delta_{ac}\delta_{bd}+\delta_{ad}\delta_{bc}-2\delta_{ab}\delta_{cd}/N\right)
\frac{\mathcal{N}_{T}^{\,2}}{|x|^{2\left(d-2+\gamma_T\right)}}\,,
\end{align}
where
\begin{align}\label{eq_normalization_Phi2_eps}
&\mathcal{N}_{\phi^2}^{\,2}=\frac{2 N}{(d-2)^2\Omega_{d-1}^2}\left[1-\varepsilon\frac{ (N+2) (\gamma_E +1+\log \pi )}{N+8}+\mO\left(\varepsilon^2\right)\right]\,,\\[0.7em]
&\mathcal{N}_{T}^{\,2}=\frac{2}{(d-2)^2\Omega_{d-1}^2}\left[1-
\varepsilon\frac{2  (\gamma_E +1+\log \pi )}{N+8}+\mO\left(\varepsilon^2\right)\right]\,,
\label{eq_normalization_T_eps}
\end{align}
with $\gamma_E$ the Euler-Mascheroni constant.

We now consider defect operators whose dimension in the epsilon expansion is close to $2$.  Excluding $SL(2,\mathbb{R})$ descendants, there are six such operators, summarized in table~\ref{TableEps2}.  As in the case of bulk operators, the degeneracy between them is lifted at one-loop order as we now discuss.

\begin{table}[t]
\centering
\begin{tabular}{c|ccccc}
 & dimension & $SO(d-2)$ rep. & $O(N-1)$  rep. & protected & comments \\ \hline
$\hat{s}_+$ & \eqref{eq_DeltaPM_epsilon}  & scalar & singlet & no &  \\
$\hat{s}_-$ & \eqref{eq_DeltaPM_epsilon} & scalar & singlet & no & only $N\geq 2$\\
$\hat{V}_{\hat{a}}$ &\eqref{eq_DeltaV_eps} & scalar & vector & no & only $N\geq 2$ \\
$\hat{T}_{\hat{a}\hat{b}}$ & \eqref{eq_DeltaT_eps} & scalar & tensor & no & only $N\geq 3$\\
$\mathbf{\nabla}\hat\phi_1$ & 2 & vector & singlet & yes & \\
$\mathbf{\nabla}\hat\phi_{\hat{a}}$ &
\eqref{eq_Dphi_epsilon} & vector & vector & no & only $N\geq 2$
\end{tabular}
\caption{Defect operators with scaling dimension close to $2$ in the epsilon expansion.}
\label{TableEps2}
\end{table}

The four  operators that transform as scalars under transverse rotations are linear combinations of the product $\hat\phi_a\hat \phi_b$ on the defect. As a reminder, this implies that to extract their anomalous dimensions we need to consider a wave-function renormalization matrix,
\begin{equation}
\hat\phi_a\hat\phi_b=\hat{Z}_{ab}^{cd}[\hat{\phi}_c\hat{\phi}_d]\,.
\end{equation}
To one-loop order, $\hat{Z}$ is given by the following expressions:
\begin{equation}\label{eq_Z_level2}
\hat{Z}_{ab}^{cd}=\delta_a^c\delta_b^d-\frac{\lambda}{48\pi^2\varepsilon}\left[\left(\delta_{ab}\delta^{cd}+2\delta_a^c\delta_b^d\right)+\frac{h^2}{2}\left(
\delta_a^c\delta_b^d+\delta_a^c\delta_b^1\delta^d_1+\delta_a^1\delta^c_1\delta_b^d\right)\right]+\mO\left(\frac{\lambda^2}{(4\pi)^4}\right)\,.
\end{equation}
It is easy to check that eq.~\eqref{eq_Z_level2} reduces to eqs. \eqref{eq_Z2_bulk1} and \eqref{eq_Z2_bulk2} for $h=0$. From eq.~\eqref{eq_Z_level2} we find the anomalous dimension matrix as follows:
\begin{equation}\label{eq_gamma_eps_general}
\hat{\gamma}=\hat{Z}^{-1}\frac{d \hat{Z}}{d M}=\hat{Z}^{-1}\left(
\beta_{\lambda}\frac{\pd \hat{Z}}{\pd \lambda}+\beta_h\frac{\pd \hat{Z}}{\pd h}\right)\,.
\end{equation}
Diagonalizing $\hat{\gamma}_{ab}^{cd}$ we then obtain the scaling dimensions via $\Delta=(d-2)+\hat{\gamma}$.  

Diagonalizing eq.~\eqref{eq_gamma_eps_general} the two scalar singlets are identified with the following linear combinations of $\hat \phi_a^2$ and $\hat\phi_1^2$:
\begin{align}\label{eq_s_def_eps}
\hat{s}_+&=\frac{N+16-\sqrt{N(N+40)+320} }{4 (N+8)}\hat\phi_a^2-\hat\phi_1^2\,,\\[0.7em]
\hat{s}_-&=
\frac{N+16+\sqrt{N(N+40)+320} }{4 (N+8)}\hat\phi_a^2-\hat\phi_1^2\,.
\end{align}
The corresponding scaling dimensions read
\begin{equation}\label{eq_DeltaPM_epsilon}
\Delta(\hat{s}_\pm)=2+\varepsilon
\frac{3N +20\pm \sqrt{N (N+40)+320}}{2 (N+8)}+\mO\left(\varepsilon^2\right)\,.
\end{equation}
Notice that we defined them so that $\Delta(\hat{s}_-)<\Delta(\hat{s}_+)$. For $N=1$ only the operator $\hat{s}_+$ exists, as can be seen by setting $N=1$ in eq.~\eqref{eq_s_def_eps} and using $\hat\phi_a^2=\hat\phi_1^2$.

For $N\geq 3$ there also exist an $O(N-1)$ vector and a traceless symmetric tensor operator:
\begin{equation}
\hat{V}_{\hat{a}}=\hat\phi_1\hat\phi_{\hat{a}}\,,\qquad
\hat{T}_{\hat{a}\hat{b}}=\hat\phi_{\hat{a}}\hat\phi_{\hat{b}}-\frac{\delta_{\hat{a}\hat{b}}}{N-1}\hat\phi_{\hat{c}}^2\,.
\end{equation}
For $N=2$ only the vector $\hat{V}_{\hat{a}}$ exists. From eq.~\eqref{eq_Z_level2} we find that their scaling dimensions are given by:
\begin{align}\label{eq_DeltaV_eps}
\Delta(\hat{V}_{\hat{a}})&=2+\varepsilon\left(1+\frac{2}{N+8}\right)+\mO\left(\varepsilon^2\right)
\,,
\\ \label{eq_DeltaT_eps}
\Delta(\hat{T}_{\hat{a}\hat{b}})&=2+\frac{2\varepsilon}{N+8}+
\mO\left(\varepsilon^2\right)
\,.
\end{align}

Finally, let us consider the two operators with dimension close to 2 with transverse unit spin:
\begin{equation}\label{eq_L2S1op_epsilon}
\mathbf{\nabla}\hat\phi_1\qquad\text{and}\qquad
\mathbf{\nabla}\hat\phi_{\hat{a}}
\,,
\end{equation}
where $\mathbf{\nabla}$ denotes the gradient in the directions transverse to the defect.  These do not mix due to their different transformation under $O(N-1)$ and are multipliticatively renormalized in a straightforward fashion. The second operator in eq.~\eqref{eq_L2S1op_epsilon} is an $O(N-1)$ vector and exists only for $N\geq 2$. The operator $\mathbf{\nabla}\phi_1$ is the \emph{displacement} operator \cite{Billo:2016cpy}; it parametrizes the breaking of translations induced by the defect and thus, similarly to the \emph{tilt} operator discussed around eq.~\eqref{eq_Ward_breaking}, has protected dimension:
\begin{equation}\label{eq_DeltaDis_epsilon}
\Delta(\mathbf{\nabla}\hat \phi_1)=2\qquad
(\text{exact})\,.
\end{equation}
 We  checked that eq.~\eqref{eq_DeltaDis_epsilon} is indeed satisfied at one-loop order.
The scaling dimension of the operator $\mathbf{\nabla}\phi_{\hat{a}}$ is computed straightforwardly and reads:
\begin{equation}\label{eq_Dphi_epsilon}
\Delta(\mathbf{\nabla}\hat\phi_{\hat{a}})=2-\frac{\varepsilon}{3}+\mO\left(\varepsilon^2\right)\,.
\end{equation}

Overall,  comparing the results \eqref{eq_DeltaPM_epsilon}, \eqref{eq_DeltaV_eps}, \eqref{eq_DeltaT_eps}, \eqref{eq_DeltaDis_epsilon} and \eqref{eq_Dphi_epsilon} we conclude that the order $\mO(\varepsilon)$ corrections lead to the following ordering among the dimensions of operators with dimension close to 2 in the epsilon expansion:
\begin{equation}\label{eq_Delta_hierarchy}
\Delta(\mathbf{\nabla}\hat\phi_{\hat{a}})<\Delta(\mathbf{\nabla}\hat\phi_{1})=2<\Delta(\hat T_{\hat{a}\hat{b}})<\Delta(\hat{s}_-)
<\Delta(\hat V_{\hat{a}})<\Delta(\hat{s}_+)\,.
\end{equation}
The extrapolation of eq.~\eqref{eq_Delta_hierarchy} to $\varepsilon=1$ together with the inequality we have previously established $\Delta(\hat\phi_{\hat{a}})=1<\Delta(\hat \phi_1)<\Delta(\mathbf{\nabla}\hat \phi_{\hat{a}})$ provides a detailed prediction for the organization of the DCFT spectrum of the lowest dimension defect operators depending on their quantum numbers under $SO(d-1)$ and $O(N-1)$.  In the future this might be used as a useful input for a numerical bootstrap analysis of this defect, along the lines of \cite{Liendo:2012hy,Gaiotto:2013nva,Gliozzi:2015qsa,Billo:2016cpy}.

Other natural observables of potential experimental interest are the expectation value of bulk operators at finite distance from the defect. As a reminder, the form of the one-point functions is fixed by conformal invariance up to a coefficient, which is the bulk-to-defect OPE coefficient \cite{Billo:2016cpy}. We computed this coefficient for the operators $\phi_a$, $\phi_a^2$, and (for $N>1$) $T_{ab}$. In terms of the renormalized operators the one-point functions read
\begin{align}
\langle\left[\phi_a\right](\mathbf{x},0)\rangle &=\delta_a^1\mathcal{N}_{\phi}\frac{a_\phi}{|\mathbf{x}|^{\frac{d-2}{2}+\gamma_\phi}}\,,\\
\langle\left[\phi_a^2\right](\mathbf{x},0)\rangle&=\mathcal{N}_{\phi^2}\frac{a_{\phi^2}}{|\mathbf{x}|^{d-2+\gamma_{\phi^2}}}\,,\\
\langle \left[T_{ab}\right](\mathbf{x},0)\rangle&=\left(\delta_a^1\delta_b^1-\delta_{ab}/N\right) 
\mathcal{N}_{T}\frac{a_T}{|\mathbf{x}|^{d-2+\gamma_T}}\,,
\end{align}
where we defined the coefficients $a_\phi,\,a_{\phi^2}$ and $a_T$ after having extracted the normalization of the two-point functions in the bulk in the absence of the defect, given in eqs. \eqref{eq_NormPhi_bulk}, \eqref{eq_normalization_Phi2_eps} and \eqref{eq_normalization_T_eps}. The epsilon expansion predictions are
\begin{align} \label{eq_Aphi_epsilon}
a^2_\phi &=\frac{N+8}{4}+\varepsilon
\frac{  (N+8)^2 \log 4+N^2-3 N-22}{8 (N+8)}
+\mO\left(\varepsilon^2\right)\,,\\[0.8em] \label{eq_S0_epsilon}
a_{\phi^2}&=\frac{N+8}{4\sqrt{2 N}}\left[1+\varepsilon
\frac{12 (N+8) \log 2  -13 N-38}{2 (N+8)^2}+\mO\left(\varepsilon^2\right)\right]\,,\\[0.8em]
a_T&=\frac{N+8}{4\sqrt{2 }}\left[1+\varepsilon
\frac{  (N+6) (N+8) \log 4+N^2-5 N-38}{2 (N+8)^2}
+\mO\left(\varepsilon^2\right)\right]\,. \label{eq_T0_epsilon}
\end{align}
The result \eqref{eq_Aphi_epsilon} for $a_{\phi}$ was previously obtained in \cite{Allais:2014fqa}.
Notice that we would need to know the $\mO(\varepsilon^2) $ correction  to the fixed-point coupling \eqref{eq_h_fix} from a 3-loop calculation to extract the $\mO(\varepsilon^2)$ corrections to the one-point functions.  This is beyond the scope of this paper.

A final comment concerns with the large $N$ limit of our results, which will be useful in the next section.  We notice in particular that from eq.~\eqref{eq_s_def_eps} $\hat{s}_{+}$ has a $1/N$ suppressed overlap with the bulk operator $\phi_a^2$, and thus it drops out from its bulk-to-defect OPE in the large $N$ limit. We also notice that in the  large $N$ limit $\Delta(\hat{s}_+)=2\Delta(\hat{\phi}_1)$, $\Delta(\hat V_{\hat{a}})=\Delta(\hat\phi_{\hat{a}})+\Delta(\hat \phi_{1})$ and $\Delta(\hat T_{\hat{a}\hat{b}})=2\Delta(\hat\phi_{\hat{a}})=2$, consistently with the interpretation of $\hat{s}_+$, $\hat{V}_{\hat{a}}$ and $\hat{T}_{\hat{a}\hat{b}}$ as double-traces that we will provide in sec.~\ref{SecLargeN}. We also notice from eqs~\eqref{eq_Aphi_epsilon} and \eqref{eq_T0_epsilon} that in the large $N$ limit $a_{\phi^2}$ does not receive corrections at order $\mO(\varepsilon)$ and that $\mathcal{N}_T a_T=\left(\mathcal{N}_{\phi}a_{\phi}\right)^2$ up to $1/N$ corrections. The latter result is compatible with the known fact that $T_{ab}$ can be thought as double-trace operators in the large $N$ limit.

\section{Large \texorpdfstring{$N$}{N} analysis}\label{SecLargeN}

\subsection{Saddle point analysis of the DCFT}

First, let us review some salient features of the large $N$ limit of the $O(N)$ Wilson-Fisher CFTs. 
One convenient description of the $O(N)$ CFTs is through a Lagrangian which describes a perturbation of $N$ free fields by an $O(N)$ symmetric quartic interaction
\begin{equation}\label{action}
S= \int d^d x \left[\frac12 (\partial\phi_a)^2+{\lambda\over N}\left(\phi_a^{2}\right)^2\right]~.
\end{equation}
The normalization of the quartic vertex, which now differs from that in eq.~\eqref{eq_BulkAction_epsilon}, contains an explicit $1/N$ factor which guarantees a smooth large $N$ limit for all correlation functions where the distances between local operators do not scale with $N$. The quartic perturbation is relevant for $d<4$ and marginally irrelevant for $d=4$. It is implicit that we are fine tuning the mass term to a fixed point, if such exists. 

To leading order in the $1/N$ expansion the computation of $\langle \phi_a(x) \phi_b(0)\rangle$ is straightforward since one can entirely ignore the quartic interaction.\footnote{The implicit fine tuning of the mass term is important since that allows us to set to zero the ``tadpole diagram'' which is not suppressed at large $N$.}
Therefore,  the corresponding two-point function is just given by the free propagator \eqref{eq_propagator_free}, that we repeat here for convenience
\begin{equation}\label{ffield}
\langle \phi_a(x) \phi_b(0)\rangle=\frac{\delta_{ab}}{(d-2)\Omega_{d-1}
|x|^{d-2}}\equiv \frac{\mathcal{N}_{\phi}^2}{|x|^{d-2}}~.
\end{equation}
Therefore $\phi_a$ in the large $N$ limit, to leading order, is a free field of scaling dimension $d/2-1$. 

The two-point function  $\langle \phi_a^{2}(x_1)\phi^{2}_b(x_2)\rangle$ has a more interesting large $N$ limit. To leading order we need to resum all the ``sausage diagrams.'' This can be done as a geometric series in momentum space (see~\cite{Moshe:2003xn} for a review). The result can be written for all distance scales, but for our purposes we are only interested in the nontrivial (infrared) bulk CFT so we quote the result at long distances
\begin{equation}\label{fundf}
\langle\phi_a^{2}(x)\phi^{2}_b(0)\rangle ={N\over \lambda^2}\sin(\pi d/2)  {2^{d-2}\Gamma(d/2-1/2)\over \pi^{\frac32}\Gamma(d/2-2) }{1\over x^4}~.
\end{equation}
Clearly then, to leading in the large $N$ limit, the operator $\phi^{2}_a$ has scaling dimension $2$ at the nontrivial $O(N)$ invariant fixed point.

Now let us consider the problem of the line operator 
\begin{equation}\label{defectlargeN}
e^{-\sqrt N h \int d\tau \phi_1}
\end{equation} 
in the $O(N)$ Wilson-Fisher fixed point. The normalization of the line operator makes it manifest that there is again a smooth large $N$ limit for distances on or from the line defect which do not scale with $N$. 
The symmetry is now explicitly broken to $O(N-1)$ due to the line operator. The coefficient $h$ is dimensionful and relevant for $d < 4$ and exactly marginal for $d = 4$ (since in $d=4$ the bulk is free, this case reduces to our discussion in section~\ref{SecWarmUp}). We are most interested in the infrared limit of this line defect, which as we argued before, must be a nontrivial DCFT due to the $g$-theorem. 

It is not trivial to evaluate the properties of the line defect~\eqref{defectlargeN} in the large $N$ limit. 
Indeed, imagine trying to compute the $g$ function. Lowering $\phi_1$ from the exponent $2L$ times would lead to a contribution of order $N^L$ from diagrams where the quartic vertex is not utilized but there are contributions of order $N^{L-1}$ and $N^{L-2}$ etc which include the quartic vertex.  Therefore to obtain a sensible result in the large $N$ limit some re-summation is required. The same comment applies to correlation functions in the presence of the defect in the large $N$ limit.

Since we will be from now on interested only in the case that the bulk is at the fixed point, we will employ the Hubbard-Stratonovich transformation and replace the bulk action~\eqref{action} with 
\begin{equation}\label{LAGb}
S= \int d^dx\left[\frac12 (\partial \phi_a)^2 + \frac12 s \phi_a \phi_a\right]~,
\end{equation}
where $s$ is a field we path integrate over and it essentially replaces $\phi_a^{2}$ in the formulation~\eqref{action} (again, see the review~\cite{Moshe:2003xn}). We find 
\begin{equation}\label{snorm}
\langle s(x) s(0) \rangle ={16\over N} \sin(\pi d/2)  {2^{d-2}\Gamma(d/2-1/2)\over \pi^{\frac32}\Gamma(d/2-2) }{1\over x^4}~.
\end{equation} 
The advantage of this formulation is that no infrared limit needs to be taken to obtain the correlators of the $O(N)$ Wilson-Fisher model.

In the formulation~\eqref{LAGb} we can model the line defect~\eqref{defectlargeN} by adding a term to the action
\begin{equation}\label{LAGd}
S= \int d^dx\left[ \frac12 (\partial \phi_a)^2 + \frac12 s \phi_a \phi_a+\sqrt N J^a \phi_a \right]~,
\end{equation}
where $J^a=\delta^a_1 h \delta^{(d-1)}(x_\perp)$.

Below we will show that the large $N$ limit of~\eqref{LAGd} can be understood by an expansion around a new saddle point and the $1/N$ corrections arise from the standard semiclassical expansion around that saddle point.  Our approach generalizes the one used to study the $O(N)$ model in the presence of a boundary \cite{PhysRevLett.38.735,Ohno:1983lma,McAvity:1995zd,Herzog:2020lel,Metlitski:2020cqy}. Our approach allows us to derive many exact results about the large $N$ limit of the line operator~\eqref{defectlargeN}.\footnote{Previous works \cite{vojta2000quantum,Liu:2021nck} studied the large $N$ limit for various models coupled to impurities, generalizing an analogous approach to the multi-channel Kondo problem \cite{tsvelick1985exact,PhysRevB.46.10812,Affleck:1995ge,PhysRevB.58.3794}.  In that setup however the impurity does not generate a non-trivial one-point function for the bulk fields to leading order in $N$, and thus the saddle point  affects only the fields on the impurity to leading order in $N$. This is different from what happens in the model \eqref{LAGd}, in which both $\phi_a$ and $s$ have non-trivial one-point function at leading order, as we show below.}

We can rewrite the action~\eqref{LAGd} as \footnote{We use the notation $J^a{1\over-\square +s} J^a=\int d^dx \int d^dy \, J^a(x)\left(\frac{1}{-\square+s}\right)(x,y)J^a(y)$.}
\begin{equation}
S= \int d^dx\left[ \frac12 \left( \phi_a+{{\sqrt N\over -\square +s}J^a}\right)(-\square +s)\left(\phi_a+{{\sqrt N\over -\square +s}J^a}\right)\right] - \frac 12 NJ^a{1\over-\square +s} J^a ~,
\end{equation}
Now we perform a formal change of variables $\phi_a=\delta \phi_a-{{\sqrt N\over -\square +s}J^a}$
to recast the action in the form 
\begin{equation}\label{newvars}
S= \int d^dx \left[\frac12  \delta\phi_a(-\square +s)\delta \phi_a\right]-\frac12NJ^a{1\over-\square +s} J^a ~,
\end{equation}
where both terms are formally of order $N$.

It is worth dwelling for a moment on the change of variables $\phi_a= \delta\phi_a-{{\sqrt N\over -\square +s}J^a}$.
Since $s(x)$ is an operator, after the change of variables, $\delta\phi_1$ is no longer a local operator necessarily obeying the usual rules for local operators in quantum field theory. Also note that after the change of variables, an apparent $O(N)$ symmetry emerges even though the defect obviously only preserves $O(N-1)$. This is again due to the new $\delta \phi_1$ field being a non-local field. Yet, this change of variables is formally allowed in the path integral. 

We can integrate out the $\phi_a$ fields exactly and obtain an effective action for the single degree of freedom $s$:
\begin{equation}\label{SD}
S_\text{eff}[s]=N \left[\frac12 \log \det (-\square +s)-\frac12J^a{1\over-\square +s} J^a\right]~.
\end{equation}

Therefore $N$ acts as $\hbar^{-1}$ as usual and we need to find the new saddle point for $s$, which replaces the usual saddle point $s=0$ in the absence of the sources.
This formalism describes the large $N$ limit of a line defect which flows from the trivial line defect ($h=0$) to a nontrivial infrared DCFT with $g<1$. Here we will not attempt to solve for this RG flow (though this should be possible in principle), but, from now on, we will instead directly look for a self-consistent solution that describes the infrared DCFT. 

A solution obeying the DCFT symmetry principles must, in the large $N$ limit, obey \begin{equation}\label{sSaddle}
\langle s\rangle={s_0\over x_\perp^2}~,
\end{equation}
where $s_0$ is a dimensionless coefficient that needs to be determined. 
Since the fluctuations of $s$ are highly suppressed in the large $N$ limit due to~\eqref{SD}, this one-point function represents the classical profile around which $s$ can be expanded with small corrections, 
\begin{equation}
s={s_0\over x_\perp^2}+\delta s~,
\end{equation}
with $\delta s$ being of order $1/\sqrt N$.  
 
The value of $s_0$ can be obtained by the following trick. Since the action~\eqref{newvars} formally restores the $O(N)$ symmetry we have $\langle\delta\phi_a\rangle=0$ in the vacuum. 
This means that in terms of the original fields,  for $a\neq 1$, we have $\langle\phi_a\rangle=0$ which of course just means that we have $O(N-1)$ symmetry while for $a=1$ we have 
\begin{equation}
\langle\phi_1\rangle=-h\sqrt N\bigl\langle {{1\over -\square +s}\delta^{(d-1)}(x_\perp)}\bigr\rangle~.
\end{equation}
Here we use the large $N$ approximation to replace, to leading order, $s$  with its VEV~${s_0\over x_\perp^2}$. Therefore we have
\begin{equation}\label{phiVEV}
\langle\phi_1\rangle=-h\sqrt N {{1\over -\square +{s_0\over x_\perp^2}} \delta^{(d-1)}(x_\perp)}+{\mathcal O}\left(N^{-1/2}\right)~.
\end{equation}
We must choose $s_0$ in such a way that the right hand side in~\eqref{phiVEV} decays as ${1\over x_\perp^{d/2-1}}$, which is forced by the conformal invariance of the bulk and the line defect. This means that 
\begin{equation}\label{phi1eom}
 \left({\partial^2 \over \partial x_\perp^2}+{d-2\over x_\perp}{\partial\over \partial x_\perp}-{s_0\over x_\perp^2}\right){1\over x_\perp^{d/2-1}}=0
 \end{equation}
has to be obeyed for $x_\perp\neq 0$. Therefore we obtain\footnote{In ref.~\cite{Allais:2014fqa} the authors also set out to analyze the line defect at large $N$ using the collective field formulation given in eq.~\eqref{SD}. Our results disagree: in ref.~\cite{Allais:2014fqa} the saddle point was not identified correctly and this problem propagated into their subsequent calculations. 

 Ref.~\cite{Allais:2014fqa} considered the same Ansatz in eq.~\eqref{sSaddle} for the saddle point, but they attempted to capture the entire defect RG flow with it. The Ansatz is too restrictive for that purpose, as it can only capture the IR DCFT. This then leads to inconsistent equations of motion: concretely they obtain $s_0=(5-2d)/4$ in their eq.~$(68)$ (they denote $s_0$ with $v$), which disagrees with our result in eq.~\eqref{snot}. The  IR limit of $\langle\phi_1\rangle$ in their eq.~$(67)$ (that they denote by $\sigma$) then does not solve the equation of motion~\eqref{phi1eom} (or eq.~$(47)$ in~\cite{Allais:2014fqa}).} 
\begin{equation}\label{snot}
s_0={(4-d)(d-2)\ov 4}~.
\end{equation}
Dividing by the normalization of the two-point function \eqref{snorm},
eq.~\eqref{snot} provides a prediction for the one-point function of $s$ in the large $N$ limit:\footnote{It is not straightforward to evaluate the normalization of the left-hand side of~\eqref{phiVEV} since $h$ is a dimensionful parameter and we need to first carefully regularize the Green's function. We will get back to evaluating the normalization of $\langle\phi_1\rangle$ soon.}
\begin{equation}\label{snot2}
\begin{split}
a_s\equiv
{s_0\over \sqrt{\langle s(\infty) s(0) \rangle}} &=\sqrt{N}\,\frac{\pi ^{3/4} (4-d) (d-2)\sqrt{\Gamma \left(\frac{d-4}{2}\right)}}{8 \sqrt{2^d \sin \left(\frac{\pi  d}{2}\right) \Gamma \left(\frac{d-1}{2}\right)}}\\
&\overset{d=3}{=}\frac{\pi\sqrt{N}}{16}\,.
\end{split}
\end{equation}

Let us compare eq.~\eqref{snot2} with the epsilon expansion result in eq.~\eqref{eq_S0_epsilon} (where we denoted the operator $s$ with $\phi^2$). Retaining terms up to order $\mO(\varepsilon)$ we have
\begin{equation}
a_s \overset{d=4-\varepsilon}{=} {\sqrt N\over 4\sqrt{2}  }+0\cdot \ep+{\mathcal O}(\varepsilon^2)~.  
\end{equation}
The first term on the right hand side can be reproduced with classical physics while to reproduce the vanishing of the linear term in $\varepsilon$, a nontrivial computation involving radiative corrections  is required. Both the $\varepsilon^0$ term and the lack of a term proportional to $\varepsilon$ were established in section~\ref{SubSecOther}, in perfect agreement with our large $N$ results.

Now that we have determined the classical profile for $s$, we can obtain more DCFT data. For $a\neq 1$, $\delta \phi_a=\phi_a$, and hence the two-point function of $\phi_{a\neq1}$ is given to leading order in the $1/N$ expansion by the Green's function for the operator $-\square +{s_0\over x_\perp^2}$.
This Green's function admits an expansion involving the modified Bessel functions of the first and second kind depending on $|\omega|x_\perp$, where $\omega$ is the frequency in the direction tangent to the line defect. From this we can read the powers of $x_\perp$ that can appear when one of the operators approaches the defect and hence the spectrum of defect operator dimensions \cite{Billo:2016cpy,Herzog:2020bqw}. We find the defect operator dimensions:  
\begin{equation}\label{tiltchannel}
\hat\Delta_{\ell}=\frac12+\sqrt{\frac14+\ell(\ell+d-3)}~,
\end{equation}
with non-negative integer $\ell$, corresponding to the transverse spin.
These are the dimensions of the primary operators that appear in the bulk-to-defect expansion of $\phi_{a\neq 1}$. Therefore, these operators are all in the vector representation of $O(N-1)$. 
For $\ell=0$ we get $\hat \Delta_0=1$ and this is very encouraging since this is identified with the tilt operator. 
This operator has dimension $\hat \Delta=1$ for all $N$ and all $d$ where the DCFT makes sense.
Another immediate consistency check is to set $d=4$, where~\eqref{tiltchannel}  gives integers for all $\ell$, which is again encouraging since the DCFT is essentially free in $d=4$ and hence the spectrum of defect operators coincides with the bulk spectrum of operators. Finally it is also simple to see that, setting $d=4-\varepsilon$, the scaling dimension \eqref{tiltchannel} for the $\ell=1$ operator matches the one-loop epsilon expansion result for the operator $\mathbf{\nabla}\hat{\phi}_{\hat a}$ obtained in eq.~\eqref{eq_Dphi_epsilon}. Of course, there also exist composite defect operators made out of products (and derivatives) of the operators corresponding to~\eqref{tiltchannel}. These behave like multi-trace operators whose dimensions are additive in the large $N$ limit. This is again in agreement with the one-loop $\varepsilon$-expansion result for the traceless symmetric operator in eq.~\eqref{eq_DeltaT_eps}, where we saw that at large $N$ its dimension is just given by the product of two tilt operators and hence has $\hat \Delta=2$.

We see that the large $N$ methods allow to determine $s_0$ and the spectrum of operators charged under $O(N-1)$ rather easily. It is more difficult to determine the normalization of the one-point function of $\phi_1$ or the spectrum of $O(N-1)$ invariant defect operators. To obtain this additional DCFT data one is required to study the saddle point of~\eqref{SD} and the Green's function for $\delta s$. A convenient way to proceed is to perform a conformal transformation and place the theory in AdS$_2\times S^{d-2}$. Indeed, these Green's functions are easier to handle in AdS$_2\times S^{d-2}$ due to the extensive recent work on AdS Green's functions. For instance,~\eqref{tiltchannel} is clearly just the spectrum of massless free fields in AdS$_2\times S^{d-2}$ which is a reflection of the fact that 
$-\square +{s_0\over x_\perp^2}$ becomes the massless Klein-Gordon operator in AdS$_2\times S^{d-2}$ due to the special value of $s_0$ in~\eqref{snot}, as we will see in detail in the next subsection.

\subsection{Mapping to \texorpdfstring{AdS$_2\times S^{d-2}$}{AdS2TimesSdminus2}}

As anticipated above, a useful perspective on the problem is provided by mapping the IR limit of the defect QFT, the DCFT of interest to AdS$_2\times S^{d-2}$  using the following Weyl transformation:
\es{Weyl}{
ds^2&=d\tau^2+dr^2+r^2 ds^2_{S^{d-2}}\\
&=r^2\le[{d\tau^2+dr^2\ov r^2}+ds^2_{S^{d-2}}\ri]\\
&\equiv r^2 d\tilde s^2\,,
}
where in the $d\tilde s^2$ is the line element of (Euclidean) AdS$_2\times S^{d-2}$ in Poincar\'e coordinates.  Similar Weyl transformations have proven to be useful to study the CFT data associated to different dimensional defects (and to boundaries) in refs.~\cite{Kapustin:2005py,Casini:2011kv,Chester:2015wao,Carmi:2018qzm,Giombi:2020rmc,Giombi:2021uae}. Of particular relevance to us are the studies of the  $O(N)$ model in the presence a boundary by mapping to AdS$_d$~\cite{Carmi:2018qzm,Giombi:2020rmc} and with a twist defect by mapping to AdS$_{d-1}\times S^1$~\cite{Giombi:2021uae}.

We can rewrite the partition function for the IR DCFT in terms of the theory in AdS:
\es{gdef}{
g&={Z[\widetilde J]\ov Z[0]}={\int Ds \ e^{-N\,S_\text{eff}[s,\widetilde J]}\ov \int Ds \ e^{-N\,S_\text{eff}[s,0]}}\,,
}
where $S_\text{eff}[s,\widetilde J]$ is the appropriate Weyl transformation of the flat space action functional in eq.~\eqref{newvars} into curved space (we will define $\widetilde J$ below). 

We can straightforwardly translate the ``bulk term'':
\es{bulkterm}{
{\rm Tr}\log\le(- \square+s \ri)\quad \to \quad {\rm Tr}\log\le(-\widetilde \square+s +{d-2\ov 4(d-1)}\widetilde R\ri)\,,
}
where $\widetilde R=-(4-d)(d-1)$  provides the conformal mass term in AdS$_2\times S^{d-2}$. The ``defect term'' from eq.~\eqref{newvars}, $J^a{1\over-\square +s} J^a $ is more problematic, since with $J^a=\delta^a_1 h \delta^{(d-1)}(x_\perp)$ it describes the full DQFT RG flow, and its transformation into curved space would be very complicated. Instead, we just want to capture the IR DCFT. In flat space we dealt with this by replacing the source $\delta^{(d-1)}(x_\perp)$ in the equation of motion for $\langle\phi_1\rangle$ with a boundary condition near the defect that enforced a nonzero $\langle\phi_1\rangle$. 

In AdS/CFT we are accustomed to dealing with exactly such a situation, as we briefly recall. Let us have  a bulk scalar field $\chi$ in AdS$_2$ with the prescribed near boundary behavior
\es{chiBdy}{
\chi(\tau,r)&\to {r^{1-\De}\ov 2\De-1}\,A(\tau) \,, \qquad (r\to0)\,,\\
\Delta&=\frac12+\sqrt{\frac14+m^2}\,.
}
While in the following $\Delta$ will denote the dimension of defect operators, we omit the hat from it to lighten the notation. Then the bulk scalar field profile is given by
\es{chiBulk}{
\chi(x)&= \le[G_{b\p}^{(\text{AdS}_2)}A\ri](x)=\int d\tau \ G_{b\p}^{(\text{AdS}_2)}(x,\tau)\,A(\tau)\,,\\
G_{b\p}^{(\text{AdS}_2)}(x,\tau)&=\lim_{r_y\to 0} r_y^{-\De}\,G_{m^2}^{(\text{AdS}_2)}(x,y)\,,
}
where $G_{b\p}^{(\text{AdS}_2)}$ is the bulk-to-boundary propagator, which can be obtained as a particular limit of the AdS$_2$ bulk-to-bulk propagator  $G_{m^2}^{(\text{AdS}_2)}=1/ (- \square_\text{AdS$_2$}+m^2)$ (defined with the usual normalizable boundary conditions).

Now let us recall the change of variables that we implemented to obtain \eqref{gdef}. The flat space prescription translates to AdS$_2\times S^{d-2}$ as follows
\es{ChangeVariables}{
\phi_1=\de\phi_1-\sqrt{N}\, {1\ov - \square+s} J \quad \to \quad \phi_1=\de\phi_1-\sqrt{N}\, G_{b\p}\, \widetilde J\,,
}
where $\widetilde J$ determines the near boundary behavior of $\phi_1$ as
\es{ChangeVariables2}{
\phi_1(\tau,r,\hat{n})&\to {r^{1-\De}\ov (2\De-1)\Om_{d-2}}\,\widetilde J\,, \qquad \Delta=\frac12+\sqrt{\frac14+s_\p -{(4-d)(d-2)/ 4}}\,,
}
with the $\Om_{d-2}$ arising from the different normalization of the bulk-to-bulk propagator from the pure AdS case and $s_\p$ the boundary value of the dynamical field $s$ and
\es{Gbb}{
G_{bb}(x,y)\equiv {1\ov -\widetilde \square+s -{(4-d)(d-2)/ 4}}\,.
}
At large $N$ the problem can be analyzed using the saddle point approximation. First, the saddle point value of $s$ is a one-point function in the IR DCFT. Scalar one-point functions are only consistent with AdS isometries (i.e. the $SL(2,\mathbb{R})$ symmetry of the DCFT), if they are constant. Therefore $s(x)=s_0$ at the saddle point. Second, the $\phi_1$ one-point function can only be constant, if $\Delta=1$,  i.e. $\phi_1$ is a massless scalar, which then implies
\es{s0result}{
s_\p= s_0={(4-d)(d-2)\ov 4}
}
in perfect agreement with \eqref{snot}.

With these preparations we are ready to write down the curved space representation of the ``defect term'' in the action:
\es{DefectTerm}{
J^a{1\over-\square +s} J^a \quad&\to\quad \widetilde J  \,G_{\p\p}\, \widetilde J\,,\\
G_{\p\p}(\tau_1,\tau_2)&\equiv \lim_{r_{1,2}\to 0} r_1^{-1} r_2^{-1}\,G_{bb}(x_1,x_2)\,,
}
with $G_{\p\p}$ the boundary to boundary propagator. The full effective action then takes the form:
\es{SeffFull}{
S_\text{eff}[s,\widetilde J]&\equiv \frac12\le[{\rm Tr}\log\le(-\widetilde \square+s +{d-2\ov 4(d-1)}\widetilde R\ri)-\widetilde J\, G_{\p\p}\,\widetilde J\ri]\,.
}
We have argued above based on symmetry that this action has a saddle point at $s(x)=s_0$ with $s_0$ given in eq.~\eqref{s0result}. Indeed,  the saddle point equation for $s$ reads:\footnote{To derive eq. \eqref{SPeq} we used the following identities
\begin{align}\nonumber
&\frac{\delta}{\delta s(x)}{\rm Tr}\log\le(-\widetilde \square+s +{d-2\ov 4(d-1)}\widetilde R\ri)=\frac{\delta}{\delta s(x)}\int d y
\log\le(-\widetilde \square+s +{d-2\ov 4(d-1)}\widetilde R\ri)(y,y)=G_{bb}(x,x)\,,\\[0.5em] \nonumber
&\frac{\delta}{\delta s(x)}\widetilde J\, G_{\p\p}\,\widetilde J
=\frac{\delta}{\delta s(x)}\int d\tau_1 \int d\tau_2\tilde{J}(\tau_1)\tilde{J}(\tau_2)
\lim_{r_{1,2}\to 0} r_1^{-1} r_2^{-1}\,\left(\frac{1}{-\widetilde \square+s 
+{d-2\ov 4(d-1)}\widetilde R}\right)(x_1,x_2)\\[0.5em]
&\hphantom{\frac{\delta}{\delta s(x)}\widetilde J\, G_{\p\p}\,\widetilde J}=-\int d\tau_1 \int d\tau_2\tilde{J}(\tau_1)\tilde{J}(\tau_2)
 G_{b\p}(\tau_1,x) G_{b\p}(x,\tau_2)\,.
\end{align}}
\es{SPeq}{
0&=G_{bb}(x,x)\vert_{s_0}+\le(\widetilde J \int d\tau\ G_{b\p}(x,\tau)\vert_{s_0} \ri)^2\,.
}
 For constant $s$ the term $G_{bb}(x,x)$ is independent of $x$: it is a tadpole diagram in AdS$_2\times S^{d-2}$. The other term is  $\langle\phi_1\rangle^2$ (to leading order in $N$), and  using
  \es{bbdyProp}{
  G_{b\p}(x,\tau)={\frac{1}{\pi\Omega_{d-2}} \le({(\tau_x-\tau)^2+r_x^2\ov r_x}\ri)^{-1}}\,,
 }
 (through explicit computation)  we can verify that it is independent of the bulk point $x=(\tau_x,r_x,\hat{n}_x)$ ($\hat{n}_x$ is a unit vector parametrizing the coordinates on $S^{d-2}$). We can then rewrite eq.~\eqref{SPeq} as
 \es{SPeqB}{
 \langle\phi_a(x)\rangle&=-\delta_a^1\sqrt{N}\,  \le({\widetilde J/ \Om_{d-2}}\ri)\,,\\
\le({\widetilde J/ \Om_{d-2}}\ri)^2&=-G_{bb}(x,x)\,.
}
We will compute $\widetilde J$ in the next subsection.

Plugging back the saddle point into the action in eq.~\eqref{gdef} and expanding in $s$ fluctuations we get: 
\es{SeffExpansion}{
S_\text{eff}[s,\widetilde J]&=\frac12\le[{\rm Tr}\log\le(-\widetilde \square\ri)-\widetilde J \,  G_{\p\p}\, \widetilde J\ri]\\
&\quad+ \frac12 \int d^dx d^dy\ \de s(x) \le[-\frac12 G_{bb}(x,y)^2-\le({\widetilde J/ \Om_{d-2}}\ri)^2 G_{bb}(x,y)\ri] \de s(y)+\mO(\de s^3)\,, 
}
where every propagator is understood to be evaluated on the saddle point. 
Note that $\de s=\mO(N^{-1/2})$, thus higher powers of $\de s$ fluctuations are indeed suppressed.

\subsection{The \texorpdfstring{$\phi_1$}{phi1} one-point function}\label{SubSecPhi1N}

\begin{figure}[!t]
\centering
\includegraphics[scale=0.8]{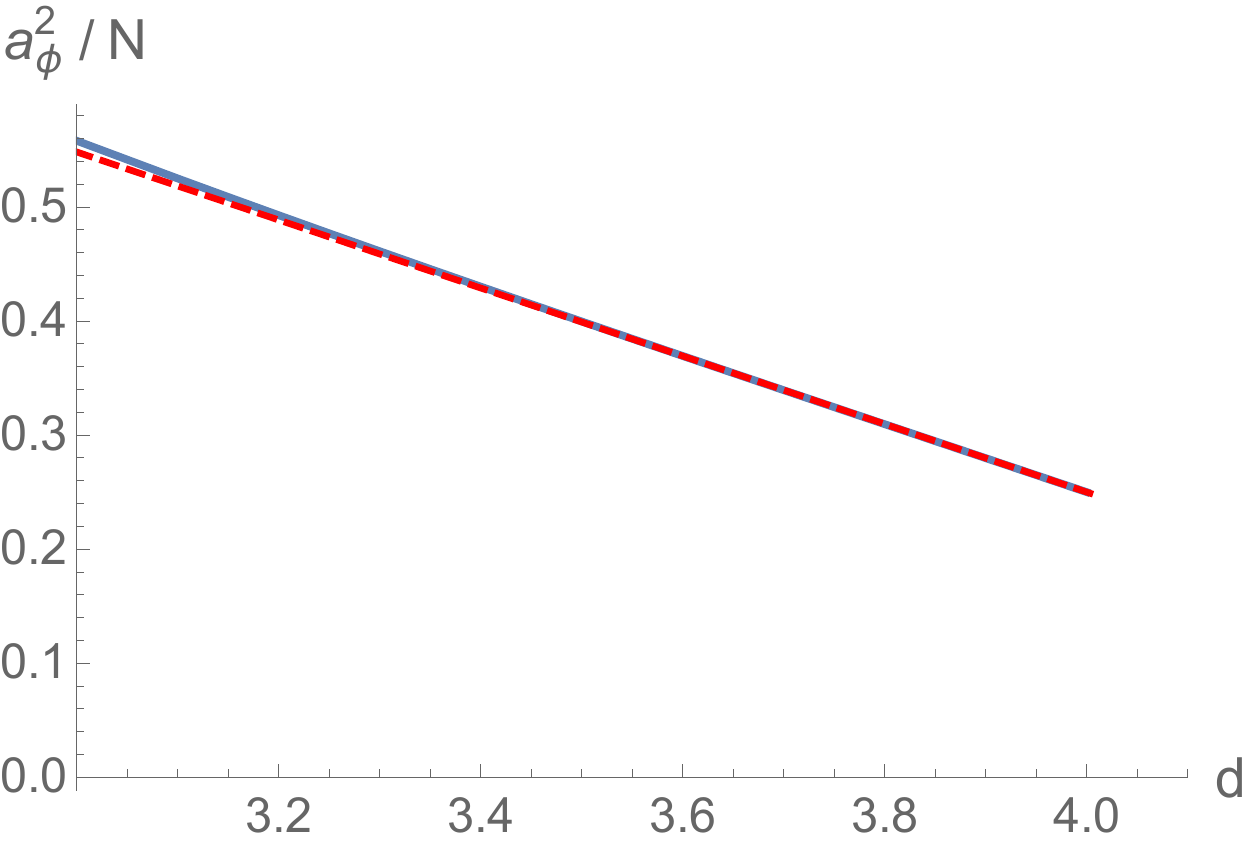}
\caption{We plot, to leading order in large $N$, the result for $a_{\phi}^2$ (blue) as a function of the dimension $d$. We compare it to the epsilon expansion result \eqref{eq_Aphi_epsilon} near $d=4$ (dashed red), finding perfect agreement.  }
\label{fig:aphi1}
\end{figure}

Next, we set out to compute $ \langle\phi_1\rangle=-\sqrt{N}\, \le({\widetilde J/ \Om_{d-2}}\ri)$ by evaluating  the tadpole diagram in AdS, namely, from  eq.~\eqref{SPeqB} we have the relation
\es{SPeq2}{
\le({\widetilde J/ \Om_{d-2}}\ri)^2&=-G_{bb}(x,x)=-{1\ov {\rm vol}\le(\text{AdS$_2\times S^{d-2}$}\ri)}{\rm Tr}\le({1\ov -\widetilde \square}\ri)\,.
}
There exist efficient techniques to evaluate the latter quantity. It will be useful below to consider its massive generalization:
\es{TrComp}{
{\rm Tr}\le({1\ov -\widetilde \square+m^2}\ri)&=\int_{-\infty}^\infty d\nu \ D(\nu) \sum_{\ell=0}^\infty d_\ell {1\ov \nu^2+{1\ov 4}+m^2+\lam_\ell}\,,\\
D(\nu)&= {{\rm vol}\le(\text{AdS$_2$}\ri)\ov 4\pi} \nu \tanh (\pi \nu)\,,\\
d_\ell&={(2\ell+q-1)\Ga\le(\ell+q-1\ri)\ov \Ga\le(q\ri)\Ga\le(\ell+1\ri)}\,,\\
\lam_\ell&=\ell(\ell+q-1)\,,
}
where $D(\nu)$ is the AdS$_2$ spectral density \cite{Carmi:2018qzm}, $q\equiv d-2$, and $d_\ell$ is the degeneracy of the eigenvalue $\lam_\ell$ of the Laplacian on $S^q$. We will work in dimensional regularization, i.e. evaluate quantities in dimensions where the integral and sum converge and then analytically continue in $d$ to the values of interest. The integrand at large $\nu$ decays as $1/\nu$, which naively leads to a divergence. However the coefficient of this divergence is $\sum_{\ell=0}^\infty d_\ell$ which is well-known to be zero in dimensional regularization.\footnote{The sum converges for $q<0$ and we analytically continue from there.} To make \eqref{TrComp} easier to manipulate, in particular to be able to exchange the $\nu$ integral with the $\ell$ sum, we consider a subtraction that vanishes in dimensional regularization (due to the identity $ \sum_{\ell=0}^\infty d_\ell=0$):
\es{Subtraction}{
{\rm Tr}\le({1\ov -\widetilde \square+m^2}\ri)&=\int_{-\infty}^\infty d\nu \ D(\nu) \sum_{\ell=0}^\infty d_\ell \le[{1\ov \nu^2+{1\ov 4}+m^2+\lam_\ell}-{1\ov \nu^2}\ri]\,,
}
where there is no singularity at $\nu=0$ since $D(\nu)\sim\nu^2$. The $\nu$ integral can be evaluated by closing the $\nu$ contour (on either the upper or lower half plains) and doing the resulting sum over residues. We obtain the result:
\es{ellSum}{
{\rm Tr}\le({1\ov -\widetilde \square+m^2}\ri)&={\rm vol}\le(\text{AdS$_2$}\ri) \sum_{\ell=0}^\infty S(\ell,q)\,\\
S(\ell,q)&=-{d_\ell\ov 2\pi} \le[\psi\le(\frac12+\sqrt{{1\ov 4}+m^2+\lam_\ell}\ri)+\ga_E+\log 4\ri]\,.
}
The sum is divergent for the regime of interest $1\leq q\leq 2$ and cannot be analytically performed. We can however find an appropriate subtraction (by expanding at large $\ell$) that makes the problem tractable.
\es{ellSum2}{
{\rm Tr}\le({1\ov -\widetilde \square+m^2}\ri)&={\rm vol}\le(\text{AdS$_2$}\ri) \le(S(0,q)+\sum_{\ell=1}^\infty \le[S(\ell,q)-\hat{S}(\ell,q)\ri]+\sum_{\ell=1}^\infty \hat{S}(\ell,q)\ri)\,,\\
\hat{S}(\ell,q)&=-{\log \ell+\ga_E+\log 4\ov \pi \,\Ga\le(q\ri)}\, \ell^{q-1}+\dots\,,
}
where the subsequent terms indicated by dots are down by integer powers of $\ell$ down to $\ell^{q-3}$. (We need these terms in the following, but we are not writing them down explicitly to avoid clutter.)
The first sum is now convergent in the dimensions of interest $1\leq q\leq 2$, while the second sum over $\hat{S}(\ell,q)$ is evaluated for $q<0$, using $\sum_{\ell=1}^\infty \ell^s=\zeta(-s)$ and $\sum_{\ell=1}^\infty \ell^s\log(s)=-\zeta'(-s)$ and then analytically continued to the physical $q$ values. Since the summand after the subtraction decays as $\ell^{q-4} \log\ell$, we can evaluate it numerically to any desired precision by approximating it with the a finite sum up to some $\ell_\text{cutoff}$, see  fig.~\ref{fig:aphi1}, where we plot $a^2_{\phi}$ for $3\leq d\leq4$ by evaluating eq.~\eqref{ellSum2} with $m^2=0$. Recall that the relation between $a^2_{\phi}$ and $\widetilde J^2$ is\footnote{Notice that the factor ${\rm vol}\le(\text{AdS$_2$}\ri)$ cancels  between the denominator and numerator in the final result.}
\es{Aanda}{
a_{\phi}^2&=N\le({\widetilde J\ov {\cal N}_\phi \,\Om_{d-2}}\ri)^2\,,\\
a_{\phi}^2\vert_{3d}&=0.558113\, N\,,
}
where ${\cal N}_\phi^2$ can be read from eq.~\eqref{ffield}. 

In fig.~\ref{fig:aphi1} we demonstrate the perfect agreement between the large $N$ and epsilon expansion results in their overlapping regime of validity. This provides a strong crosscheck for both methods.\footnote{We have also verified the agreement analytically expanding the argument $S(\ell,q)-\hat{S}(\ell,q)$ of the convergent sum in eq.~\eqref{ellSum2} in $q=2-\varepsilon$ and comparing the coefficients with eq.~\eqref{eq_Aphi_epsilon}.} Another check for the large $N$ computation is the following. For the trivial defect we can repeat the above computation with $s_0=0$ that corresponds to $m^2=-{(4-d)(d-2)/ 4}$. For the trivial defect we expect that one-point functions vanish, i.e. $\widetilde J=0$, and we indeed recover this from our computation.\footnote{We have checked this vanishing analytically for arbitrary dimensions $d$.}

\subsection{The \texorpdfstring{$g$}{g}-function}\label{SubSecNg}

With the above ingredients it is now straightforward to compute the leading order $g$ coefficient.  Due to the large $N$ limit being described by a saddle point, and due to the fact that $g<1$ follows from the $g$-theorem, we expect that the $g$ function is exponentially small in the large $N$ limit.

Let us first consider the ``defect term''  from eq.~\eqref{SeffExpansion}. We consider AdS$_2$ in global coordinates, so that its conformal boundary is an $S^1$  (appropriate for a circular defect in flat space), and we get
\es{bdyg}{
{N\ov 2} \widetilde J\, G_{\p\p}\, \widetilde J&= \frac{N\widetilde{J}^2}{\Omega_{d-2}} \int_0^{2\pi} d\phi {1\over 4\sin^2(\phi/2) }\,.
}
The integral is divergent.  We encountered an analogous contribution in sec.~\ref{SecWarmUp} from an exactly marginal defect deformation in free scalar theory;\footnote{In \eqref{freeg} we have to take $d\to 4$ to arrive at this result.} therefore this integral must be set to zero.

Having dealt with the boundary term above, we turn our attention to the bulk contribution ${\rm Tr}\log\le(-\widetilde \square\ri)$ to  $g$. To evaluate it, we will need to give meaning to (the infinite) ${\rm vol}\le(\text{AdS$_2$}\ri)$. Let us write AdS$_2$ in global coordinates:
\es{AdS2Global}{
d\tilde s^2=d\rho^2+\sinh^2\rho d\phi^2\,,
}
and put the boundary at (large) radial cutoff $\rho=\rho_c$, we get
\es{bdyLength}{
&{\rm vol}\le(\p\text{AdS$_2$}\ri)=2\pi\le[{e^{\rho_c}\ov 2}+0+\mO\le(e^{-\rho_c}\ri)\ri]\,,\\
&{\rm vol}\le(\text{AdS$_2$}\ri)=2\pi \int_0^{\rho_c} d\rho \ \sinh \rho=2\pi\le[{e^{\rho_c}\ov 2}-1+\mO\le(e^{-\rho_c}\ri)\ri]\,,\\
&\implies \qquad {\rm vol}\le(\text{AdS$_2$}\ri)={\rm vol}\le(\p\text{AdS$_2$}\ri)-2\pi+O\le(e^{-\rho_c}\ri)\,.
}
We can absorb any contribution proportional to ${\rm vol}\le(\p\text{AdS$_2$}\ri)$ into a defect cosmological constant counterterm, thus for the computation of $g$ we can use the regularized volume of AdS$_2$ familiar from prior literature~\cite{Casini:2011kv,Klebanov:2011uf}:
\es{AdS2vol}{
 \qquad {\rm vol}\le(\text{AdS$_2$}\ri)\vert_\text{reg}=-2\pi\,.
}

\begin{figure}[t!]
\centering
\includegraphics[scale=0.8]{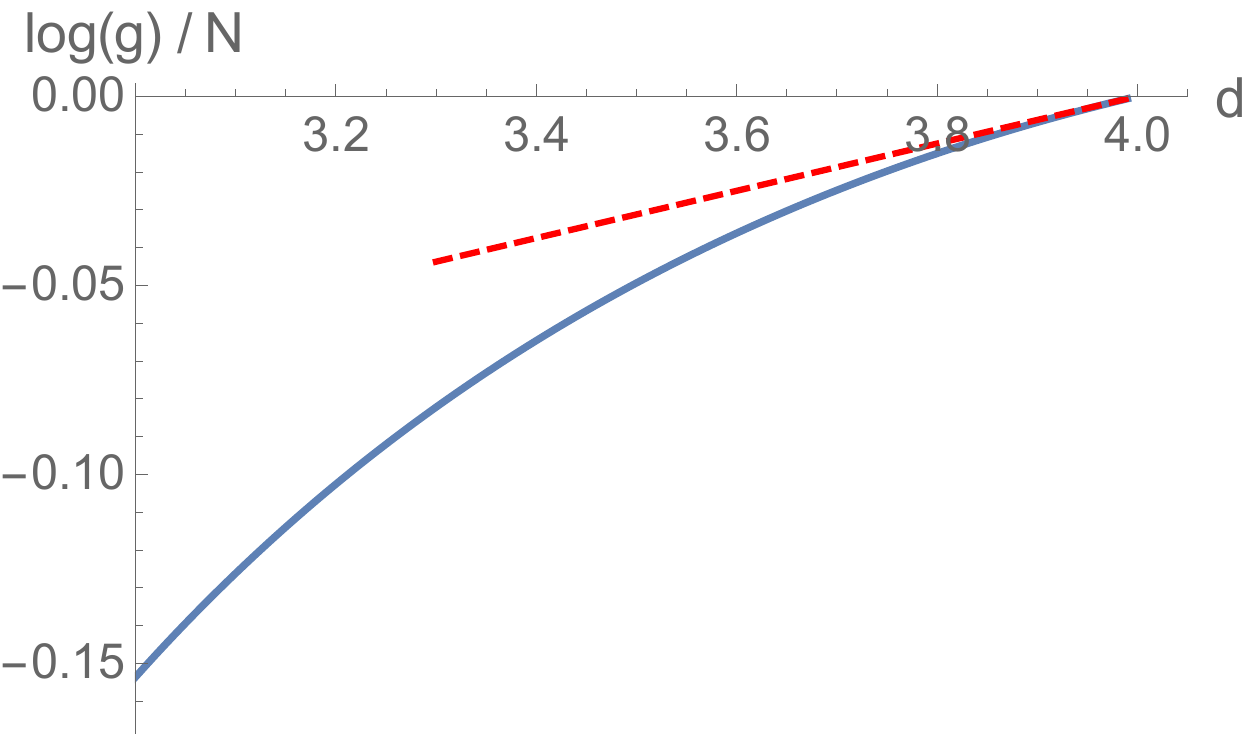}
\caption{We plot the leading order in large $N$ result for $\log g/N$ (blue) as a function of the dimension $d$. We compare it to the epsilon expansion result \eqref{eq_logG_1loopFix} near $d=4$ (dashed red), finding perfect agreement.  }
\label{fig:logg}
\end{figure}

Recalling that $g$ is the difference of free energies between the DCFT of interest and the trivial defect theory, we have:
\es{gleading}{
\log g&=-{N\ov2}\le[{\rm Tr}\log\le(-\widetilde \square\ri)-{\rm Tr}\log\le(-\widetilde \square-{(4-d)(d-2)\ov 4}\ri)\ri]+O\le(N^0\ri)\\
&={N\ov2}\int_{-{(4-d)(d-2)/ 4}}^0 dm^2\  {\rm Tr}\le({1\ov -\widetilde \square+m^2}\ri)\,,
}
where the lower boundary of the integral corresponds to $m^2=-{(4-d)(d-2)/ 4}$,  the AdS mass corresponding to the trivial defect (as discussed above), while $m^2=0$ is the value appropriate for the nontrivial DCFT. Happily, we can directly integrate $S(\ell,q)$ from eq.~\eqref{ellSum2} to obtain
\es{Sint}{
\log g&=N\sum_{\ell=0}^\infty T(\ell,q)\\
T(\ell,q)&={d_\ell}\le[\psi^{(-2)}\le(\frac12+\sqrt{{1\ov 4}+\lam_\ell}\ri)-\sqrt{{1\ov 4}+\lam_\ell}\,\log\Ga\le(\frac12+\sqrt{{1\ov 4}+\lam_\ell}\ri)\ri.\\
&\quad\le.-\psi^{(-2)}\le(\ell+{q\ov 2}\ri)+\le(\ell+{q-1\ov 2}\ri)\,\log\Ga\le(\ell+{q\ov 2}\ri) -{q(2-q)(\ga_E+\log 4)\ov 8}\ri]\,,
}
where $\psi^{(n)}(z)$ is the $n$th polygamma function and we used that $\frac12+\sqrt{{(q-1)^2\ov 4}+\lam_\ell}=\ell+{q\ov 2}$.\footnote{For $0<q<1$ and $\ell=0$ this formula involves choosing the nonobvious branch of the square root. This choice ensures that the subtracted contribution is correctly quantized and corresponds to the trivial defect; a topic that we plan to return to in the future.}
The evaluation of the sum proceeds in complete analogy with what we encountered in eq.~\eqref{ellSum2}. The asymptotics $\hat{T}(\ell,q)$ can be obtained by simply integrating $\hat{S}(\ell,q)$ with respect to $m^2$. Beyond remarking that the summand $T(\ell,q)$ vanishes for $d=4$ ($q=2$) giving $\log g=0$,\footnote{The vanishing of $T(\ell,q)$ in $d=4$ in itself does not prove the vanishing of $\log g$, we also have to check that summing up $\hat{T}(\ell,q)$ does not produce a nonzero contribution.} we omit the details of this evaluation and just plot the final result in fig.~\ref{fig:logg}. From the plot it is evident that we get perfect agreement with the epsilon expansion result \eqref{eq_logG_1loopFix} near $d=4$.\footnote{We have also verified this agreement analytically by expanding $T(\ell,q)-\hat{T}(\ell,q)$ around $q=2$ and performing the sum analytically. }  The most physically interesting result is 
\es{g3d}{
\log g\vert_{3d}&=- 0.153673\,N+\mO\le(N^0\ri)\,.
}

\subsection{The DCFT spectrum}\label{SubSecNspectrum}

In this section we consider the spectrum of single-trace $O(N-1)$ singlet operators,  focusing on operators with zero transverse spin.

We shall proceed similarly in spirit to the discussion above eq.~\eqref{tiltchannel} and extract the operator spectrum from the propagator of the $\delta s$ fluctuation. From the action in eq.~\eqref{SeffExpansion} we infer that, neglecting an overall factor of $N$, this is given by the inverse of the following function 
\begin{equation}\label{eq_Ginv_deltaS}
G_{\delta s\,\delta s}^{-1}(x,y)=-\frac12 G_{bb}(x,y)^2-\le({\widetilde J/ \Om_{d-2}}\ri)^2 G_{bb}(x,y)\,,
\end{equation}
where $G_{bb}(x,y)$ is defined in eq.~\eqref{SPeq}.  The first term in eq.~\eqref{eq_Ginv_deltaS} is diagrammatically represented as a bubble diagrams, qualitatively analogous to the one giving the propagator \eqref{snorm} in the absence of the defect (but with a different $s_0$). 

The bulk to bulk propagator $G_{bb}(x,y)$ can be written explicitly as a sum over KK modes on the sphere in terms of Gegenbauer polynomials and using the spectral decomposition for the AdS$_2$ propagator\footnote{See e.g.  \cite{Cornalba:2007fs,Cornalba:2008qf,Penedones:2010ue,Giombi:2017hpr,Carmi:2018qzm} for details on the AdS spectral decomposition.}
\begin{equation}\label{eq_Gbb_spectral}
G_{bb}(x,y)={1\ov {\rm vol}\le({\cal M}\ri)}\int_{-\infty}^{\infty} d\nu\  D(\nu)\sum_{\ell=0}^{\infty} d_\ell\, \frac{1}{\nu^2+\frac14+\lambda_\ell}\, 
\Omega_{\nu}\left(X,Y\right)\, {\cal C}_{\ell}^{\left(\frac{q-1}{2}\right)}(\hat{n}_x\cdot\hat{n}_y)\,,
\end{equation}
where ${\cal M}=\text{AdS$_2\times S^{q}$}$, we introduced a rescaled Gegenbauer polynomial as ${\cal C}_{\ell}^{(m)}(x)\equiv C_{\ell}^{(m)}(x)/C_{\ell}^{(m)}(1)$, and $\Omega_{\nu}(X,Y)$ is the harmonic function in AdS
\begin{equation}
-\Delta_X^{(AdS_2)}\Omega_\nu(X,Y)=
\left(\nu^2+\frac{1}{4}\right)\Omega_\nu(X,Y)\,,
\end{equation}
normalized so that $\Omega_\nu(X,X)=1.$\footnote{This normalization differs from the usually adopted one, see e.g. \cite{Carmi:2018qzm}.} (Notice that the coincident point limit is regular.) The detailed expression of the harmonic function and other details on AdS spectral decomposition can be found, e.g. in \cite{Carmi:2018qzm}. In the following we will not need its explicit form. 

An important fact about eq.~\eqref{eq_Gbb_spectral} is that the integrand in eq.~\eqref{eq_Gbb_spectral} has poles at 
\begin{equation}\label{eq_nu_pole0}
\nu=\pm i \left(\Delta_{\ell}-\frac{1}{2}\right)\,,
\end{equation}
where $\Delta_{\ell}$ coincides with the scaling dimension for the $O(N-1)$ vector defect operators given in eq.~\eqref{tiltchannel}. This is not an accident. Indeed, similarly to the familiar K\"all{\'e}n-Lehmann decomposition, the complex poles in the AdS spectral decomposition are in one to one correspondence with the masses of the exchanged particles. Equivalently, these are related to the spectrum of boundary (or in this case defect) operators via eq.~\eqref{eq_nu_pole0}. This fact will be important in what follows.

At least in principle, it is then clear how to extract the spectrum of scaling dimensions for the defect operators exchanged in the bulk to defect channel of $\delta s$ from eq.~\eqref{eq_Ginv_deltaS}. This can be done by decomposing the inverse propagator in eq.~\eqref{eq_Ginv_deltaS} similarly to eq.~\eqref{eq_Gbb_spectral}
\begin{equation}\label{Gbb_formal1}
G_{\delta s\,\delta s}^{-1}(x,y)={1\ov {\rm vol}\le({\cal M}\ri)}\int_{-\infty}^{\infty} d\nu\  D(\nu)\sum_{\ell=0}^{\infty} d_\ell\, B_{\ell}(\nu)\, 
\Omega_{\nu}\left(X,Y\right)\, {\cal C}_{\ell}^{\left(\frac{q-1}{2}\right)}(\hat{n}_x\cdot\hat{n}_y)\,,
\end{equation}
where $B_{\ell}(\nu)$ are the appropriate spectral densities. Formally, these can be found inverting eq.~\eqref{Gbb_formal1} as follows \cite{Cornalba:2008qf,Giombi:2017hpr,Carmi:2018qzm}:
\begin{equation}\label{eq_inversion}
B_{\ell}(\nu)=\int d^dx\sqrt{g(x)}\, G_{\delta s\,\delta s}^{-1}(x,y)\, \Omega_{\nu}(X,Y)\, {\cal C}_{\ell}^{\left(\frac{q-1}{2}\right)}(\hat{n}_x\cdot\hat{n}_y)\,.
\end{equation}
The complex zeros of $B_{\ell}(\nu)$ then correspond to the scaling dimensions of defect operators with transverse spin $\ell$.

In the following we will focus on the spectrum of scalar operators, since based on the results in sec.~\ref{SecEpsilon} we expect the lowest dimension irrelevant $O(N-1)$ singlet operator to have zero transverse spin. To this aim, we set $\ell=0$ in eq.~\eqref{eq_inversion} and get
\begin{equation}\label{eq_B0_first}
\begin{split}
B_0\left( \nu\right)&=\int d^dx\sqrt{g(x)}\left[-\frac{1}{2}G_{bb}(x,y)^2-\le({\widetilde J/ \Om_{d-2}}\ri)^2 G_{bb}(x,y)\right]\Omega_{\nu}(X,Y) \\
&=-\frac12 \int d^dx\sqrt{g(x)}\, G_{bb}(x,y)^2\,\Omega_{\nu}(X,Y)
+\frac{\widetilde J^2}{\Om_{q}^2\,\Delta(\Delta-1)}\,,
\end{split}
\end{equation}
where in the second line we have set $\nu=\pm i\left(\Delta-\frac12\right)$ and we used eq.~\eqref{eq_Gbb_spectral} to extract the spectral density of the homogeneous mode of the free massless propagator.  

We will soon show how to evaluate the integral of the square propagator in the second line of eq.~\eqref{eq_B0_first} by taking advantage of recent remarkable developments in the study of AdS loop diagrams (see e.g.  \cite{Fitzpatrick:2010zm,Penedones:2010ue,Fitzpatrick:2011hu,Giombi:2017hpr,Carmi:2018qzm,Carmi:2019ocp}).  However,  it is instructive to first discuss  how to recover the epsilon expansion one-loop result for the scaling dimension of the operator $\hat{\phi}_1$ given in eq.~\eqref{eq_gamma1_epsilon}.   Remarkably, this can be done without knowing the detailed form of the spectral decomposition of the bubble diagram. To this aim, notice that due to the short distance behavior of the propagator $G_{bb}(x,y)\sim 1/|x-y|^{d-2}$ the first integral in eq.~\eqref{eq_B0_first} displays a short-distance logarithmic divergence in $d=4$ (recall that $\Omega_{\nu}(X,X)=1$). This implies that in $d=4-\varepsilon$ with $\varepsilon>0$ the result displays a $1/\varepsilon$ pole, whose coefficient can be extracted by approximating the bulk-to-bulk propagator $G_{bb}(x,y)$ with the flat space one \eqref{ffield}. As a result we rewrite eq.~\eqref{eq_B0_first} as:
\begin{equation}\label{eq_B0_epsilon}
B_0\left(\pm i\Big(\Delta-\frac{1}{2}\Big)\right)=-\frac{1}{16\pi^2\varepsilon}+\frac{\widetilde J^2}{16\pi^2\,\Delta(\Delta-1)}+\text{regular}\,.
\end{equation}
More precisely, we shall see later that the terms neglected in eq.~\eqref{eq_B0_epsilon} are indeed regular for sufficiently generic $\Delta$ and \eqref{eq_B0_epsilon} always holds for $\Delta<2$. Then we can solve eq.~\eqref{eq_B0_epsilon} by choosing $\Delta$ such that the second term cancels the pole for $\varepsilon\rightarrow 0$. Using that $\widetilde J^2=1+\mO\left(\varepsilon\right)$ from eqs.~\eqref{Aanda} and \eqref{eq_Aphi_epsilon}, we thus conclude that a solution to $B_0\left(\pm i(\Delta-1/2)\right)=0$ reads:
\begin{equation}\label{eq_Delta1_Nepsilon}
\Delta=1+\varepsilon+\mO\left(\varepsilon^2\right)\,,
\end{equation}
where the $\mO\left(\varepsilon^2\right)$ corrections are determined by requiring that the regular part in eq.~\eqref{eq_B0_epsilon} also cancels. Eq.~\eqref{eq_Delta1_Nepsilon} perfectly agrees with the diagrammatic result \eqref{eq_gamma1_epsilon} for $\Delta(\hat{\phi}_1)$. Notice that at this stage it is not obvious how eq.~\eqref{eq_B0_epsilon} may admit different solutions, corresponding to the other operators in the bulk to defect OPE of $s$.  We will discuss the resolution of this apparent paradox after providing a more precise analysis of the spectral density $B_0(\nu)$.

To find the spectral density \eqref{eq_B0_first} for arbitrary $d$ and $\Delta$ we use a result of ref.~\cite{Carmi:2018qzm}. There, the authors derived an expression for the spectral density for the AdS bubble diagram corresponding to the product of two propagators with equal masses. More precisely, denoting with $G_{m^2}^{(\text{AdS}_2)}$ the AdS$_2$ scalar propagator with mass $m$, they found:
\begin{equation}\label{eq_CarmiFormula}
\int d^2X\left[G_{m^2}^{(\text{AdS}_2)}(X,Y)\right]^2\Omega_{\nu}(X,Y)=\tilde{B}_{m^2}(\nu)\,,
\end{equation}
where, setting $m^2=\tilde{\Delta}(\tilde{\Delta}-1)$,
\begin{align}\label{eq:finalBnu}
 \begin{aligned}
 \frac{4(4\pi)^{\frac{1}{2}}\tilde{B}_{m^2}(\nu)}{\Gamma(\tilde{\Delta})^2\Gamma\left(2\tilde{\Delta}-\frac{1}{2}\right)}=
& \left\{\Gamma\left(\tilde{\Delta}-\frac{1+2i\nu}{4}\right)\,{}_5\tilde{F}_{4}\left[\begin{array}{c}\{\frac{1}{2},\tilde{\Delta},\tilde{\Delta},\tilde{\Delta}-\frac{1+2i\nu}{4},2\tilde{\Delta}-\frac{1}{2}\}\\\{\tilde{\Delta}+\frac{1}{2},\tilde{\Delta}+\frac{1}{2},\tilde{\Delta}+\frac{3-2i\nu}{4},2\tilde{\Delta}\}\end{array};1\right]\right.\\
+& \left. \Gamma\left(\tilde{\Delta}-\frac{1-2i\nu}{4}\right)\,{}_5\tilde{F}_{4}\left[\begin{array}{c}\{\frac{1}{2},\tilde{\Delta},\tilde{\Delta},\tilde{\Delta}-\frac{1-2i\nu}{4},2\tilde{\Delta}-\frac{1}{2}\}\\\{\tilde{\Delta}+\frac{1}{2},\tilde{\Delta}+\frac{1}{2},\tilde{\Delta}+\frac{3+2i\nu}{4},2\tilde{\Delta}\}\end{array};1\right]\right\}\,.
 \end{aligned}
 \end{align}
In eq.~\eqref{eq:finalBnu} $_5\tilde{F}_4$ is the regularized hypergeometric function. 
It can be shown that the asymptotic behavior of eq.~\eqref{eq:finalBnu} at large $m^2$ and fixed $\nu$ is:
\begin{equation}\label{eq_Bnu_asymp}
\tilde{B}_{m^2}(\nu)\overset{m^2\rightarrow\infty}{\sim}\frac{1}{4\pi m^2}\left[1+\mO\left(\frac{1}{m}\right)\right]\,.
\end{equation}
This property will be important in what follows.

With eq.~\eqref{eq:finalBnu} we finally have all the ingredients to study the zeroes of the $\ell=0$ spectral density in eq.~\eqref{Gbb_formal1}. Using eq.~\eqref{eq_CarmiFormula} and the orthogonality of Gegenbauer's polynomials,\footnote{More precisely we use:
\begin{equation}
\int d^q\hat{n}_x\ {\cal C}_{\ell}^{\left(\frac{q-1}{2}\right)}(\hat{n}_x\cdot\hat{n}_y) \, {\cal C}_{p}^{\left(\frac{q-1}{2}\right)}(\hat{n}_x\cdot\hat{n}_z)
=\delta_{\ell p}\,{\Omega_q\ov d_{\ell}}\, {\cal C}_{\ell}^{\left(\frac{q-1}{2}\right)} (\hat{n}_y\cdot\hat{n}_z)\,.
\end{equation}}
we can write the contribution from the squared AdS$_2\times S^q$ propagator in eq.~\eqref{eq_Ginv_deltaS} as an infinite sum over KK modes on the sphere and obtain the following result:
\begin{equation}\label{eq_MasterEq}
B_0\left(\pm i\Big(\Delta-\frac{1}{2}\Big)\right)=\frac{\widetilde J^2}{\Om_{q}^2\,\Delta(\Delta-1)}-\frac{1}{2}\sum_{\ell=0}^{\infty}\frac{d_{\ell}}{\Omega_q}\tilde{B}_{\lambda_{\ell}}\left(\pm i\Big(\Delta-\frac{1}{2}\Big)\right)\,.
\end{equation}
The zeros of eq.~\eqref{eq_MasterEq} then correspond to the scaling dimensions of single-trace scalar defect operators which are neutral under both the $SO(d-1)$ rotation group and the $O(N-1)$ internal group.  Notice that thanks to the property \eqref{eq_Bnu_asymp} the sum in eq.~\eqref{eq_MasterEq} is convergent for $d<4$ (equivalently $q<2$), while it diverges logarithmically in $d=4$, in agreement with the discussion leading to  eq.~\eqref{eq_B0_epsilon}.\footnote{It is also possible to reproduce eq.~\eqref{eq_B0_epsilon} from \eqref{eq_MasterEq} by replacing $\tilde{B}_{\lambda_{\ell}}$ with its asymptotic behavior \eqref{eq_Bnu_asymp} and using the formula for the degeneracy $d_{\ell}$ in eq.~\eqref{TrComp}.}

\begin{figure}[t!]
\centering
\includegraphics[scale=0.8]{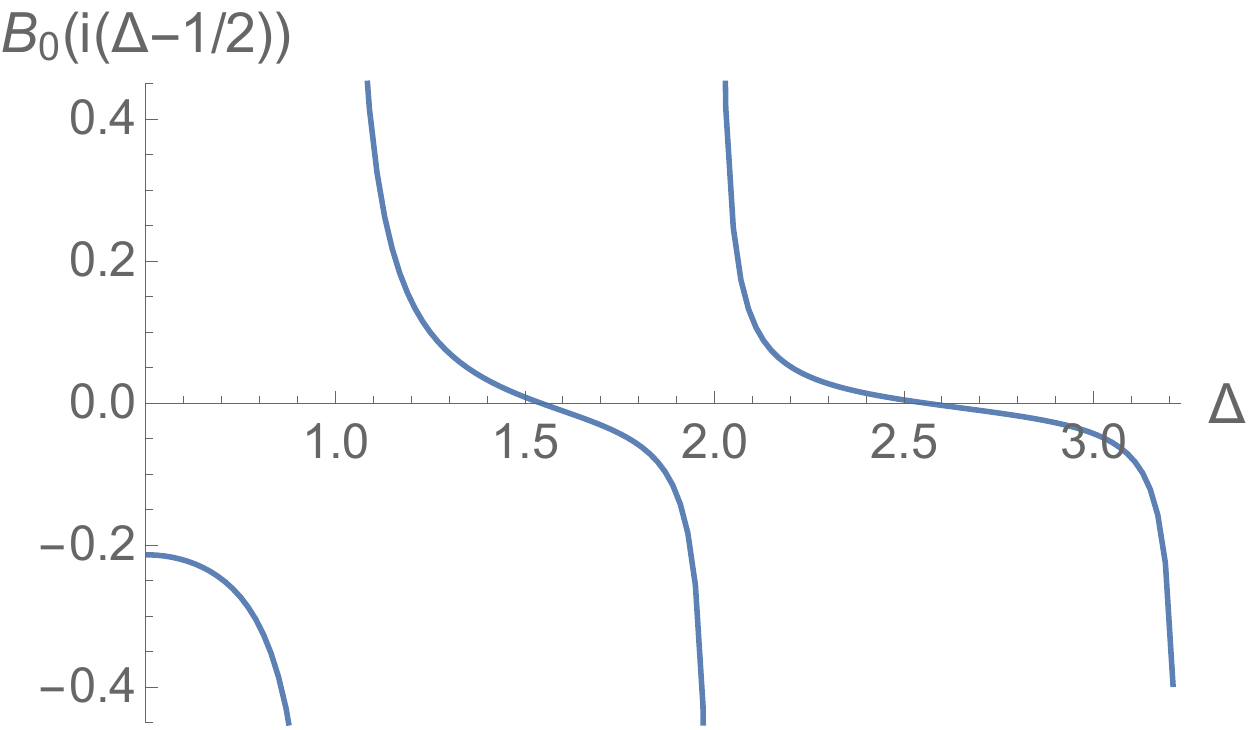}
\caption{Plot of the $\ell=0$ spectral density in eq.~\eqref{eq_MasterEq} for $1/2<\De\leq 3$. }
\label{fig:B0}
\end{figure}

It is now simple to find the zeroes of eq.~\eqref{eq_MasterEq} numerically in $d=3$. We first restrict our attention to operators with $\Delta<3$.  For $1/2\leq \Delta\leq 3$ the spectral density is shown explicitly in fig.~\ref{fig:B0}; notice that the spectral density is invariant under the shadow reflection $\Delta\leftrightarrow 1-\Delta$, hence the range $\Delta>1/2$ fully specifies it. We do not find any zero for $\Delta\leq 1$, in agreement with the expected stability of the DCFT. As the plot clearly shows, we find two zeros for $1\leq \Delta\leq 3$, corresponding to the scaling dimensions of the two lowest dimension singlets exchanged by $s$, denoted $\hat{\phi}_1$ and $\hat{s}_-$ in the notation of sec.~\ref{SecEpsilon}. We find their scaling dimensions to be:
\es{spectrum}{
\Delta(\hat{\phi}_1)\vert_{3d}&=1.541728082+\mO\le(1/N\ri)\,,\\
\Delta(\hat{s}_{-})\vert_{3d}&=2.556469028+\mO\le(1/N\ri)\,.
}

\begin{figure}[t!]
\centering
\includegraphics[scale=0.8]{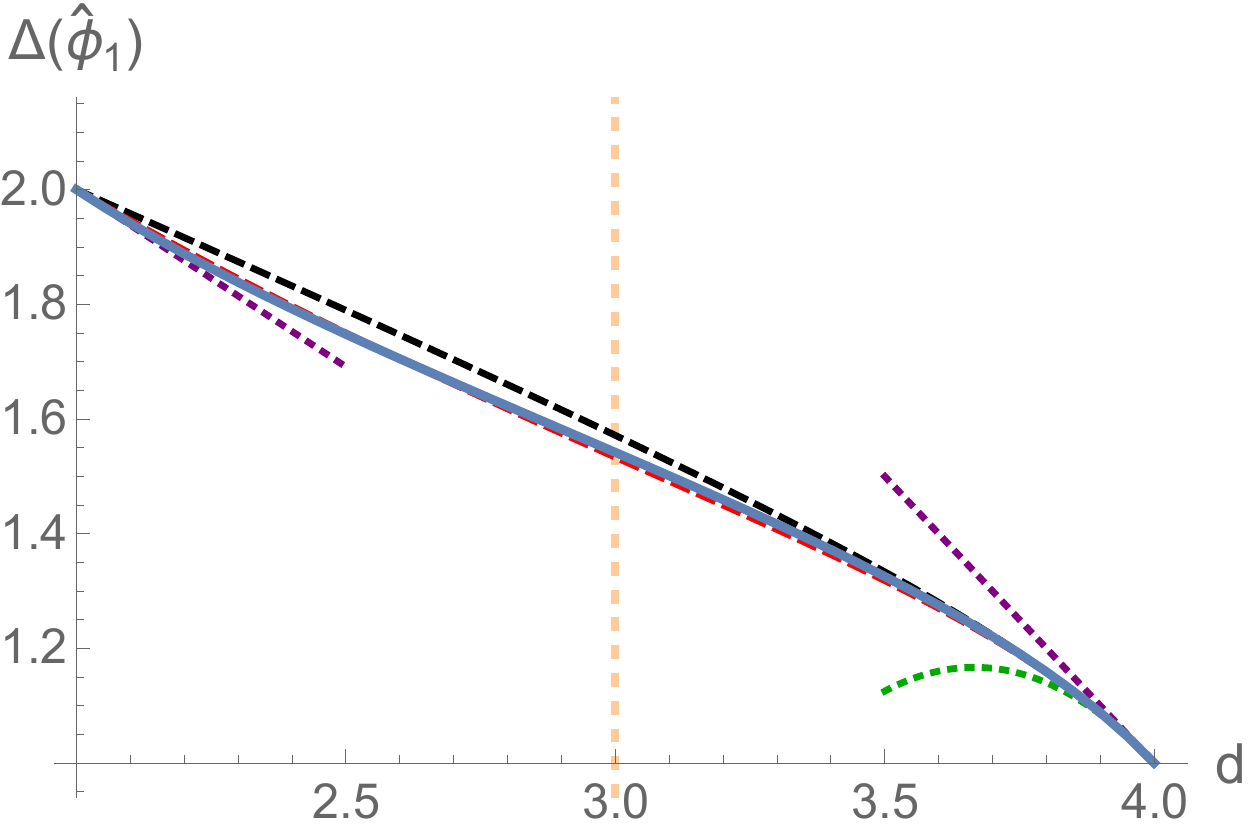}
\caption{We plot the leading order in large $N$ result for $\Delta(\hat{\phi}_1)$ (blue) as a function of the dimension $d$. We compare it to the one loop (dotted purple) and two loop (dotted green)   $4-\ep$ expansion results near $d=4$, finding perfect agreement. We also included the expansion of the function in $2+\tilde\ep$ (dotted purple) near $d=2$ from eq.~\eqref{spectrum2pepd}. We also plot the Pad\'e$_{[1,2]}$ and Pad\'e$_{[2,1]}$ approximants (red and dashed) that provide very good approximations to the true answer in all dimensions.}
\label{fig:Delta1}
\end{figure}

In figure~\ref{fig:Delta1} we plot the value for $\Delta(\hat{\phi}_1)$ for $2\leq d<4$ and compare it with the two-loop epsilon expansion result \eqref{eq_gamma1_epsilon}, finding an agreement. We also compare the numerical result with the large $N$ limit of the  Pad\'e$_{[1,2]}$ and Pad\'e$_{[2,1]}$ approximants that were obtained in sec.~\ref{SubSecD2} from the two-loop epsilon expansion result combined with $\Delta(\hat{\phi}_1)\vert_{2d}=2$. We see that this only $\mO(2\%)$ off from the correct result in $d=3$.

We end with a couple of further results. First, we show that $\Delta(\hat{\phi}_1)\vert_{2d}=2$, as was argued based on the $d=2+\tilde\ep$  (or $q=\tilde{\ep}$) expansion in sec.~\ref{SubSecD2}, see also fig.~\ref{fig:Delta1}. For this, we first have to compute ${\widetilde J^2}$.  Using 
\es{dell2}{
d_0=1\,,\qquad d_1=1+\tilde\ep\,,\qquad d_{\ell>1}=\mO(\tilde\ep)\,,
}
in eq.~\eqref{ellSum2}, we obtain
\es{firstterm}{
{\widetilde J^2}={\Om_q\ov \pi\, \tilde \ep }+ \mO(\tilde\ep)\,.
}
This then determines the first term in  eq.~\ref{eq_MasterEq}: it diverges as $1/\tilde \ep $ for $\De\approx 2$. To find the zero of $B_0\left( i\Big(\Delta-\frac{1}{2}\Big)\right)$, we have to compensate this divergence with a pole. To find this pole, we use the following additional property of eq.~\eqref{eq:finalBnu}:  $\tilde{B}_{m^2}(\nu)$ has simple poles at $\nu=\pm i\left(2\tilde{\Delta}+2 k-\frac12\right) $ for every $k\in\mathbb{N}$. These are associated with the propagation of two-particle (double-trace) states in the bubble diagram and are the AdS$_2$ analog of the two-particle production branch cut in flat space. This property is made manifest by the following rewriting of eq.~\eqref{eq:finalBnu} as an infinite sum \cite{Carmi:2018qzm}:
\begin{equation}\label{eq_Bnu_poles}
\tilde{B}_{m^2}(\nu)=\sum_{k=0}^{\infty}
\frac{2 \tilde{\Delta} +2 k-\frac{1}{2}}{\nu ^2+\left(2 \tilde{\Delta} +2 k-\frac{1}{2}\right)^2}\,
\frac{\left(\frac{1}{2}\right)_k \Gamma (k+\tilde{\Delta} )^2 \Gamma \left(k+2 \tilde{\Delta} -\frac{1}{2}\right)}{ \sqrt{4 \pi } \Gamma (k+1) \Gamma \left(k+\tilde{\Delta}  +\frac{1}{2}\right)^2 \Gamma (k+2 \tilde{\Delta} )}\,.
\end{equation}

After a little thought we conclude that we only need to focus on $\ell=0,1$ with $k=0$ from \eqref{eq_Bnu_poles} that give poles at $\De=2$ and $\De=\frac12+\sqrt{\frac14+\tilde\ep}$ respectively. Then solving for the zero of $B_0\left( i\Big(\Delta-\frac{1}{2}\Big)\right)$ perturbatively gives
\es{spectrum2pepd}{
\Delta(\hat{\phi}_1)=2-{\sqrt{5}-1\ov2}\,\tilde \ep+ \mO(\tilde\ep^2)\,.
}
This result validates the assumption we made in sec.~\ref{SubSecD2} that in the $2+\tilde\ep$ expansion $\Delta(\hat{\phi}_1)=2+ \mO(\tilde\ep)$. We leave it for future work to obtain eq.~\eqref{spectrum2pepd} from a perturbative $2+\tilde\ep$ expansion computation (together with its finite $N$ counterpart).

Second, we go back to the $4-\varepsilon$ expansion and show how to solve for the spectrum of operators other than $\hat{\phi}_1$.
  We saw in eq.~\eqref{eq_B0_epsilon} that the contribution from the AdS$_2\times S^q$ bubble diagram to the spectral density results in a $1/\varepsilon$ pole, associated with the large $\ell$ behavior of the sum in eq.~\eqref{eq_MasterEq}. To solve the equation $B_0(\nu)=0$ we thus need to cancel this pole.
  
Similarly to the previous discussion on the limit $q\rightarrow 0$, from eq. \eqref{eq_Bnu_poles} we see that, for $\Delta$ sufficiently larger than one, we can cancel the $1/\varepsilon$ pole in eq.~\eqref{eq_B0_epsilon}, if we go with the  scaling dimension sufficiently close to the one of the double-trace poles. Using that the masses that appear in eq.~\eqref{eq_MasterEq} are  given $\lambda_{\ell}=\ell(\ell+1)+\mO(\varepsilon)$ leading to $\tilde{\Delta}=\ell+1+\mO(\varepsilon)$, we find that the physical scaling dimensions in $d=4-\varepsilon$ are given by:\footnote{Notice that for $p>0$ multiple elements of the sum in eq.~\eqref{eq_MasterEq} diverge for $\Delta$ given by eq.~\eqref{eq_Delta_k}.}
\begin{equation}\label{eq_Delta_k}
\Delta=2+2p+\mO\left(\varepsilon\right)
\,,\qquad
\forall \;p\in\mathds{N}\,.
\end{equation}
The operators in eq.~\eqref{eq_Delta_k} are naturally identified with what we obtain by distributing derivatives among the two fields in the product $\phi_a\phi_a$, such that the final operator has zero transverse spin. In fact, a more careful analysis including the first $\mO(\varepsilon)$ allows to solve for the subleading correction in eq.~\eqref{eq_Delta_k} and distinguish between classically degenerate operators. For $p=0$ we find a unique solution with scaling dimension:
\begin{equation}\label{eq_DeltaSminusN}
\Delta=2+\varepsilon+\mO\left(\varepsilon^2\right)\,.
\end{equation}
The corresponding operator is naturally identified with $\hat{s}_-$ defined in eq.~\eqref{eq_s_def_eps}, and indeed the result eq.~\eqref{eq_DeltaSminusN} is in perfect agreement with the large $N$ limit of the epsilon expansion one in eq.~\eqref{eq_DeltaPM_epsilon}.\footnote{The operator $\hat{s}_+$ discussed in sec.~\ref{SubSecOther} instead does not appear to leading order in $N$ in the bulk to defect OPE of $s$ and it is identified with a double trace of $\hat{\phi}_1$, in agreement with $\Delta(\hat{s}_+)=2\Delta(\hat{\phi}_1)$ in the large $N$ limit (see eq.~\eqref{eq_DeltaPM_epsilon}).} For $p=1$ we instead find two solutions with 
\begin{equation}
\Delta=4+\frac{\varepsilon}{6}\left(1\pm\sqrt{5}\right)+\mO\left(\varepsilon^2\right)\,.
\end{equation}
In general we find $p+1$ distinct solutions which reduce to eq.~\eqref{eq_Delta_k} in the limit $\varepsilon\rightarrow 0$. This nicely agrees with the fact that the number of different defect primary operators of the form $\pd^m\phi_a\pd^{2p-m}\phi_a$ and zero transverse spin is indeed $p+1$. 

\begin{figure}[t!]
\centering
\includegraphics[scale=0.8]{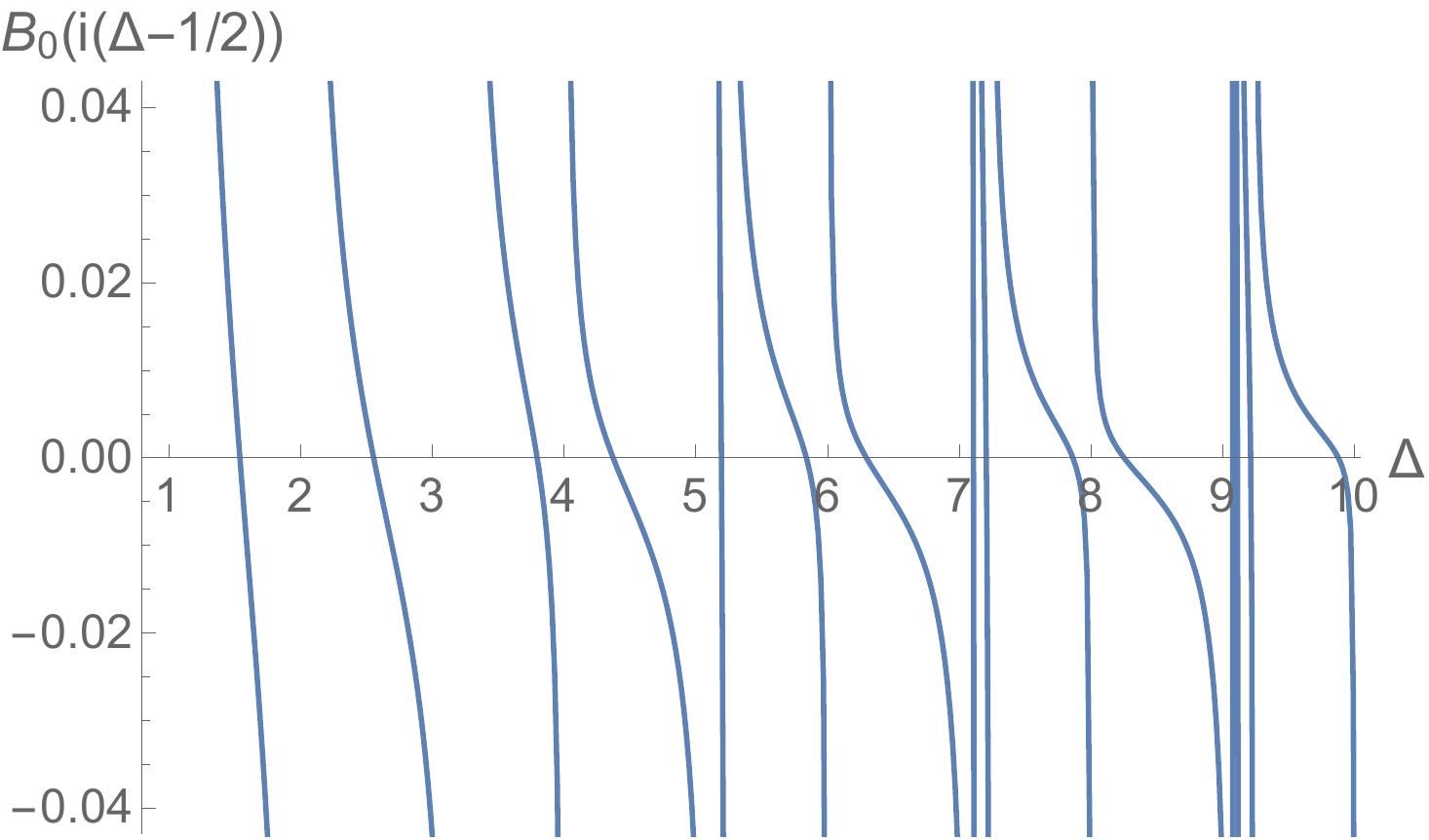}
\caption{Plot of the $\ell=0$ spectral density in eq.~\eqref{eq_MasterEq} for $1<\De\leq 10$.  }
\label{fig:B0Long}
\end{figure}

Our last comment concerns with the spectrum of higher-dimension operators in $d=3$. To this aim we notice that the discussion around eq.~\eqref{eq_Bnu_poles} implies that the physical spectral density \eqref{eq_MasterEq} has poles for $\Delta=2\Delta_{\ell}+2k$ for arbitrary nonnegative integers $\ell$ and $k$, where $\Delta_{\ell}$ is given in eq.~\eqref{tiltchannel}, as well as for $\Delta=1$ due to the first term. In between two subsequent poles we experimentally find that there is a unique zero. For $\Delta\leq 10$ this is demonstrated in fig.~\ref{fig:B0Long}.  Notice that, since $\Delta_{\ell}\approx \ell+\frac12+\mO\left(\ell^{-1}\right)$ for large $\ell$, this observation predicts that the dimensions of heavy operators  are asymptotically given by $\Delta\approx 2\mathbb{N}+1$.  

\section{Outlook}\label{sec:outlook}

In this work we studied the symmetry breaking defect \eqref{WFl} in the critical $O(N)$ models, both in the epsilon and the large $N$ expansion. Our results are compatible within each other and with Monte Carlo simulations. We already summarized them in the Introduction and we will not repeat them here. Instead, we would like to comment here on future research directions.

First of all it is certainly possible to obtain higher order predictions in the $\varepsilon$-expansion and thus more precise determinations for bulk and defect correlators.  As already mentioned, it would be interesting to study the $d\rightarrow 2$ limit of the DCFT for $N\geq 2$ more carefully in the future, in particular within the $d=2+\tilde{\varepsilon}$ expansion \cite{PhysRevLett.36.691,PhysRevB.14.3110,Hikami_1978,Giombi:2020rmc}. This analysis might provide additional data to construct Pad\'e approximations in $2<d\leq 4$.  It should also be possible to extract the scaling dimensions of defect operators with nonzero transverse spin at large $N$ by generalizing the approach in sec. \ref{SubSecNspectrum}. 

One might also approach the problem of determining the infrared data for a symmetry breaking defect in the $O(N)$ model within the numerical conformal bootstrap \cite{Rattazzi:2008pe,Poland:2018epd}, along the lines of \cite{Liendo:2012hy,Gaiotto:2013nva,Gliozzi:2015qsa,Billo:2016cpy}. The results of this paper and the stability of the DCFT$_\text{IR}$ might provide a useful input in this context.\footnote{We thank E.Lauria for discussions on this point.}

The relevance of symmetry breaking defects for Monte Carlo studies also motivates looking at different models. An obvious candidate is provided by a defect of the form \eqref{WFl} in the Gross-Neveu model. This setup has been considered in quantum Monte Carlo simulations in \cite{Assaad:2013xua}.

A related interesting setup is provided by wedges obtained from the intersection of two boundaries. A particularly relevant case for magnets and liquid mixtures arises for the Ising bulk theory when the surfaces are in the normal universality class \cite{Cardy_1983,PhysRevE.60.5163}. This setup was studied in the $\varepsilon$-expansion in \cite{Cardy_1983} and in Monte Carlo in \cite{1998EPJB....5..805P}. Even when both the bulk and the boundary are critical, there is a nontrivial RG flow on the edge analogous to the one that we considered in this paper. At long distances this leads to a fixed point invariant under the one-dimensional $SL(2,\mathbb{R})$ conformal group \cite{Cardy_1983,Antunes:2021qpy}. 
It would be interesting to explore this system further, perhaps developing a large $N$ approach.

Finally we hope that our work will foster further numerical simulations of symmetry breaking defects, allowing for more comparisons with field theoretical predictions in the future. Eventually, it would be interesting to explicitly validate the predictions discussed in this work against experiments.

\section*{Acknowledgements}

We thank A. Antunes,  L. Delacr\'etaz, A. Gimenez-Grau, S. Giombi, T. Hartman, E. Lauria,  L. Rastelli, S. Sachdev,  A. Schwimmer, S. Shao, S. Theisen and B. Van Rees for useful discussions.  We are particularly grateful to Avia Raviv-Moshe for collaboration at the initial stages of this project.
GC is supported by the Simons Foundation (Simons Collaboration on the Non-perturbative Bootstrap) grants 488647 and 397411. MM and ZK are supported in part by the Simons Foundation grant 488657 (Simons Collaboration on the Non-Perturbative Bootstrap) and the BSF grant no. 2018204. 

\appendix

\section{The defect away from bulk criticality}\label{AppAway}

As explained in the introduction, we are particularly interested in systems whose universality class is described by an $O(N)$ $ (\phi_a^2)^2$ field theory in three spacetime dimensions:
\begin{equation}\label{eq_Away_S}
S=\int d^3x\left[\frac{1}{2}(\pd\phi_a)^2+\frac{m^2}{2}\phi_a^2
+\frac{\lambda}{4!}\left(\phi_a^2\right)^2\right]\,.
\end{equation}
We will consider the effects of a defect physically representing a magnetic field localized around a few lattice sites of the quantum lattice system.  Choosing for definiteness the magnetic field in the direction ``$1$'', we can model this via the following defect:
\begin{equation}\label{eq_defect}
D_h=\exp\left(-h\int_D d\tau \,\phi_1\right)\,,
\end{equation}
where the integration is over the worldline of the defect and $h$ represents the magnetic field strength. The defect is at $\mathbf{x}=0$ in coordinates $x^\mu=\left(\tau, \mathbf{x}\right)$ (in Euclidean signature). The defect \eqref{eq_defect} breaks  $O(N)$ explicitly to $O(N-1)$ for $N\geq 2$ and fully breaks the $\mathbb{Z}_2$ symmetry for $N=1$. While our interest in the main text is towards the bulk critical point, in this appendix we provide some general considerations on the effect of the defect for a  bulk in the spontaneously broken phase (ordered phase).

Obviously, in the gapped phase, schematically associated with $m^2> 0$  in the action \eqref{eq_Away_S}, the IR dynamics of the bulk is trivial, and thus so is the defect.

In the ordered phase, which is associated with $m^2<0$ at weak coupling $\lambda/m\ll 1$, the bulk spontaneously breaks the internal group as $O(N)\rightarrow O(N-1)$.   Assuming the absence of additional perturbations, the order parameter naturally aligns in the direction of the magnetic field:
\begin{equation}\label{eq_App_phi1}
\langle\phi_a\rangle\propto- \text{sgn}(h) \delta_a^1\,f\,,
\end{equation}
where $f^2\sim \frac{m^2}{\lambda}$ at weak coupling and the factor $\text{sgn}(h)$ stresses that $\phi_a$ and the magnetic field align so as to minimize the energy. For $N=1$ the system is again gapped in the IR and the situation is qualitatively analogous to the unbroken phase. For $N>1$, instead the theory contains $N-1$ massless Goldstone bosons which give rise to nontrivial IR dynamics.

To analyze the setup quantitatively, we consider the low energy effective theory for the Goldstone modes. To this aim we parametrize the $O(N)/O(N-1)$ coset space in terms of the $N-1$ Goldstone fields $\pi_{\hat{a}}$ as
\begin{equation}\label{eq_coset}
\Omega=\exp\left[i T_{1\hat{a}} \pi_{\hat{a}}\right]\,,
\end{equation}
where $\hat{a}=2,\ldots,N$ and $\{T_{ab}\}$ denote the generators of the $O(N)$ group. In the low energy effective theory we naturally represent the bulk order parameter $\phi_a$ in terms of the Goldstone fields as $\phi_a\sim f\times R_F\left[\Omega\right]_{a1} $ where $R_F[\Omega]$ denotes the fundamental representation of the coset matrix in eq.~\eqref{eq_coset}. 
It is then straightforward to construct the most general effective field theory (EFT) for the system \eqref{eq_Away_S} in the presence of the defect \eqref{eq_defect}. To the lowest order in derivatives it reads:\footnote{Notice that the value of the parameter $h$ in the EFT \eqref{eq_EFT_broken1} does not need to coincide with $h$ in the  UV appearing in eq.~\eqref{eq_defect}. In eq.~\eqref{eq_EFT_broken1} we neglect for simplicity additional $O(N-1)$ invariant defect operators, e.g. $(R_F\left[\Omega\right]_{11})^n$, whose effect is only to renormalize the coupling $\alpha$ in eq.~\eqref{eq_EFT_broken}. }
\begin{equation}\label{eq_EFT_broken1}
S_{EFT}=\int d^3x\frac{f^2}{2}(D_\mu\pi_{\hat{a}})^2-
|h\,f|\int_D d\tau \,R_F\left[\Omega\right]_{11}+\ldots\,,
\end{equation}
where $D_\mu\pi_{\hat{a}}$ denotes the nonlinear covariant derivative obtained from the Maurer-Cartan form associated with the coset \eqref{eq_coset} \cite{Callan:1969sn,Coleman:1969sm}. The absolute value in eq.~\eqref{eq_EFT_broken1} follows from the factor $\text{sgn}(h)$ in eq.~\eqref{eq_App_phi1}. In practice all the nonlinear terms in eq.~\eqref{eq_EFT_broken1} are irrelevant and we can use the linearizations $D_\mu\pi_{\hat{a}}\simeq \pd_\mu\pi_{\hat{a}}$ and $R_F\left[\Omega\right]_{11}\simeq 1-\frac{1}{2}\pi_{\hat{a}}^2$. Therefore,  the IR physics is described by $N-1$ decoupled free fields with a marginal quadratic perturbation on the defect that breaks their emergent shift symmetry, which is the low energy remnant of the action of the broken generators.  Neglecting a field-independent constant contribution, the action reads:
\begin{equation}\label{eq_EFT_broken}
S_{EFT}\simeq  \int d^3x\frac{1}{2}(\pd_\mu\pi_{\hat{a}})^2+\frac{\alpha}{2}\int_D d\tau\, \pi_{\hat{a}}^2\,,\qquad
\alpha\equiv\left|\frac{h}{f}\right|>0\,,
\end{equation}
where we canonically normalized the fields and introduced a positive coupling $\alpha$ on the defect.

It is now simple to perform an RG analysis of the model \eqref{eq_EFT_broken}. Despite the theory being quadratic, the coupling $\alpha$ displays a non-trivial running.  
\begin{figure}[t]
   \centering
     \begin{subfigure}[t]{0.4\textwidth}
         \centering
         \includegraphics[width=0.6\textwidth]{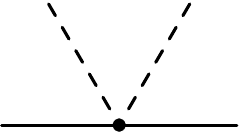}
        \caption{}\label{fig:DiagramsAlpha0}
     \end{subfigure}
    \hspace*{0.2cm}
     \begin{subfigure}[t]{0.4\textwidth}
         \centering
         \includegraphics[width=0.6\textwidth]{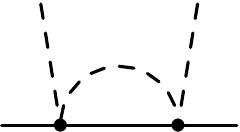}
        \caption{}\label{fig:DiagramsAlpha1}
     \end{subfigure}
\caption{Diagrams contributing to the renormalization of $\alpha$. 
Scalar propagators are represented by dashed lines, the defect is denoted by a solid line.  
An insertion of the vertex $\alpha$ is represented in fig.~\ref{fig:DiagramsAlpha0}. Fig.~\ref{fig:DiagramsAlpha1} is the diagram contributing to its renormalization.}
\label{fig:DiagramsAlpha}
\end{figure}
To one loop order this arises from the diagrams displayed in figure~\ref{fig:DiagramsAlpha}. Working in dimensional regularization,  and denoting the bare coupling with $\alpha_0$, we find the following 1-loop renormalization in the minimal subtraction scheme:
\begin{equation}
\alpha_0=M^{3-d}\left(\alpha+\frac{\delta \alpha}{3-d}\right)\,,\qquad
\delta \alpha=\frac{\alpha^2}{2\pi}\,,
\end{equation}
where $M$ is the sliding scale. We conclude that the $3d$ beta function of $\alpha$ reads:
\begin{equation}\label{eq_beta_gamma_away}
\frac{d \alpha}{d\log M}\equiv \beta_\alpha=\alpha\frac{d \delta \alpha}{d \alpha}-\delta \alpha=\frac{\alpha^2}{2\pi}\,.
\end{equation}
From eq.~\eqref{eq_beta_gamma_away} we thus infer that $\alpha$ is marginally irrelevant and flows to zero. This implies that the defect becomes transparent in the deep IR limit. The absence of a nontrival fixed point for the defect coupling is in beautiful agreement with the results of \cite{Lauria:2020emq}, where the authors argued that a free massless scalar does not admit any nontrivial conformal line defect in $d=3$.\footnote{We thank Edoardo Lauria and Balt Van Rees for discussions on this point.}

The IR triviality of defect does not prevent correlation functions from receiving important logarithmic corrections, as typical in the presence of a marginal paramater (see e.g. \cite{Metlitski:2020cqy}). For instance,  from the beta function \eqref{eq_beta_gamma_away} we easily infer that the one point function $\pi_{\hat{a}}^2$ at far distance from the defect decays as
\begin{equation}\label{eq_pi2_away}
\langle \pi_{\hat{a}}^2(0,\mathbf{x})\rangle
\stackrel{|\mathbf{x}|\rightarrow\infty}{=}-\frac{N-1}{8|\mathbf{x}|\log\left(|\mathbf{x}|M\right)}\,,
\end{equation}
to leading logarithmic accuracy.  Similarly,  we find logarithmic corrections to correlators of the defect operators $\pi_{\hat{a}}$ and $\pi_{\hat{a}}^2$. Using that these have anomalous dimensions, respectively, $\gamma_\pi=\frac{\alpha}{2\pi}$ and $\gamma_{\pi^2}=\frac{\alpha}{\pi}$, a straightforward application of Callan-Symanzik equation unveils the following long distance limit for the renormalized two-point functions:\footnote{A similar result for the two-point function of $\pi_{\hat{a}}$ was previously obtained in \cite{Lauria:2020emq} via explicit resummation of the perturbative series in momentum space.}
\begin{equation}
\langle \pi_{\hat{a}}(\tau,\mathbf{0})\pi_{\hat{b}}(0,\mathbf{0})\rangle
\overset{\tau\rightarrow\infty}{\sim}
\frac{\delta_{\hat{a}\hat{b}}}{|\tau|\log^2\left(|\tau|M\right)}\,,\qquad
\langle \pi_{\hat{a}}^2(\tau,\mathbf{0})\pi_{\hat{b}}^2(0,\mathbf{0})\rangle
\overset{\tau\rightarrow\infty}{\sim}
\frac{2(N-1)}{|\tau|^2\log^4\left(|\tau|M\right)}.
\end{equation}
These results are related to correlators of the order parameter through the (linearized) matching conditions
\begin{equation}
\phi_1\propto f\left(1-\frac{1}{2}\pi_{\hat{a}}^2/f^2\right)\,,\qquad
\phi_{\hat{a}}\propto \pi_{\hat{a}}\,.
\end{equation}
In particular eq. \eqref{eq_pi2_away} determines the rate at which the bulk order parameter approaches the value in the absence of the defect at long distances and might therefore be relevant in Monte Carlo simulations along the lines of \cite{Assaad:2013xua,2017PhRvB..95a4401P}.

We end this section with few comments. First we notice that the result in eq.~\eqref{eq_beta_gamma_away} is one-loop exact since the model \eqref{eq_EFT_broken} is quadratic. We also remark that eq.~\eqref{eq_beta_gamma_away} holds also when $\alpha<0$ in eq.~\eqref{eq_EFT_broken}, corresponding to an anti-alignment between the magnetic field and the order parameter. In this case there is an instability and the bulk reorders to align with the external magnetic field. This instability is reflected in the fact that eq.~\eqref{eq_beta_gamma_away} now implies that the coupling runs towards large values.

\section{One-loop correction to \texorpdfstring{$g$}{g} in the epsilon expansion}\label{App_G_eps}

In eq.~\eqref{eq_g_epsilon1} we are instructed to evaluate the integral
\es{eq_g_bulk_intPRE_App}{
\mathcal{I}&=-\lim_{\ep\to0}\int d^4[\tau]\int d^{2-\ep}x_\bot \,d^2x_\parallel \,\prod_{i=1}^4\frac{1}{4\pi^2\left[x_\bot+x_\parallel-x_\parallel(\tau_i)\right]^{2-\ep}}\,.
}
We can start by performing the $\tau$ integrals. These are factorized, and are evaluated easily:
\es{tauIntegrals}{
\int d\tau \ \frac{1}{\left[x_\bot+x_\parallel-x_\parallel(\tau)\right]^{2-\ep}}=\frac{2\pi}{\le[{z^2+(r+1)^2}\ri]^{1-\ep/2}}\,\, {}_2F_1\le(\frac12,1-\frac{\ep}{2},1;\frac{4r}{{z^2+(r+1)^2}}\ri)\,,
}
where we set the defect radius $R=1$ and introduced the notations $x_\bot^2=z^2,\, x_\parallel^2=r^2$. Plugging this result back into \eqref{eq_g_bulk_intPRE_App} (and performing the trivial integrals over angles in the $x_\bot$ and $x_\parallel$ planes), we end up with a two-dimensional integrals over $z$ and $r$. 
In $4d$ this integral is divergent from the collision of the bulk vertex with the defect operator insertions, i.e. from the regime $z\sim\abs{r-1}\ll1$. Dimensional regularization is supposed to regulate this divergence. The resulting integrand is however too complicated to integrate in $4-\ep$ dimensions.\footnote{The integral is significantly easier to evaluate in a cutoff scheme, where we impose a cutoff in the $x_\parallel$ directions. We obtain a linear perimeter divergence and a finite piece, the latter is half of what we get in \eqref{eq_g_bulk_intPRE_App2} in dimensional regularization. One could of course do the full computation in~\eqref{eq_g_epsilon1} in a cutoff scheme with the same final result for $\log g$, but to do so would necessitate the recomputation of the leading contribution  and of the renormalization of the coupling in the cutoff scheme.}

We can make progress by inventing a simple subtraction that makes the integrand well-behaved for $0\leq \ep< \ep_c$ with some nonzero $ \ep_c$. Then we can safely take the $\ep\to 0$ limit of the subtracted integrand, and are left with the task of evaluating the integral of the subtraction itself in $4-\ep$ dimensions, a much more feasible task. In formulas:
\es{eq_g_bulk_intPRE_App12}{
\mathcal{I}&=\lim_{\ep\to0}\int dz\, dr\ f_\ep(z,r)\\
&=\lim_{\ep\to0}\le[\int dz\, dr\ \le(f_\ep(z,r)-\hat{f}_\ep(z,r)\ri)+ \int dz\, dr\ \hat{f}_\ep(z,r)\ri]\\
&=\int dz\, dr\ \le(f_0(z,r)-\hat{f}_0(z,r)\ri)+\lim_{\ep\to0} \int dz\, dr\ \hat{f}_\ep(z,r)\,.
}
A minimal candidate for $\hat{f}_\ep$ is of the form:
\es{fhatMin}{
\hat{f}^\text{(min)}_\ep(z,r)=A(\ep)\frac{(r-2)z^{1-\ep}}{\le[{z^2+(r-1)^2}\ri]^{2(1-\ep)}}+B(\ep)\frac{z^{1-\ep}}{\le[{z^2+(r-1)^2}\ri]^{3/2(1-\ep)}}\,,
}
where the coefficients $A(\ep)$ and $B(\ep)$ can be obtained from the expansion of \eqref{tauIntegrals}. $\hat{f}^\text{(min)}_\ep$ subtracts the divergences from the integrand, but itself is not integrable for any $\ep$, hence it is not suitable for our purposes. A simple fix for this issue is to modify $\hat{f}^\text{(min)}_\ep$ as
\es{fhat}{
\hat{f}_\ep(z,r)=\frac{\hat{f}^\text{(min)}_\ep(z,r)}{ 1+(r-1)^2}\,.
}
This function has the same singularity structure as $\hat{f}^\text{(min)}_\ep$, but has improved large $r$ behavior,  and is simple enough to allow for the evaluation of its integral in fractional dimensions. We emphasize that our choice of $\hat{f}_\ep$ is by no means unique. The integral $ \int dz\, dr\ \hat{f}_\ep(z,r)$ with the choice of $\hat{f}_\ep$ in eq.~\ref{fhat} converges for $1/3<\ep<2/3$. We evaluate its integral for these $\ep$'s and then analytically continue to $\ep\to0$ with the result
\es{SubtValue}{
\lim_{\ep\to0} \int dz\, dr\ \hat{f}_\ep(z,r)=\frac{3\pi+\log 4 +20}{512 \pi^2}\,.
}
The subtracted integrand in $4d$ reads
\es{SubtValue2}{
\int dz\, dr\ \le(f_0(z,r)-\hat{f}_0(z,r)\ri)&=-\frac{1}{4\pi^2}\int dz\, dr\, \le[\frac{rz}{ \le(r^4+2r^2(z^2-1)+(z^2+1)^2\ri)^2}\ri.\\
&\le.\quad+\frac{(r-2)z}{ 16 \le(r(r-2)+2\ri)\le((r-1)^2+z^2\ri)^2 }\ri]\\
&=-\frac{3\pi+\log 4 -12}{ 512 \pi^2}\,,
}
leading to the final result:
\es{eq_g_bulk_intPRE_App2}{
\mathcal{I}&=\frac{1}{ 16\pi^2}\,,
}
that we quote  in eq.~\eqref{eq_epsilonInt2} in the main text.

\bibliography{Biblio}

\providecommand{\href}[2]{#2}\begingroup\raggedright\begin{thebibliography}{10}

\bibitem{Wilson:1974sk}
K.~G. Wilson, \emph{{Confinement of Quarks}},
  \href{https://doi.org/10.1103/PhysRevD.10.2445}{\emph{Phys. Rev. D}
  {\bfseries 10} (1974) 2445}.

\bibitem{Chang:2018iay}
C.-M. Chang, Y.-H. Lin, S.-H. Shao, Y.~Wang and X.~Yin, \emph{{Topological
  Defect Lines and Renormalization Group Flows in Two Dimensions}},
  \href{https://doi.org/10.1007/JHEP01(2019)026}{\emph{JHEP} {\bfseries 01}
  (2019) 026} [\href{https://arxiv.org/abs/1802.04445}{{\ttfamily
  1802.04445}}].

\bibitem{Affleck:1995ge}
I.~Affleck, \emph{{Conformal field theory approach to the Kondo effect}},
  {\emph{Acta Phys. Polon. B} {\bfseries 26} (1995) 1869}
  [\href{https://arxiv.org/abs/cond-mat/9512099}{{\ttfamily
  cond-mat/9512099}}].

\bibitem{Affleck:1991tk}
I.~Affleck and A.~W.~W. Ludwig, \emph{{Universal noninteger 'ground state
  degeneracy' in critical quantum systems}},
  \href{https://doi.org/10.1103/PhysRevLett.67.161}{\emph{Phys. Rev. Lett.}
  {\bfseries 67} (1991) 161}.

\bibitem{Friedan:2003yc}
D.~Friedan and A.~Konechny, \emph{{On the boundary entropy of one-dimensional
  quantum systems at low temperature}},
  \href{https://doi.org/10.1103/PhysRevLett.93.030402}{\emph{Phys. Rev. Lett.}
  {\bfseries 93} (2004) 030402}
  [\href{https://arxiv.org/abs/hep-th/0312197}{{\ttfamily hep-th/0312197}}].

\bibitem{Casini:2016fgb}
H.~Casini, I.~Salazar~Landea and G.~Torroba, \emph{{The g-theorem and quantum
  information theory}},
  \href{https://doi.org/10.1007/JHEP10(2016)140}{\emph{JHEP} {\bfseries 10}
  (2016) 140} [\href{https://arxiv.org/abs/1607.00390}{{\ttfamily
  1607.00390}}].

\bibitem{Cuomo:2021rkm}
G.~Cuomo, Z.~Komargodski and A.~Raviv-Moshe, \emph{{Renormalization Group Flows
  on Line Defects}},  \href{https://arxiv.org/abs/2108.01117}{{\ttfamily
  2108.01117}}.

\bibitem{Beccaria:2017rbe}
M.~Beccaria, S.~Giombi and A.~Tseytlin, \emph{{Non-supersymmetric Wilson loop
  in $ \mathcal{N} $ = 4 SYM and defect 1d CFT}},
  \href{https://doi.org/10.1007/JHEP03(2018)131}{\emph{JHEP} {\bfseries 03}
  (2018) 131} [\href{https://arxiv.org/abs/1712.06874}{{\ttfamily
  1712.06874}}].

\bibitem{Kobayashi:2018lil}
N.~Kobayashi, T.~Nishioka, Y.~Sato and K.~Watanabe, \emph{{Towards a
  $C$-theorem in defect CFT}},
  \href{https://doi.org/10.1007/JHEP01(2019)039}{\emph{JHEP} {\bfseries 01}
  (2019) 039} [\href{https://arxiv.org/abs/1810.06995}{{\ttfamily
  1810.06995}}].

\bibitem{Assaad:2013xua}
F.~F. Assaad and I.~F. Herbut, \emph{{Pinning the order: the nature of quantum
  criticality in the Hubbard model on honeycomb lattice}},
  \href{https://doi.org/10.1103/PhysRevX.3.031010}{\emph{Phys. Rev. X}
  {\bfseries 3} (2013) 031010}
  [\href{https://arxiv.org/abs/1304.6340}{{\ttfamily 1304.6340}}].

\bibitem{2017PhRvB..95a4401P}
F.~{Parisen Toldin}, F.~F. {Assaad} and S.~{Wessel}, \emph{{Critical behavior
  in the presence of an order-parameter pinning field}},
  \href{https://doi.org/10.1103/PhysRevB.95.014401}{\emph{Phys. Rev. B}
  {\bfseries 95} (2017) 014401}
  [\href{https://arxiv.org/abs/1607.04270}{{\ttfamily 1607.04270}}].

\bibitem{sachdev2008quantum}
S.~Sachdev, \emph{Quantum magnetism and criticality},
  \href{https://doi.org/10.1038/nphys894}{\emph{Nature Physics} {\bfseries 4}
  (2008) 173} [\href{https://arxiv.org/abs/0711.3015}{{\ttfamily 0711.3015}}].

\bibitem{LAW2001159}
B.~M. Law, \emph{Wetting, adsorption and surface critical phenomena},
  \href{https://doi.org/https://doi.org/10.1016/S0079-6816(00)00025-3}{\emph{Progress
  in Surface Science} {\bfseries 66} (2001) 159}.

\bibitem{2003svcm.book..237F}
M.~E. {Fisher} and P.-G. {de Gennes}, \emph{{Ph\'enom\`enes aux parois dans un
  m\'elange binaire critique}}, pp.~237--241.
\newblock World Scientific, 2003.
\newblock 10.1142/9789812564849\_0025.

\bibitem{Ebadi:2020ldi}
S.~Ebadi et~al., \emph{{Quantum phases of matter on a 256-atom programmable
  quantum simulator}},
  \href{https://doi.org/10.1038/s41586-021-03582-4}{\emph{Nature} {\bfseries
  595} (2021) 227} [\href{https://arxiv.org/abs/2012.12281}{{\ttfamily
  2012.12281}}].

\bibitem{2014arXiv1412.3449A}
A.~{Allais}, \emph{{Magnetic defect line in a critical Ising bath}},
  \href{https://arxiv.org/abs/1412.3449}{{\ttfamily 1412.3449}}.

\bibitem{PhysRevLett.84.2180}
A.~Hanke, \emph{Critical adsorption on defects in ising magnets and binary
  alloys}, \href{https://doi.org/10.1103/PhysRevLett.84.2180}{\emph{Phys. Rev.
  Lett.} {\bfseries 84} (2000) 2180}.

\bibitem{Allais:2014fqa}
A.~Allais and S.~Sachdev, \emph{{Spectral function of a localized fermion
  coupled to the Wilson-Fisher conformal field theory}},
  \href{https://doi.org/10.1103/PhysRevB.90.035131}{\emph{Phys. Rev. B}
  {\bfseries 90} (2014) 035131}
  [\href{https://arxiv.org/abs/1406.3022}{{\ttfamily 1406.3022}}].

\bibitem{sachdev1999quantum}
S.~Sachdev, C.~Buragohain and M.~Vojta, \emph{Quantum impurity in a nearly
  critical two-dimensional antiferromagnet}, {\emph{Science} {\bfseries 286}
  (1999) 2479} [\href{https://arxiv.org/abs/cond-mat/0004156}{{\ttfamily
  cond-mat/0004156}}].

\bibitem{vojta2000quantum}
M.~Vojta, C.~Buragohain and S.~Sachdev, \emph{Quantum impurity dynamics in
  two-dimensional antiferromagnets and superconductors},
  \href{https://doi.org/10.1103/PhysRevB.61.15152}{\emph{Physical Review B}
  {\bfseries 61} (2000) 15152}
  [\href{https://arxiv.org/abs/cond-mat/9912020}{{\ttfamily
  cond-mat/9912020}}].

\bibitem{PhysRevB.61.4041}
A.~M. Sengupta, \emph{Spin in a fluctuating field: The bose(+fermi) kondo
  models}, \href{https://doi.org/10.1103/PhysRevB.61.4041}{\emph{Phys. Rev. B}
  {\bfseries 61} (2000) 4041}
  [\href{https://arxiv.org/abs/cond-mat/9707316}{{\ttfamily
  cond-mat/9707316}}].

\bibitem{Sachdev:2001ky}
S.~Sachdev, \emph{{Static hole in a critical antiferromagnet: Field theoretic
  renormalization group}},
  \href{https://doi.org/10.1016/S0921-4534(01)00198-8}{\emph{Physica C}
  {\bfseries 357} (2001) 78}
  [\href{https://arxiv.org/abs/cond-mat/0011233}{{\ttfamily
  cond-mat/0011233}}].

\bibitem{Sachdev:2003yk}
S.~Sachdev and M.~Vojta, \emph{{Quantum impurity in an antiferromagnet:
  Nonlinear sigma model theory}},
  \href{https://doi.org/10.1103/PhysRevB.68.064419}{\emph{Phys. Rev. B}
  {\bfseries 68} (2003) 064419}
  [\href{https://arxiv.org/abs/cond-mat/0303001}{{\ttfamily
  cond-mat/0303001}}].

\bibitem{PhysRevLett.96.036601}
S.~Florens, L.~Fritz and M.~Vojta, \emph{Kondo effect in bosonic spin liquids},
  \href{https://doi.org/10.1103/PhysRevLett.96.036601}{\emph{Phys. Rev. Lett.}
  {\bfseries 96} (2006) 036601}
  [\href{https://arxiv.org/abs/cond-mat/0507188}{{\ttfamily
  cond-mat/0507188}}].

\bibitem{PhysRevB.75.224420}
S.~Florens, L.~Fritz and M.~Vojta, \emph{Boundary quantum criticality in models
  of magnetic impurities coupled to bosonic baths},
  \href{https://doi.org/10.1103/PhysRevB.75.224420}{\emph{Phys. Rev. B}
  {\bfseries 75} (2007) 224420}
  [\href{https://arxiv.org/abs/cond-mat/0703609}{{\ttfamily
  cond-mat/0703609}}].

\bibitem{Liu:2021nck}
S.~Liu, H.~Shapourian, A.~Vishwanath and M.~A. Metlitski, \emph{{Magnetic
  impurities at quantum critical points: Large-N expansion and connections to
  symmetry-protected topological states}},
  \href{https://doi.org/10.1103/PhysRevB.104.104201}{\emph{Phys. Rev. B}
  {\bfseries 104} (2021) 104201}
  [\href{https://arxiv.org/abs/2104.15026}{{\ttfamily 2104.15026}}].

\bibitem{Billo:2013jda}
M.~Bill\'o, M.~Caselle, D.~Gaiotto, F.~Gliozzi, M.~Meineri and R.~Pellegrini,
  \emph{{Line defects in the 3d Ising model}},
  \href{https://doi.org/10.1007/JHEP07(2013)055}{\emph{JHEP} {\bfseries 07}
  (2013) 055} [\href{https://arxiv.org/abs/1304.4110}{{\ttfamily 1304.4110}}].

\bibitem{Gaiotto:2013nva}
D.~Gaiotto, D.~Mazac and M.~F. Paulos, \emph{{Bootstrapping the 3d Ising twist
  defect}}, \href{https://doi.org/10.1007/JHEP03(2014)100}{\emph{JHEP}
  {\bfseries 03} (2014) 100} [\href{https://arxiv.org/abs/1310.5078}{{\ttfamily
  1310.5078}}].

\bibitem{Bianchi:2021snj}
L.~Bianchi, A.~Chalabi, V.~Proch\'azka, B.~Robinson and J.~Sisti,
  \emph{{Monodromy defects in free field theories}},
  \href{https://doi.org/10.1007/JHEP08(2021)013}{\emph{JHEP} {\bfseries 08}
  (2021) 013} [\href{https://arxiv.org/abs/2104.01220}{{\ttfamily
  2104.01220}}].

\bibitem{Giombi:2021uae}
S.~Giombi, E.~Helfenberger, Z.~Ji and H.~Khanchandani, \emph{{Monodromy Defects
  from Hyperbolic Space}},  \href{https://arxiv.org/abs/2102.11815}{{\ttfamily
  2102.11815}}.

\bibitem{Gimenez-Grau:2021wiv}
A.~Gimenez-Grau and P.~Liendo, \emph{{Bootstrapping Monodromy Defects in the
  Wess-Zumino Model}},  \href{https://arxiv.org/abs/2108.05107}{{\ttfamily
  2108.05107}}.

\bibitem{Soderberg:2021kne}
A.~S\"oderberg, \emph{{Fusion of conformal defects in four dimensions}},
  \href{https://doi.org/10.1007/JHEP04(2021)087}{\emph{JHEP} {\bfseries 04}
  (2021) 087} [\href{https://arxiv.org/abs/2102.00718}{{\ttfamily
  2102.00718}}].

\bibitem{Rodriguez-Gomez:2022gbz}
D.~Rodriguez-Gomez, \emph{{A Scaling Limit for Line and Surface Defects}},
  \href{https://arxiv.org/abs/2202.03471}{{\ttfamily 2202.03471}}.

\bibitem{Fitzpatrick:2010zm}
A.~L. Fitzpatrick, E.~Katz, D.~Poland and D.~Simmons-Duffin, \emph{{Effective
  Conformal Theory and the Flat-Space Limit of AdS}},
  \href{https://doi.org/10.1007/JHEP07(2011)023}{\emph{JHEP} {\bfseries 07}
  (2011) 023} [\href{https://arxiv.org/abs/1007.2412}{{\ttfamily 1007.2412}}].

\bibitem{Penedones:2010ue}
J.~Penedones, \emph{{Writing CFT correlation functions as AdS scattering
  amplitudes}}, \href{https://doi.org/10.1007/JHEP03(2011)025}{\emph{JHEP}
  {\bfseries 03} (2011) 025} [\href{https://arxiv.org/abs/1011.1485}{{\ttfamily
  1011.1485}}].

\bibitem{Fitzpatrick:2011hu}
A.~L. Fitzpatrick and J.~Kaplan, \emph{{Analyticity and the Holographic
  S-Matrix}}, \href{https://doi.org/10.1007/JHEP10(2012)127}{\emph{JHEP}
  {\bfseries 10} (2012) 127} [\href{https://arxiv.org/abs/1111.6972}{{\ttfamily
  1111.6972}}].

\bibitem{Giombi:2017hpr}
S.~Giombi, C.~Sleight and M.~Taronna, \emph{{Spinning AdS Loop Diagrams: Two
  Point Functions}}, \href{https://doi.org/10.1007/JHEP06(2018)030}{\emph{JHEP}
  {\bfseries 06} (2018) 030}
  [\href{https://arxiv.org/abs/1708.08404}{{\ttfamily 1708.08404}}].

\bibitem{Carmi:2018qzm}
D.~Carmi, L.~Di~Pietro and S.~Komatsu, \emph{{A Study of Quantum Field Theories
  in AdS at Finite Coupling}},
  \href{https://doi.org/10.1007/JHEP01(2019)200}{\emph{JHEP} {\bfseries 01}
  (2019) 200} [\href{https://arxiv.org/abs/1810.04185}{{\ttfamily
  1810.04185}}].

\bibitem{Carmi:2019ocp}
D.~Carmi, \emph{{Loops in AdS: From the Spectral Representation to Position
  Space}}, \href{https://doi.org/10.1007/JHEP06(2020)049}{\emph{JHEP}
  {\bfseries 06} (2020) 049}
  [\href{https://arxiv.org/abs/1910.14340}{{\ttfamily 1910.14340}}].

\bibitem{Kapustin:2005py}
A.~Kapustin, \emph{{Wilson-'t Hooft operators in four-dimensional gauge
  theories and S-duality}},
  \href{https://doi.org/10.1103/PhysRevD.74.025005}{\emph{Phys. Rev. D}
  {\bfseries 74} (2006) 025005}
  [\href{https://arxiv.org/abs/hep-th/0501015}{{\ttfamily hep-th/0501015}}].

\bibitem{Lauria:2020emq}
E.~Lauria, P.~Liendo, B.~C. Van~Rees and X.~Zhao, \emph{{Line and surface
  defects for the free scalar field}},
  \href{https://doi.org/10.1007/JHEP01(2021)060}{\emph{JHEP} {\bfseries 01}
  (2021) 060} [\href{https://arxiv.org/abs/2005.02413}{{\ttfamily
  2005.02413}}].

\bibitem{Nishioka:2021uef}
T.~Nishioka and Y.~Sato, \emph{{Free energy and defect $C$-theorem in free
  scalar theory}}, \href{https://doi.org/10.1007/JHEP05(2021)074}{\emph{JHEP}
  {\bfseries 05} (2021) 074}
  [\href{https://arxiv.org/abs/2101.02399}{{\ttfamily 2101.02399}}].

\bibitem{Gomis:2015yaa}
J.~Gomis, P.-S. Hsin, Z.~Komargodski, A.~Schwimmer, N.~Seiberg and S.~Theisen,
  \emph{{Anomalies, Conformal Manifolds, and Spheres}},
  \href{https://doi.org/10.1007/JHEP03(2016)022}{\emph{JHEP} {\bfseries 03}
  (2016) 022} [\href{https://arxiv.org/abs/1509.08511}{{\ttfamily
  1509.08511}}].

\bibitem{Schwimmer:2018hdl}
A.~Schwimmer and S.~Theisen, \emph{{Moduli Anomalies and Local Terms in the
  Operator Product Expansion}},
  \href{https://doi.org/10.1007/JHEP07(2018)110}{\emph{JHEP} {\bfseries 07}
  (2018) 110} [\href{https://arxiv.org/abs/1805.04202}{{\ttfamily
  1805.04202}}].

\bibitem{Schwimmer:2019efk}
A.~Schwimmer and S.~Theisen, \emph{{Osborn Equation and Irrelevant Operators}},
  \href{https://doi.org/10.1088/1742-5468/ab3284}{\emph{J. Stat. Mech.}
  {\bfseries 1908} (2019) 084011}
  [\href{https://arxiv.org/abs/1902.04473}{{\ttfamily 1902.04473}}].

\bibitem{Friedan:2012jk}
D.~Friedan, A.~Konechny and C.~Schmidt-Colinet, \emph{{Lower bound on the
  entropy of boundaries and junctions in 1+1d quantum critical systems}},
  \href{https://doi.org/10.1103/PhysRevLett.109.140401}{\emph{Phys. Rev. Lett.}
  {\bfseries 109} (2012) 140401}
  [\href{https://arxiv.org/abs/1206.5395}{{\ttfamily 1206.5395}}].

\bibitem{Collier:2021ngi}
S.~Collier, D.~Mazac and Y.~Wang, \emph{{Bootstrapping Boundaries and Branes}},
   \href{https://arxiv.org/abs/2112.00750}{{\ttfamily 2112.00750}}.

\bibitem{Kleinert:2001ax}
H.~Kleinert and V.~Schulte-Frohlinde, \emph{{Critical properties of
  $\phi^4$-theories}}. World Scientific, 2001,
  \href{https://doi.org/10.1142/4733}{10.1142/4733}.

\bibitem{Collins:1984xc}
J.~C. Collins, \emph{{Renormalization}: {An Introduction to Renormalization,
  The Renormalization Group, and the Operator Product Expansion}}, vol.~26 of
  \emph{Cambridge Monographs on Mathematical Physics}. Cambridge University
  Press, Cambridge, 1986,
  \href{https://doi.org/10.1017/CBO9780511622656}{10.1017/CBO9780511622656}.

\bibitem{Pelissetto:2000ek}
A.~Pelissetto and E.~Vicari, \emph{{Critical phenomena and renormalization
  group theory}},
  \href{https://doi.org/10.1016/S0370-1573(02)00219-3}{\emph{Phys. Rept.}
  {\bfseries 368} (2002) 549}
  [\href{https://arxiv.org/abs/cond-mat/0012164}{{\ttfamily
  cond-mat/0012164}}].

\bibitem{Poland:2018epd}
D.~Poland, S.~Rychkov and A.~Vichi, \emph{{The Conformal Bootstrap: Theory,
  Numerical Techniques, and Applications}},
  \href{https://doi.org/10.1103/RevModPhys.91.015002}{\emph{Rev. Mod. Phys.}
  {\bfseries 91} (2019) 015002}
  [\href{https://arxiv.org/abs/1805.04405}{{\ttfamily 1805.04405}}].

\bibitem{Cuomo:2022xgw}
G.~Cuomo, Z.~Komargodski, M.~Mezei and A.~Raviv-Moshe, \emph{{Spin Impurities,
  Wilson Lines and Semiclassics}},
  \href{https://arxiv.org/abs/2202.00040}{{\ttfamily 2202.00040}}.

\bibitem{Gubser:2008yx}
S.~S. Gubser, A.~Nellore, S.~S. Pufu and F.~D. Rocha, \emph{{Thermodynamics and
  bulk viscosity of approximate black hole duals to finite temperature quantum
  chromodynamics}},
  \href{https://doi.org/10.1103/PhysRevLett.101.131601}{\emph{Phys. Rev. Lett.}
  {\bfseries 101} (2008) 131601}
  [\href{https://arxiv.org/abs/0804.1950}{{\ttfamily 0804.1950}}].

\bibitem{Cuomo:2021cnb}
G.~Cuomo, M.~Mezei and A.~Raviv-Moshe, \emph{{Boundary conformal field theory
  at large charge}}, \href{https://doi.org/10.1007/JHEP10(2021)143}{\emph{JHEP}
  {\bfseries 10} (2021) 143}
  [\href{https://arxiv.org/abs/2108.06579}{{\ttfamily 2108.06579}}].

\bibitem{PhysRevB.25.331}
A.~C. Brown, \emph{Critical properties of an altered ising model},
  \href{https://doi.org/10.1103/PhysRevB.25.331}{\emph{Phys. Rev. B} {\bfseries
  25} (1982) 331}.

\bibitem{Oshikawa:1996ww}
M.~Oshikawa and I.~Affleck, \emph{{Defect lines in the Ising model and boundary
  states on orbifolds}},
  \href{https://doi.org/10.1103/PhysRevLett.77.2604}{\emph{Phys. Rev. Lett.}
  {\bfseries 77} (1996) 2604}
  [\href{https://arxiv.org/abs/hep-th/9606177}{{\ttfamily hep-th/9606177}}].

\bibitem{Oshikawa:1996dj}
M.~Oshikawa and I.~Affleck, \emph{{Boundary conformal field theory approach to
  the critical two-dimensional Ising model with a defect line}},
  \href{https://doi.org/10.1016/S0550-3213(97)00219-8}{\emph{Nucl. Phys. B}
  {\bfseries 495} (1997) 533}
  [\href{https://arxiv.org/abs/cond-mat/9612187}{{\ttfamily
  cond-mat/9612187}}].

\bibitem{Cardy:1989ir}
J.~L. Cardy, \emph{{Boundary Conditions, Fusion Rules and the Verlinde
  Formula}}, \href{https://doi.org/10.1016/0550-3213(89)90521-X}{\emph{Nucl.
  Phys. B} {\bfseries 324} (1989) 581}.

\bibitem{Ishibashi:1988kg}
N.~Ishibashi, \emph{{The Boundary and Crosscap States in Conformal Field
  Theories}}, \href{https://doi.org/10.1142/S0217732389000320}{\emph{Mod. Phys.
  Lett. A} {\bfseries 4} (1989) 251}.

\bibitem{Cardy:2004hm}
J.~L. Cardy, \emph{{Boundary conformal field theory}},
  \href{https://arxiv.org/abs/hep-th/0411189}{{\ttfamily hep-th/0411189}}.

\bibitem{Giombi:2020rmc}
S.~Giombi and H.~Khanchandani, \emph{{CFT in AdS and boundary RG flows}},
  \href{https://doi.org/10.1007/JHEP11(2020)118}{\emph{JHEP} {\bfseries 11}
  (2020) 118} [\href{https://arxiv.org/abs/2007.04955}{{\ttfamily
  2007.04955}}].

\bibitem{Wallace_1975}
D.~J. Wallace and R.~K.~P. Zia, \emph{Harmonic perturbations of generalized
  heisenberg spin systems},
  \href{https://doi.org/10.1088/0022-3719/8/6/014}{\emph{Journal of Physics C:
  Solid State Physics} {\bfseries 8} (1975) 839}.

\bibitem{Calabrese:2002bm}
P.~Calabrese, A.~Pelissetto and E.~Vicari, \emph{{Multicritical phenomena in
  O(n(1)) + O(n(2)) symmetric theories}},
  \href{https://doi.org/10.1103/PhysRevB.67.054505}{\emph{Phys. Rev. B}
  {\bfseries 67} (2003) 054505}
  [\href{https://arxiv.org/abs/cond-mat/0209580}{{\ttfamily
  cond-mat/0209580}}].

\bibitem{Billo:2016cpy}
M.~Bill\`o, V.~Gon\c{c}alves, E.~Lauria and M.~Meineri, \emph{{Defects in
  conformal field theory}},
  \href{https://doi.org/10.1007/JHEP04(2016)091}{\emph{JHEP} {\bfseries 04}
  (2016) 091} [\href{https://arxiv.org/abs/1601.02883}{{\ttfamily
  1601.02883}}].

\bibitem{Liendo:2012hy}
P.~Liendo, L.~Rastelli and B.~C. van Rees, \emph{{The Bootstrap Program for
  Boundary CFT$_d$}},
  \href{https://doi.org/10.1007/JHEP07(2013)113}{\emph{JHEP} {\bfseries 07}
  (2013) 113} [\href{https://arxiv.org/abs/1210.4258}{{\ttfamily 1210.4258}}].

\bibitem{Gliozzi:2015qsa}
F.~Gliozzi, P.~Liendo, M.~Meineri and A.~Rago, \emph{{Boundary and Interface
  CFTs from the Conformal Bootstrap}},
  \href{https://doi.org/10.1007/JHEP05(2015)036}{\emph{JHEP} {\bfseries 05}
  (2015) 036} [\href{https://arxiv.org/abs/1502.07217}{{\ttfamily
  1502.07217}}].

\bibitem{Moshe:2003xn}
M.~Moshe and J.~Zinn-Justin, \emph{{Quantum field theory in the large N limit:
  A Review}}, \href{https://doi.org/10.1016/S0370-1573(03)00263-1}{\emph{Phys.
  Rept.} {\bfseries 385} (2003) 69}
  [\href{https://arxiv.org/abs/hep-th/0306133}{{\ttfamily hep-th/0306133}}].

\bibitem{PhysRevLett.38.735}
A.~J. Bray and M.~A. Moore, \emph{Critical behavior of a semi-infinite system:
  $n$-vector model in the large-$n$ limit},
  \href{https://doi.org/10.1103/PhysRevLett.38.735}{\emph{Phys. Rev. Lett.}
  {\bfseries 38} (1977) 735}.

\bibitem{Ohno:1983lma}
K.~Ohno and Y.~Okabe, \emph{{The 1/N Expansion for the N Vector Model in the
  SemiInfinite Space}}, \href{https://doi.org/10.1143/PTP.70.1226}{\emph{Prog.
  Theor. Phys.} {\bfseries 70} (1983) 1226}.

\bibitem{McAvity:1995zd}
D.~M. McAvity and H.~Osborn, \emph{{Conformal field theories near a boundary in
  general dimensions}},
  \href{https://doi.org/10.1016/0550-3213(95)00476-9}{\emph{Nucl. Phys. B}
  {\bfseries 455} (1995) 522}
  [\href{https://arxiv.org/abs/cond-mat/9505127}{{\ttfamily
  cond-mat/9505127}}].

\bibitem{Herzog:2020lel}
C.~P. Herzog and N.~Kobayashi, \emph{{The $O(N)$ model with $\phi^6$ potential
  in ${\mathbb R}^2 \times {\mathbb R}^+$}},
  \href{https://doi.org/10.1007/JHEP09(2020)126}{\emph{JHEP} {\bfseries 09}
  (2020) 126} [\href{https://arxiv.org/abs/2005.07863}{{\ttfamily
  2005.07863}}].

\bibitem{Metlitski:2020cqy}
M.~A. Metlitski, \emph{{Boundary criticality of the O(N) model in d = 3
  critically revisited}},  \href{https://arxiv.org/abs/2009.05119}{{\ttfamily
  2009.05119}}.

\bibitem{tsvelick1985exact}
A.~Tsvelick and P.~Wiegmann, \emph{Exact solution of the multichannel kondo
  problem, scaling, and integrability},
  \href{https://doi.org/10.1007/BF01017853}{\emph{Journal of Statistical
  Physics} {\bfseries 38} (1985) 125}.

\bibitem{PhysRevB.46.10812}
V.~J. Emery and S.~Kivelson, \emph{Mapping of the two-channel kondo problem to
  a resonant-level model},
  \href{https://doi.org/10.1103/PhysRevB.46.10812}{\emph{Phys. Rev. B}
  {\bfseries 46} (1992) 10812}.

\bibitem{PhysRevB.58.3794}
O.~Parcollet, A.~Georges, G.~Kotliar and A.~Sengupta, \emph{Overscreened
  multichannel $\mathrm{SU}(n)$ kondo model: Large-$n$ solution and conformal
  field theory}, \href{https://doi.org/10.1103/PhysRevB.58.3794}{\emph{Phys.
  Rev. B} {\bfseries 58} (1998) 3794}
  [\href{https://arxiv.org/abs/cond-mat/9711192}{{\ttfamily
  cond-mat/9711192}}].

\bibitem{Herzog:2020bqw}
C.~P. Herzog and A.~Shrestha, \emph{{Two point functions in defect CFTs}},
  \href{https://doi.org/10.1007/JHEP04(2021)226}{\emph{JHEP} {\bfseries 04}
  (2021) 226} [\href{https://arxiv.org/abs/2010.04995}{{\ttfamily
  2010.04995}}].

\bibitem{Casini:2011kv}
H.~Casini, M.~Huerta and R.~C. Myers, \emph{{Towards a derivation of
  holographic entanglement entropy}},
  \href{https://doi.org/10.1007/JHEP05(2011)036}{\emph{JHEP} {\bfseries 05}
  (2011) 036} [\href{https://arxiv.org/abs/1102.0440}{{\ttfamily 1102.0440}}].

\bibitem{Chester:2015wao}
S.~M. Chester, M.~Mezei, S.~S. Pufu and I.~Yaakov, \emph{{Monopole operators
  from the $4-\epsilon$ expansion}},
  \href{https://doi.org/10.1007/JHEP12(2016)015}{\emph{JHEP} {\bfseries 12}
  (2016) 015} [\href{https://arxiv.org/abs/1511.07108}{{\ttfamily
  1511.07108}}].

\bibitem{Klebanov:2011uf}
I.~R. Klebanov, S.~S. Pufu, S.~Sachdev and B.~R. Safdi, \emph{{Renyi Entropies
  for Free Field Theories}},
  \href{https://doi.org/10.1007/JHEP04(2012)074}{\emph{JHEP} {\bfseries 04}
  (2012) 074} [\href{https://arxiv.org/abs/1111.6290}{{\ttfamily 1111.6290}}].

\bibitem{Cornalba:2007fs}
L.~Cornalba, \emph{{Eikonal methods in AdS/CFT: Regge theory and multi-reggeon
  exchange}},  \href{https://arxiv.org/abs/0710.5480}{{\ttfamily 0710.5480}}.

\bibitem{Cornalba:2008qf}
L.~Cornalba, M.~S. Costa and J.~Penedones, \emph{{Eikonal Methods in AdS/CFT:
  BFKL Pomeron at Weak Coupling}},
  \href{https://doi.org/10.1088/1126-6708/2008/06/048}{\emph{JHEP} {\bfseries
  06} (2008) 048} [\href{https://arxiv.org/abs/0801.3002}{{\ttfamily
  0801.3002}}].

\bibitem{PhysRevLett.36.691}
E.~Br\'ezin and J.~Zinn-Justin, \emph{Renormalization of the nonlinear
  $\ensuremath{\sigma}$ model in $2+\ensuremath{\epsilon}$
  dimensions---application to the heisenberg ferromagnets},
  \href{https://doi.org/10.1103/PhysRevLett.36.691}{\emph{Phys. Rev. Lett.}
  {\bfseries 36} (1976) 691}.

\bibitem{PhysRevB.14.3110}
E.~Br\'ezin and J.~Zinn-Justin, \emph{Spontaneous breakdown of continuous
  symmetries near two dimensions},
  \href{https://doi.org/10.1103/PhysRevB.14.3110}{\emph{Phys. Rev. B}
  {\bfseries 14} (1976) 3110}.

\bibitem{Hikami_1978}
S.~Hikami and E.~Brezin, \emph{Three-loop calculations in the two-dimensional
  non-linear $\sigma$ model},
  \href{https://doi.org/10.1088/0305-4470/11/6/015}{\emph{Journal of Physics A:
  Mathematical and General} {\bfseries 11} (1978) 1141}.

\bibitem{Rattazzi:2008pe}
R.~Rattazzi, V.~S. Rychkov, E.~Tonni and A.~Vichi, \emph{{Bounding scalar
  operator dimensions in 4D CFT}},
  \href{https://doi.org/10.1088/1126-6708/2008/12/031}{\emph{JHEP} {\bfseries
  12} (2008) 031} [\href{https://arxiv.org/abs/0807.0004}{{\ttfamily
  0807.0004}}].

\bibitem{Cardy_1983}
J.~L. Cardy, \emph{Critical behaviour at an edge},
  \href{https://doi.org/10.1088/0305-4470/16/15/026}{\emph{Journal of Physics
  A: Mathematical and General} {\bfseries 16} (1983) 3617}.

\bibitem{PhysRevE.60.5163}
A.~Hanke, M.~Krech, F.~Schlesener and S.~Dietrich, \emph{Critical adsorption
  near edges}, \href{https://doi.org/10.1103/PhysRevE.60.5163}{\emph{Phys. Rev.
  E} {\bfseries 60} (1999) 5163}.

\bibitem{1998EPJB....5..805P}
M.~{Pleimling} and W.~{Selke}, \emph{{Critical phenomena at edges and
  corners}}, \href{https://doi.org/10.1007/s100510050506}{\emph{European
  Physical Journal B} {\bfseries 5} (1998) 805}
  [\href{https://arxiv.org/abs/cond-mat/9801320}{{\ttfamily
  cond-mat/9801320}}].

\bibitem{Antunes:2021qpy}
A.~Antunes, \emph{{Conformal bootstrap near the edge}},
  \href{https://doi.org/10.1007/JHEP10(2021)057}{\emph{JHEP} {\bfseries 10}
  (2021) 057} [\href{https://arxiv.org/abs/2103.03132}{{\ttfamily
  2103.03132}}].

\bibitem{Callan:1969sn}
C.~G. Callan, Jr., S.~R. Coleman, J.~Wess and B.~Zumino, \emph{{Structure of
  phenomenological Lagrangians. 2.}},
  \href{https://doi.org/10.1103/PhysRev.177.2247}{\emph{Phys. Rev.} {\bfseries
  177} (1969) 2247}.

\bibitem{Coleman:1969sm}
S.~R. Coleman, J.~Wess and B.~Zumino, \emph{{Structure of phenomenological
  Lagrangians. 1.}},
  \href{https://doi.org/10.1103/PhysRev.177.2239}{\emph{Phys. Rev.} {\bfseries
  177} (1969) 2239}.

\end{thebibliography}\endgroup
	\bibliographystyle{JHEP.bst}

\end{document}